\documentclass[preprint2]{aastex}
\makeatletter
\renewcommand\appendix{%
 \par 
 \setcounter{section}{\z@}%
 \setcounter{subsection}{\z@}%
 \setcounter{equation}{\z@}%
 \def\thesection{\Alph{section}}%
 \def\theequation{%
  \thesection\arabic{equation}%
 }%
 \@addtoreset{equation}{section}%
 \appendix@figtab@defs 
 \def\section{%
  \@startsection 
   {section}{1}{\z@}%
   {5ex\@plus.5ex}{1ex\@plus.2ex}{\normalsize\bfseries}%
   }%
}%
\makeatother

\begin{document}

\title{Mu- and Tau-Neutrino Spectra Formation in Supernovae}

\author{Georg~G.~Raffelt} 
\affil{Max-Planck-Institut f\"ur Physik 
(Werner-Heisenberg-Institut)\\
F\"ohringer Ring 6, 80805 M\"unchen, Germany}

\begin{abstract}
  The $\mu$- and $\tau$-neutrinos emitted from a proto-neutron star
  are produced by nucleonic bremsstrahlung $NN\to NN\nu\bar\nu$ and
  pair annihilation $e^+e^-\to \nu\bar\nu$, reactions which freeze out
  at the ``energy sphere.''  Before escaping from there to infinity
  the neutrinos diffuse through the ``scattering atmosphere,'' a layer
  where their main interaction is elastic scattering on nucleons $\nu
  N\to N\nu$.  If these collisions are taken to be iso-energetic as in
  all numerical supernova simulations, the neutrino flux spectrum
  escaping to infinity depends only on the medium temperature $T_{\rm
  ES}$ and the thermally averaged optical depth $\bar\tau_{\rm ES}$ at
  the energy sphere.  For $\bar\tau_{\rm ES}=10$--50 one finds for the
  spectral flux temperature of the escaping neutrinos $T_{\rm
  flux}=0.5$--$0.6\,T_{\rm ES}$.  Including energy exchange (nucleon
  recoil) in $\nu N\to N\nu$ can shift $T_{\rm flux}$ both up or
  down. $\Delta T_{\rm flux}$ depends on $\bar\tau_{\rm ES}$, on the
  scattering atmosphere's temperature profile, and on $T_{\rm ES}$.
  Based on a numerical study we find that for typical conditions
  $\Delta T_{\rm flux}/T_{\rm flux}$ is between $-10\%$ and $-20\%$,
  and even for extreme parameter choices does not exceed $-30\%$. The
  exact value of $\Delta T_{\rm flux}/T_{\rm flux}$ is surprisingly
  insensitive to the assumed value of the nucleon mass, i.e.\ the
  exact efficiency of energy transfer between neutrinos and nucleons
  is not important as long as it can occur at all. Therefore,
  calculating the $\nu_\mu$ and $\nu_\tau$ spectra does not seem to
  require a precise knowledge of the nuclear medium's dynamical
  structure functions.
\end{abstract}

\keywords{diffusion --- neutrinos --- supernovae: general}


\section{Introduction}

\label{sec:Introduction}

The role of $\mu$- and $\tau$-neutrinos and anti-neutrinos in a
supernova (SN) core is rather different from that of $\nu_e$ and
$\bar\nu_e$. After collapse, a huge amount of electron lepton number
is trapped, leading to a large $\nu_e$ chemical potential, but no
$\mu$- or $\tau$-lepton number is present.  The complete absence of
$\tau$-leptons and the scarcity of muons implies that $\nu_\mu$,
$\bar\nu_\mu$, $\nu_\tau$, and $\bar\nu_\tau$ interact primarily by
neutral-current processes while $\nu_e$ and $\bar\nu_e$ also interact
by more efficient charged-current reactions.  The neutrino energies
emitted from a SN core are a few tens of~MeV, too low to produce muons
or $\tau$-leptons, so that again $\nu_\mu$ and $\nu_\tau$ interact
only by neutral-current reactions with the matter above the SN core or
in terrestrial detectors.  On the other hand, $\nu_e$ and $\bar\nu_e$
have charged-current reactions so that, for example, $\bar\nu_e p\to n
e^+$ provides the dominant SN neutrino signal in a water Cherenkov
detector.  Moreover, the $\nu_e$ and $\bar\nu_e$ flux is thought to be
responsible for re-heating the stalled shock in the delayed explosion
scenario. Later, these neutrinos regulate the $n/p$ ratio in the hot
medium above the neutron star and thus govern the nucleosynthesis
processes which take place in this region.  Little wonder that in
numerical studies far more attention has been paid to the treatment of
$\nu_e$ and $\bar\nu_e$ transport than to the other flavors.

It is timely to take a fresh look at the $\nu_\mu$ and $\nu_\tau$
spectra formation problem because the physical interest in the
flavor-dependent spectra is now much greater than it was a few years
ago. The construction of a neutral-current neutrino observatory is
considered where the $\nu_\mu$ and $\nu_\tau$ fluxes from a future
galactic SN provide the dominant signal (Smith 1997; Boyd \& Murphy
2001).  Further, neutrino oscillations can partly swap the
flavor-dependent spectra produced at the source, especially if $\nu_e$
has a large mixing angle with the other flavors. A vast number of
papers has been devoted to various consequences of flavor oscillations
on SN neutrinos and their detection (e.g.\ Raffelt~1996; Lunardini \&
Smirnov 2001), a topic of serious concern at a time when the
phenomenon of flavor oscillations appears to be experimentally
established.

Moreover, there has been much progress in the numerical treatment of
neutrino transport. New algorithms have been developed to effectively
solve the Boltzmann collision equation in those critical SN regions
where the neutrino mean free path is neither long nor short relative
to the important geometric scales (Burrows et~al.\ 2000; Mezzacappa
et~al.\ 2001; Rampp \& Janka 2000).  With vastly increased CPU power
one is beginning to perform simulations where a reliable transport
scheme is self-consistently coupled with the hydrodynamic evolution so
that the calculated multi-flavor neutrino fluxes and spectra will
depend only on the adopted input physics.

Our present approach is complementary to these global numerical
simulations. We study $\nu_\mu$ and $\nu_\tau$ transport and spectra
formation in the framework of the simplest possible model that
incorporates enough of the essential physics to mimic the full
problem.  This approach allows us to ascertain the significance (or
irrelevance) of microscopic input-physics variations that may modify
the spectra. The nontrivial insights gathered from this study can
serve as a basis for a more informed choice about the micro-physics
that should be implemented in a full simulation.  In addition, our
approach has the pedagogical benefit of providing a simple and
transparent framework for understanding the crucial physics.

\begin{figure}[t]
\columnwidth=6.5cm
\plotone{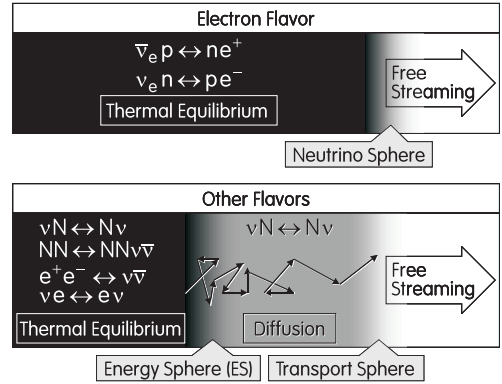}
\caption{\label{fig:spheres}Schematic picture of neutrino spectra
formation in the atmosphere of a SN core.}
\end{figure}

The spectra formation problem is schematically illustrated in
Fig.~\ref{fig:spheres}. The electron (anti-)neutrinos are kept in
thermal equilibrium by beta processes up to a radius usually referred
to as the ``neutrino sphere.'' Beyond this radius the neutrinos stream
off freely, their spectrum representing the medium temperature at the
neutrino sphere. Of course, this picture is crude because the
interaction cross section varies as $\epsilon^2$ (neutrino energy
$\epsilon$) so that different energy groups decouple at different
radii and thus at different medium temperatures.

The other flavors interact with the medium primarily by
neutral-current collisions on nucleons $\nu N\leftrightarrow N\nu$, a
reaction which is sub-dominant for the electron flavor. The nucleon
mass $m=938~{\rm MeV}$ is much larger than the relevant temperatures
which are around $T=10~{\rm MeV}$ so that energy exchange between
neutrinos and nucleons is inefficient.  However, nucleon-nucleon
bremsstrahlung $NN\leftrightarrow NN\nu\bar\nu$ as well as the
leptonic processes $e^+e^-\leftrightarrow\nu\bar\nu$ and $\nu
e\leftrightarrow e\nu$ allow for the exchange of energy and the
creation or destruction of neutrino pairs and thus keep neutrinos in
local thermal equilibrium up to a radius where these reactions freeze
out, the ``energy sphere.''  However, the neutrinos are still trapped
by $\nu N\leftrightarrow N\nu$ up to the ``transport sphere'' whence
they stream freely.  Between the energy and transport spheres,
neutrinos propagate by diffusion.  This region plays the role of a
{\it scattering atmosphere}.

\begin{figure}[t]
\columnwidth=6.5cm
\plotone{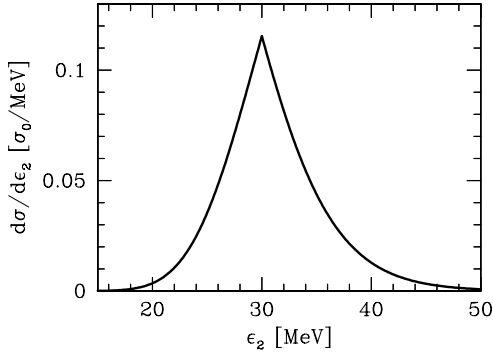}
\caption{\label{fig:srecoil} Distribution of final-state energies
  $\epsilon_2$ of a neutrino with initial energy $\epsilon_1=30~{\rm
    MeV}$, scattering on non-degenerate nucleons in thermal equilibrium
  with $T=10~{\rm MeV}$. Details of how to calculate this
  plot are described in
  Appendix~\protect\ref{sec:ScatteringCrossSection}.}
\end{figure}

In all numerical simulations of SN neutrino transport the neutrino
collisions in the scattering atmosphere were treated as iso-energetic
so that the energy $\epsilon_2$ of the outgoing neutrino in $\nu N\to
N \nu$ was set equal to the energy $\epsilon_1$ of the initial state.
The main motivation for this approximation was its numerical
simplicity and the lack of a compelling interest in details of the
emerging $\nu_\mu$ and $\nu_\tau$ spectra. It is clear, however, that
iso-energetic collisions are not a particularly good approximation. In
Fig.~\ref{fig:srecoil} we show the distribution of final-state
energies $\epsilon_2$ when $\epsilon_1=30~{\rm MeV}$ and the medium
temperature is $10~{\rm MeV}$. A typical nucleon velocity is then
about 20\% of the speed of light so that it is not surprising that
even after a single collision the neutrino energy is considerably
smeared out. Since neutrinos interact many times in the scattering
atmosphere, and since the medium temperature decreases between the
energy and transport spheres, there can be a significant downward
adjustment of the neutrino energies (Janka et~al.\ 1996; Hannestad \&
Raffelt 1998). The main purpose of the present paper is to provide a
conceptual understanding and a quantitative estimate of the magnitude
of this effect.

\begin{figure}
\columnwidth=6.5cm
\plotone{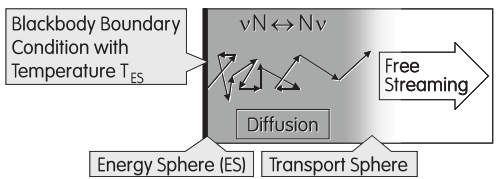}
\caption{\label{fig:bbflux}Schematic picture of 
  our simplified treatment of the scattering atmosphere.  $T_{\rm ES}$
  is the medium temperature at the energy sphere.}
\end{figure}

To address this problem we simplify the model of
Fig.~\ref{fig:spheres}.  The very concept of an ``energy sphere''
suggests that one should think of it as a source of thermal neutrinos
which subsequently diffuse through the scattering atmosphere.  Taking
this concept literally amounts to the simple picture illustrated in
Fig.~\ref{fig:bbflux}. One no longer worries about detailed processes
like $NN$ bremsstrahlung to thermalize the neutrinos, but directly
feeds a thermal flux into the scattering atmosphere.

Section~II of our paper is devoted to showing that this simple picture
actually provides a surprisingly accurate representation of the
spectra formation problem. The neutrinos streaming off the transport
sphere then have fluxes and spectra which depend only on the
temperature $T_{\rm ES}$ and the thermally averaged transport optical
depth $\bar\tau_{\rm ES}$ at the energy sphere which here coincides
with the bottom of the scattering atmosphere.

As a next step in Section~III we study a scattering atmosphere with a
blackbody boundary condition at the bottom and with iso-energetic $\nu
N$ collisions as the only neutrino interaction channel.  We derive an
explicit relationship between $T_{\rm ES}$ and the spectral flux
temperature $T_{\rm flux}$ of the escaping neutrinos as a function of
$\bar\tau_{\rm ES}$.  Comparing with full-scale numerical simulations
indicates that this exceedingly simple model accounts for the main
features of the $\nu_\mu$ and $\nu_\tau$ spectra.

Then in Section~IV we include nucleon recoils in this model.  We
consider different types of temperature profiles to estimate the shift
of the flux temperature and identify the critical parameters which
govern $\Delta T_{\rm flux}$.

In Section~V we summarize and discuss our findings. Many technical
details, especially regarding our implementation of neutrino-nucleon
interactions with recoil energy transfer and $NN$ bremsstrahlung,
are documented in a series of Appendices.


\section{Energy Sphere as a Blackbody Surface}

The neutrino sphere, transport sphere, or energy sphere are only
approximate concepts because of the energy dependence of the neutrino
cross sections. Still, we presently argue that the concept of the
energy sphere as a well-defined blackbody surface at the bottom of the
scattering atmosphere is much better suited to understand the neutrino
spectra than one may have hoped.

To get started we study a practical example.  We have performed a
$\nu_\mu$ and $\nu_\tau$ transport calculation with the simple Monte
Carlo code described in Appendix~\ref{sec:NumericalCode}.  We use
plane-parallel geometry with the medium temperature and density
following the power laws
\begin{equation}\label{eq:powerlaw}
\rho=\rho_0\,\left(\frac{r_0}{r}\right)^p\quad\hbox{and}\quad
T=T_0\,\left(\frac{r_0}{r}\right)^q,
\end{equation}
where for the present example $p=10$ and $q=10/4$. The other
parameters are $r_0=10~{\rm km}$, $\rho_0=2\times10^{14}~{\rm
  g~cm^{-3}}$ and $T_0=31.66~{\rm MeV}$.

The neutrinos interact by iso-energetic $\nu N$ scattering and by $NN$
bremsstrahlung.  This choice of micro-physics is orthogonal to most
previous studies which ignored bremsstrahlung, but included
$e^+e^-\leftrightarrow\nu\bar\nu$ and $\nu e\leftrightarrow e\nu$.  In
this traditional approach spectra formation is conceptually even more
complicated because the number-changing process
$e^+e^-\leftrightarrow\nu\bar\nu$ freezes out more deeply than the
energy-changing process $\nu e\leftrightarrow e\nu$ so that in
addition to the transport and energy spheres there is a ``number
sphere'' (Suzuki 1989). Later it was recognized that $NN$
bremsstrahlung is more important than $e^+e^-$ annihilation (Suzuki
1991, 1993).  Hannestad \& Raffelt (1998), Thompson, Burrows \&
Horvath (2000), and Burrows et~al.\ (2000) confirm the importance of
$NN$ bremsstrahlung.  Moreover, Hannestad \& Raffelt (1998) find that
the freeze-out spheres for bremsstrahlung and for $\nu
e\leftrightarrow e\nu$ roughly coincide in their example of a SN core.
Burrows et~al.\ (2000) find that the $\nu_\mu$ spectrum depends
sensitively on the assumed strength of $NN$ bremsstrahlung, suggesting
that $\nu_\mu$ thermalization is not vastly dominated by $\nu e$
scattering.

Therefore, reducing the thermalization processes to $NN$
bremsstrahlung likely captures the dominant effect. As this process
allows for both the change of neutrino number and energy, the concept
of an energy sphere or ``thermalization depth'' is unambiguous, but of
course energy dependent.  Ignoring $\nu e$ scattering has the
additional benefit that it becomes self-consistent to ignore the
chemical composition of the medium. While $\nu N$ scattering and $NN$
bremsstrahlung do depend on the chemical composition, this effect is
weak for $\nu N$ scattering.  While it may be strong for $NN$
bremsstrahlung, the existing calculations of this process are
uncertain to perhaps a factor of~2 (see
Appendix~\ref{sec:Bremsstrahlung}) so that one would not gain much by
modeling the composition dependence.

Once the micro physics has been fixed it is straightforward to locate
the energy sphere by virtue of a classic argument.  Assume the
neutrinos have only two interaction channels, i.e.\ iso-energetic
scattering with the inverse transport mean free path (mfp)
$\lambda_T^{-1}$ and a reaction with the mfp $\lambda_E^{-1}$ which
changes the energy by a large amount and thus leads to quick
thermalization.  In this situation the optical depth $\int
dr\,\lambda_E^{-1}$ for energy exchange is not relevant because
neutrinos with a large $\lambda_T^{-1}$ are trapped and thus have a
greater chance of participating in an energy-changing reaction. This
reasoning leads to the optical depth for thermalization (Shapiro \&
Teukolsky 1983)
\begin{equation}\label{eq:thermalizationdepth}
\tau_{\rm therm}(r)=\int_r^\infty\!\! dr'\,
\sqrt{\lambda_E^{-1}(r')\,
\left[\lambda_T^{-1}(r')+\lambda_E^{-1}(r')\right]}.
\end{equation}
The radius $r_{\rm ES}$ of the energy sphere is implied by
\begin{equation}\label{eq:thermalizationdepth2}
\tau_{\rm therm}(r_{\rm ES})=\frac{2}{3}.
\end{equation}
Averaging $\tau_{\rm therm}$ over a thermal neutrino spectrum at the
local medium temperature leads in our example to 
$r_{\rm ES}=13.14~{\rm km}$
for the location of the energy sphere
with a temperature $T_{\rm ES}=16~{\rm MeV}$.

In order to discuss non-equilibrium neutrino fluxes and spectra we
need to characterize the neutrino distribution function by a few
intuitive parameters.  One is the {\it spectral temperature\/} $T_*$
which we define as
\begin{equation}\label{eq:spectralT}
T_*\equiv\frac{\langle\epsilon\rangle}{3}=
\frac{\int_0^\infty d\epsilon\,\epsilon 
\int_{-1}^{+1}d\mu f(\epsilon,\mu)}
{3\int_0^\infty d\epsilon 
\int_{-1}^{+1}d\mu f(\epsilon,\mu)}.
\end{equation}
Here, $f(\epsilon,\mu)$ is the neutrino distribution function with
$\epsilon$ the energy and $\mu$ the cosine of the angle between the
neutrino momentum and the radial direction.  (Even for our
plane-parallel geometry we will often say ``radial'' when we mean
perpendicular to the medium layer.)  We always ignore Pauli blocking,
i.e.\ the Fermi-Dirac nature of neutrinos.  The Boltzmann collision
equation is then linear in $f$, a considerable numerical
simplification which is justified because the $\nu_\mu$ and $\nu_\tau$
distributions do not have chemical potentials in a SN core.  Moreover,
in the scattering atmosphere their distribution function is diluted,
further reducing the relevance of phase-space blocking. For our study
the simplification of ignoring blocking effects far outweighs the
small loss of precision.  Boltzmann statistics implies that in
equilibrium neutrinos follow the distribution function
$f=e^{-\epsilon/T}$ so that $\langle\epsilon\rangle=3 T$. Inverting
this relationship for the non-equilibrium case leads to our definition
of~$T_*$.

\begin{figure}
\columnwidth=6.5cm
\plotone{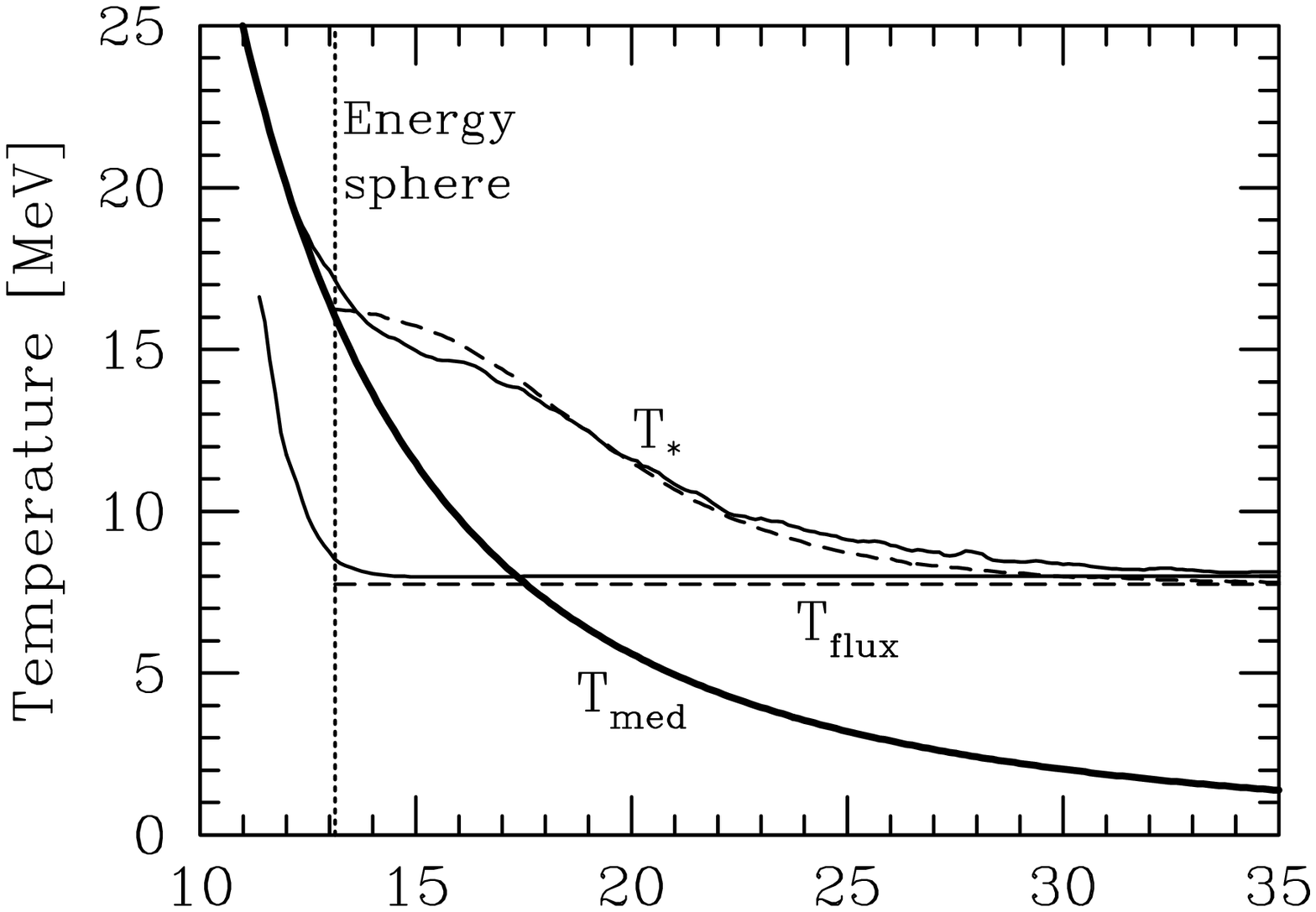}\\
\plotone{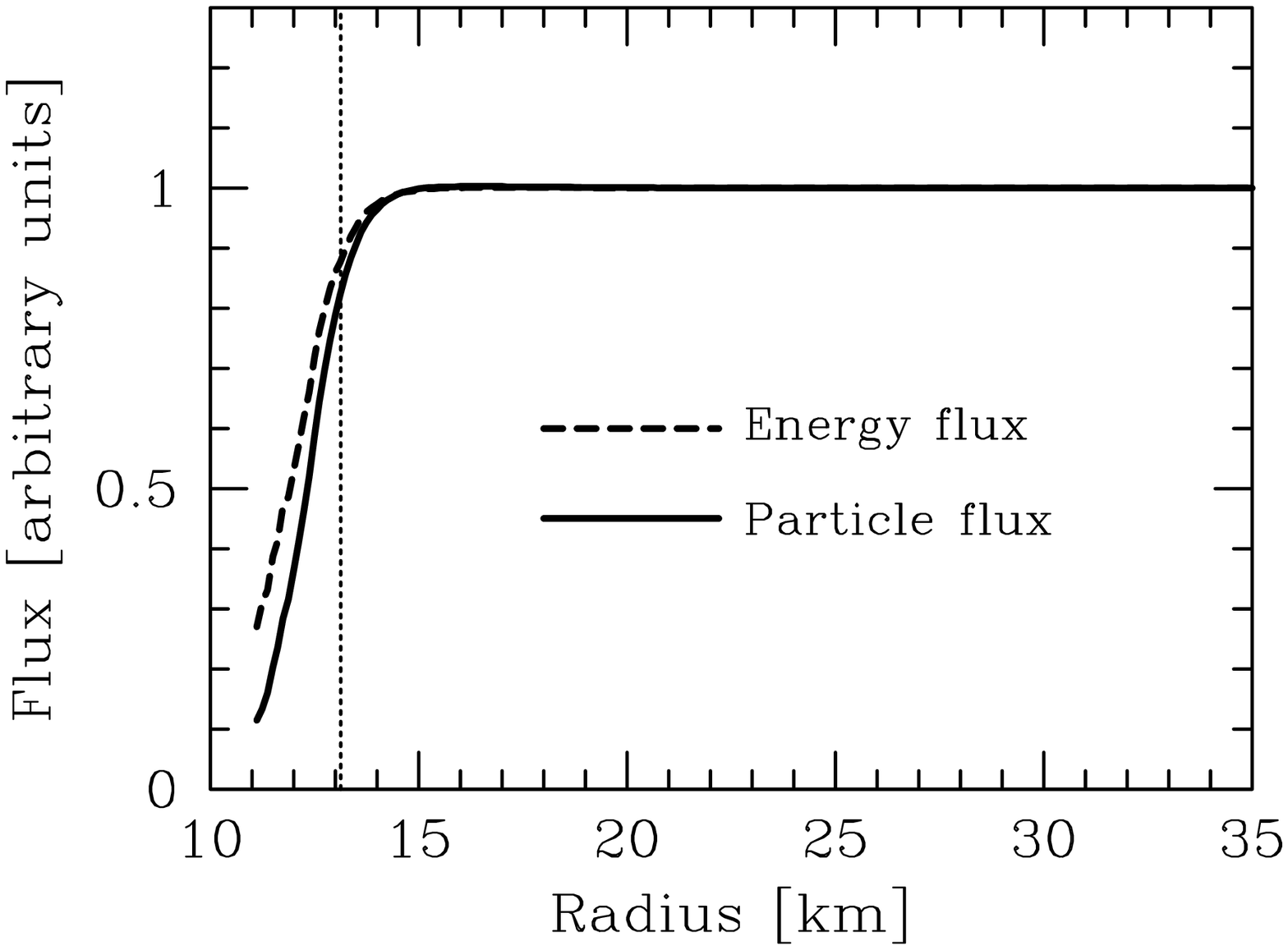}
\caption{\label{fig:profile}Numerical example for $\nu_\mu$ and
$\nu_\tau$ transport as described in the text. The inner boundary
condition was at 10~km.  The thin dashed lines in the upper panel
refer to the case without $NN$ bremsstrahlung where a blackbody
boundary condition with $T_{\rm ES}=16~{\rm MeV}$ is used at the
energy sphere.}
\end{figure}

In Fig.~\ref{fig:profile} we show the $T_*$ profile for our example as
a solid line.  (The slight wiggles arise from the limited statistics
of the numerical Monte Carlo calculation.)  We observe that $T_*$
coincides with the medium temperature $T_{\rm med}$ up to the energy
sphere; for larger radii the two curves separate.  $T_*$ still drops
considerably until it reaches an asymptotic value which is about half
of the energy-sphere temperature $T_{\rm ES}$. This behavior is
understood by the $\epsilon^2$ dependence of the scattering cross
section which causes higher-energy neutrinos to be trapped more
effectively than lower-energy ones.  Hence, the escaping neutrino flux
must be biased towards lower energies.

Therefore, one should carefully distinguish between the spectral
temperature $T_*$ as defined in Eq.~(\ref{eq:spectralT}) and the {\it
  spectral flux temperature\/}
\begin{equation}\label{eq:fluxT}
T_{\rm flux}\equiv
\frac{\int_0^\infty d\epsilon\,\epsilon 
\int_{-1}^{+1}d\mu\,\mu\,f(\epsilon,\mu)}
{3\int_0^\infty d\epsilon 
\int_{-1}^{+1}d\mu\,\mu\,f(\epsilon,\mu)}\,.
\end{equation}
This is $\langle\epsilon\rangle/3$ of those neutrinos which actually
stream. Note that apart from overall factors $2\pi\int_{-1}^{+1}
d\mu\,\mu\,f(\epsilon,\mu)$ represents the particle flux. $T_{\rm
flux}$ does not include the isotropic part of the distribution which
determines $T_*$ when the neutrinos are trapped.  At large distances
where neutrinos stream freely we have $f(\epsilon,\mu)=0$ for backward
directions so that $T_*$ and $T_{\rm flux}$ are identical up to small
differences caused by the exact angular distribution.

In Fig.~\ref{fig:profile} we show the $T_{\rm flux}$ profile for our
numerical example.  At large radii indeed $T_*\approx T_{\rm flux}$.
Moreover, $T_{\rm flux}$ stays constant throughout the scattering
atmosphere because no energy exchange with the medium is possible.
Flux conservation in the scattering atmosphere is apparent from the
lower panel where we show the particle and energy fluxes which both
remain constant from roughly the energy sphere outward.

It appears that the separation of the $T_*$ and $T_{\rm med}$
profiles, as well as the saturation of $T_{\rm flux}$ and the
particle and energy fluxes all happen within a narrow range of radii
around the energy sphere.  This observation motivates us to repeat the
simulation with the lower boundary now at $r_{\rm ES}$, switching off
$NN$ bremsstrahlung entirely, and injecting neutrinos with a
thermal spectrum corresponding to $T_{\rm ES}=16~{\rm MeV}$, i.e.\ 
enforcing a neutrino blackbody surface at $r_{\rm ES}$.  The
resulting temperature profiles are shown in
Fig.~\ref{fig:profile} as dashed lines.  $T_{\rm flux}$ is
now strictly horizontal.  Likewise, the energy and particle
fluxes are now strictly conserved so that their profiles
are horizontal lines not shown in this plot.

The $T_*$ and $T_{\rm flux}$ profiles are surprisingly close to the
previous case where neutrinos thermalized by $NN$ bremsstrahlung.
Thus, the concept of an energy sphere as a blackbody surface looks
encouragingly well justified.

One reason for the extraordinary ``sharpness'' of the energy sphere in
our example is the unusual behavior of the neutrino absorption rate by
inverse $NN$ bremsstrahlung; Eq.~(\ref{eq:bremsmfp}) implies
approximately $\lambda_{\rm brems}^{-1}\propto \epsilon^{-1}$.  This
is one of the few neutrino reaction rates which {\it decrease\/} with
energy; the scattering rate $\lambda_T^{-1}$ increases
as~$\epsilon^2$.  With $\lambda_T^{-1}+\lambda_{\rm brems}^{-1}
\approx \lambda_T^{-1}$ we find that the effective thermalization rate
under the integral in Eq.~(\ref{eq:thermalizationdepth}) varies
as~$\epsilon^{1/2}$.  Therefore, different energy groups decouple
thermally at similar radii.

Another reason for a sharply defined energy-sphere temperature is that
the density as a function of radius falls much more steeply than the
temperature.  Therefore, as a function of optical depth the
temperature varies slowly.  In our example $\lambda_T^{-1}\propto\rho$
and approximately $\lambda_{\rm brems}^{-1} \propto\rho^2 T^3$ so that
$(\lambda_{\rm brems}^{-1}\lambda_T^{-1})^{1/2} \propto\rho^{3/2}
T^{3/2}$ or $\tau_{\rm therm}\propto (r_0/r)^{3(p+q)/2-1}$ so that
$T\propto \tau_{\rm therm}^w$ with $w=2q/[3(p+q)-2]$.  In our explicit
example this is $T\propto \tau_{\rm therm}^{10/71}$.  Therefore, the
energy spheres for the different energy groups are at similar
temperatures.

We thus expect that a blackbody boundary condition at the bottom of
the scattering atmosphere provides us with a reasonably accurate
understanding of the $\nu_\mu$ and $\nu_\tau$ spectra formation
problem. On the other hand, for the problem of hydrostatically
self-consistent atmospheres with multi-flavor neutrino transport
(e.g.\ Schinder \& Shapiro 1983) it would be physically more
appropriate to use a prescribed neutrino energy flux from the inner
star; the atmosphere would then be allowed to adjust in order to
transport exactly this flux. In our study, on the other hand, we
always assume a fixed background model where the neutrino fluxes are
determined by externally prescribed density and temperature profiles
as well as the physical assumptions about their interaction with the
medium.  In that sense our study is similar in spirit to the
Monte-Carlo studies of Janka \& Hillebrandt (1989a,b) or the more
recent work of Burrows et al.\ (2000). We do not wish to imply that in
a real SN core the neutrino fluxes are determined by the temperature
at the base of the scattering atmosphere unless this temperature
corresponds to a self-consistent stellar structure.


\section{Spectral Modification by the Scattering Atmosphere}

\label{sec:ScatteringAtmosphere}

\subsection{Flux Transmission} 

In our simple picture where the scattering atmosphere is irradiated by
a thermal neutrino flux from below, part of that flux will be
reflected, part of it transmitted.  The reflected flux will be
re-absorbed because a blackbody surface is by definition a perfect
absorber.  To determine the emerging $\nu_\mu$ and $\nu_\tau$ flux it
is therefore enough to calculate the energy-dependent flux
transmission factor, i.e.\ the fraction of the primary flux that is
transmitted without being reflected back into the radiating surface.

For sufficiently low energies the scattering atmosphere is transparent
so that the transmission factor is unity.  In the opposite limit where
the neutrino mfp is short compared with the geometric dimension of the
layer (diffusion limit), the Boltzmann collision equation can be
solved analytically (Appendix~\ref{sec:DiffusionLimit}), leading for a
plane-parallel geometry to a transmission ratio~of
\begin{equation}
s(\tau_{\rm ES})=\frac{4}{3\tau_{\rm ES}}.
\end{equation}
Here, $\tau_{\rm ES}$ is 
the {\it transport optical depth\/} of the energy sphere
\begin{equation}\label{eq:tansportdepth}
\tau_{\rm ES}\equiv\int_{r_{\rm ES}}^\infty 
dr\, n_B(r)\int_{-1}^{+1}d\cos\theta\,
(1-\cos\theta)\,\frac{d\sigma}{d\cos\theta},
\end{equation}
where $n_B$ is the number density of scatterers (``baryons''),
$d\sigma/d\cos\theta$ the differential scattering cross section, and
$r_{\rm ES}$ the lower boundary of the scattering atmosphere, i.e.\
the energy sphere.

In the intermediate regime $\tau_{\rm ES}={\cal O}(1)$ we solve the
Boltzmann collision equation numerically. To do so we have to commit
ourselves to a specific form of the differential scattering cross
section. We use the ``dipole formula''
$d\sigma/d\cos\theta=\sigma_0\,(1+b\cos\theta)/2$ where $-1\leq
b\leq+1$ and the total cross section is $\sigma_0$. When expressed in
terms of the transport optical depth $\tau_{\rm ES}$, the flux
transmission is independent of $b$ for small and large $\tau_{\rm
ES}$.  For intermediate $\tau_{\rm ES}$ the variation with $b$ is less
than $\pm 1\%$, i.e.~for all practical intents and purposes it is
enough to use the transport optical depth as the only relevant
parameter.

\begin{figure}[t]
\columnwidth=6.5cm
\plotone{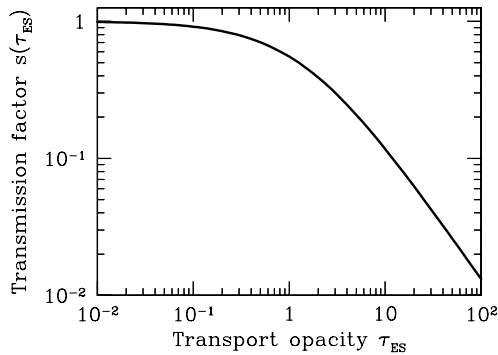}
\caption{\label{fig:transmission}Flux transmission factor
  for a plane-parallel scattering atmosphere with transport optical 
  depth $\tau_{\rm ES}$.}
\end{figure}

The numerical transmission factor for $b=0$ is shown in
Fig.~\ref{fig:transmission}; for other $b$-values it is within the
line-width of this curve. An analytic approximation is
\begin{eqnarray}\label{eq:approxsuppression}
s(\tau_{\rm ES})&=&\left[1+\frac{3\,\tau_{\rm ES}}{4}\right]^{-1}
\nonumber\\
&&\kern-5em{}
\times
\left[1-\frac{0.033}{1+1.5\,(\ell_\tau+0.17)^2+0.5\,(\ell_\tau+0.32)^6}
\right],\nonumber\\
\end{eqnarray}
where $\ell_\tau\equiv\log_{10}(\tau_{\rm ES})$. The first bracket alone
overestimates the numerical result by less than 4\% in the entire
range $0<\tau_{\rm ES}<\infty$.  
The second bracket improves the approximation
to better than $\pm0.05\%$.

For a spherically symmetric geometry these results need to be
modified. In Appendix~\ref{sec:DiffusionLimit} we show that in the
diffusion limit the transmission factor is the same as for a
plane-parallel geometry, provided one includes a factor $(r_{\rm
ES}/r)^2$ under the integral in Eq.~(\ref{eq:tansportdepth}). The
resulting quantity $\tau^*_{\rm ES}$, of course, is no longer the
transport optical depth, but for the purposes of flux transmission it
plays an analogous role.  For the spherical case we have not
investigated the intermediate regime $\tau^*_{\rm ES}={\cal O}(1)$,
but assume that the plane-parallel interpolation formula
Eq.~(\ref{eq:approxsuppression}) applies with reasonable accuracy.

\subsection{Spectral Modification}

Next we study the flux and spectrum of the transmitted neutrinos if
the primary flux is thermal at a temperature $T_{\rm ES}$.  We
characterize the ``thickness'' of the plane-parallel scattering
atmosphere by $\bar\tau_{\rm ES}$, defined as the transport optical
depth at the energy sphere, averaged over a Maxwell-Boltzmann spectrum
at $T_{\rm ES}$.  Taking the scattering cross section to vary as
$\epsilon^2$ we note that $\langle \epsilon^2\rangle=12\,T_{\rm ES}^2$
so that $\bar\tau_{\rm ES}$ amounts to the transport optical depth at
the fixed energy $\epsilon=\sqrt{12}\,T_{\rm ES}\approx3.4641\,T_{\rm
ES}$.  Put another way, we use $\tau_{\rm ES}(\epsilon)=(\bar\tau_{\rm
ES}/12)(\epsilon/T_{\rm ES})^2$ to calculate the energy-dependent
transmission factor.

\begin{figure}[t]
\columnwidth=6.5cm
\plotone{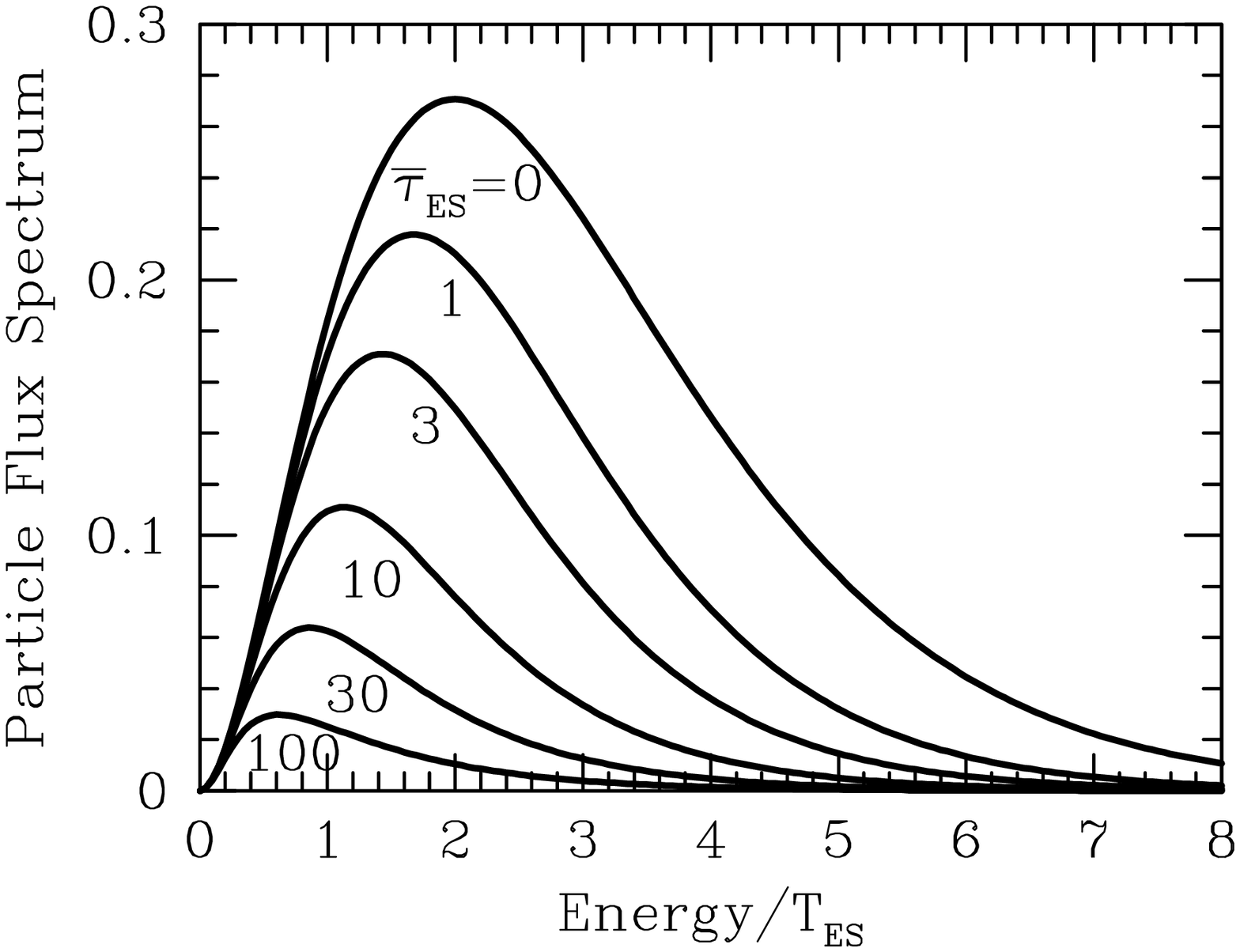}
\caption{\label{fig:supspec1}
  Transmitted flux for the indicated values of $\bar\tau_{\rm ES}$.}
\bigskip\bigskip
\columnwidth=6.5cm
\plotone{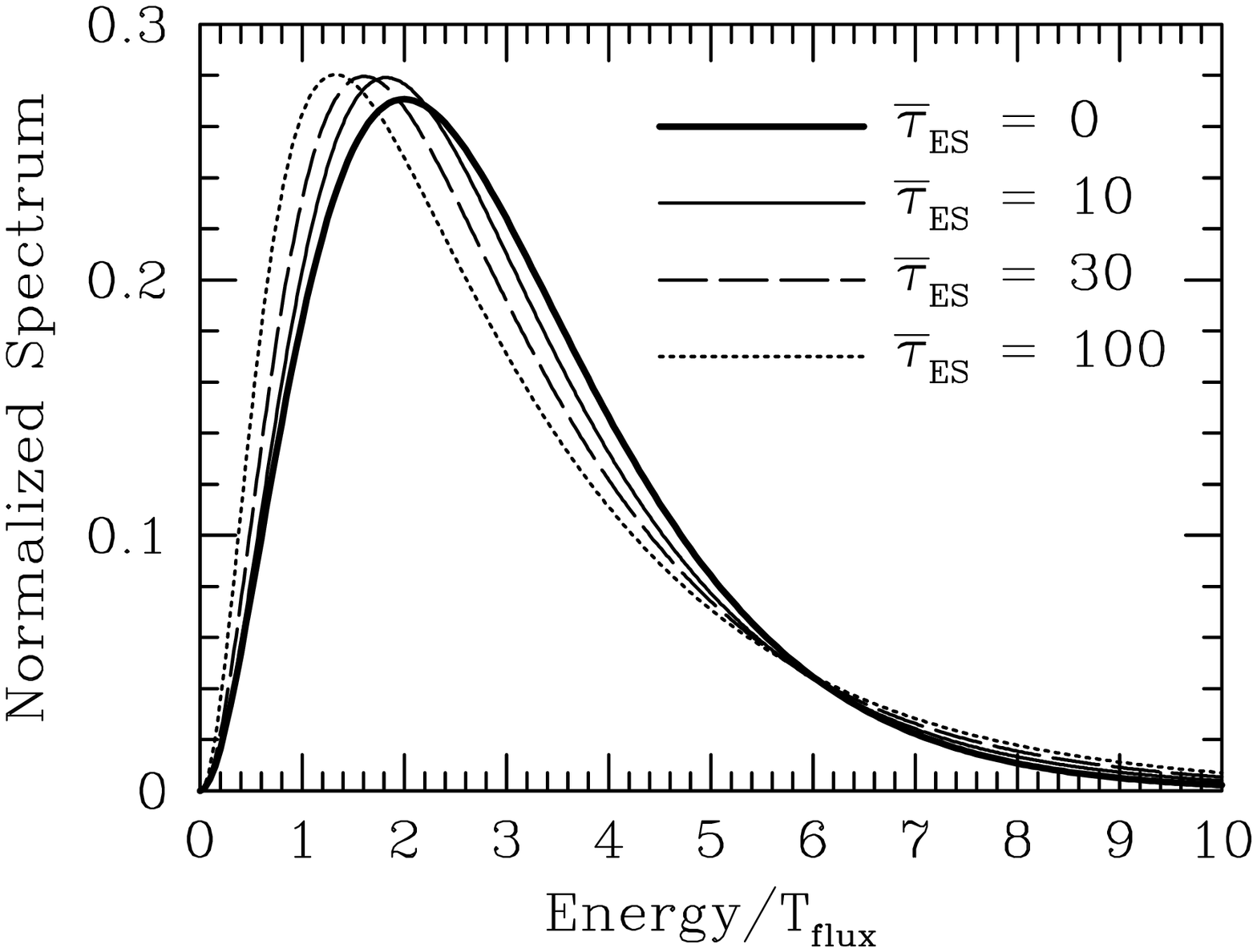}
\caption{\label{fig:supspec2}
  Transmitted flux (normalized) for the indicated values of 
  $\bar\tau_{\rm ES}$
  where the neutrino energy is in units of $T_{\rm flux}$.}
\end{figure}

In Fig.~\ref{fig:supspec1} we show the energy-dependent transmitted
flux for several values of $\bar\tau_{\rm ES}$.  One parameter to
characterize these spectra is their flux temperature $T_{\rm flux}$ as
defined in Eq.~(\ref{eq:fluxT}). In Fig.~\ref{fig:supspec2} we show
the same spectra after re-scaling the horizontal axis with the
appropriate $T_{\rm flux}$ and normalizing them.  This representation
shows how closely these functions mimic true thermal spectra.

\begin{figure}[b]
\columnwidth=6.9cm
\plotone{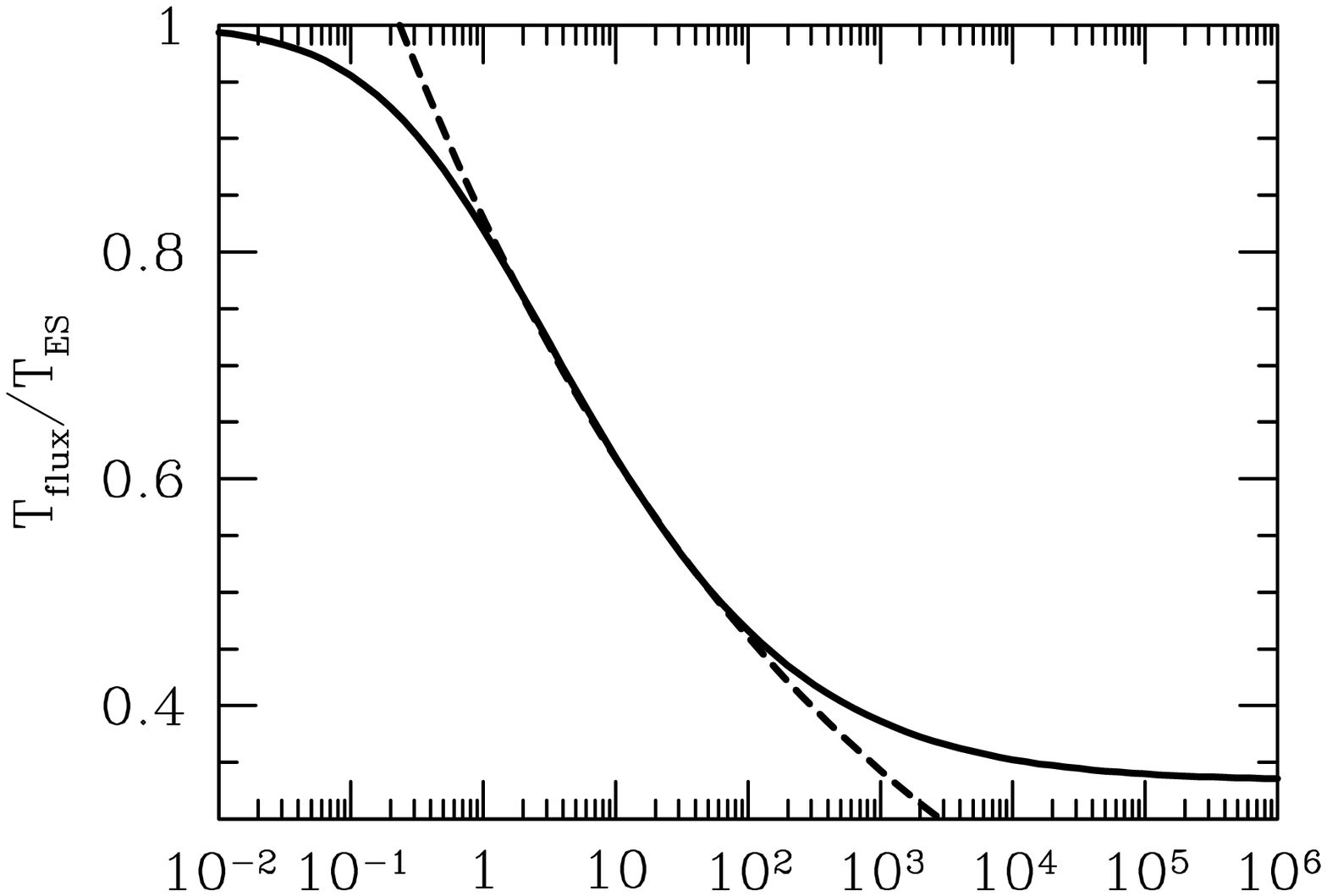}\\
\plotone{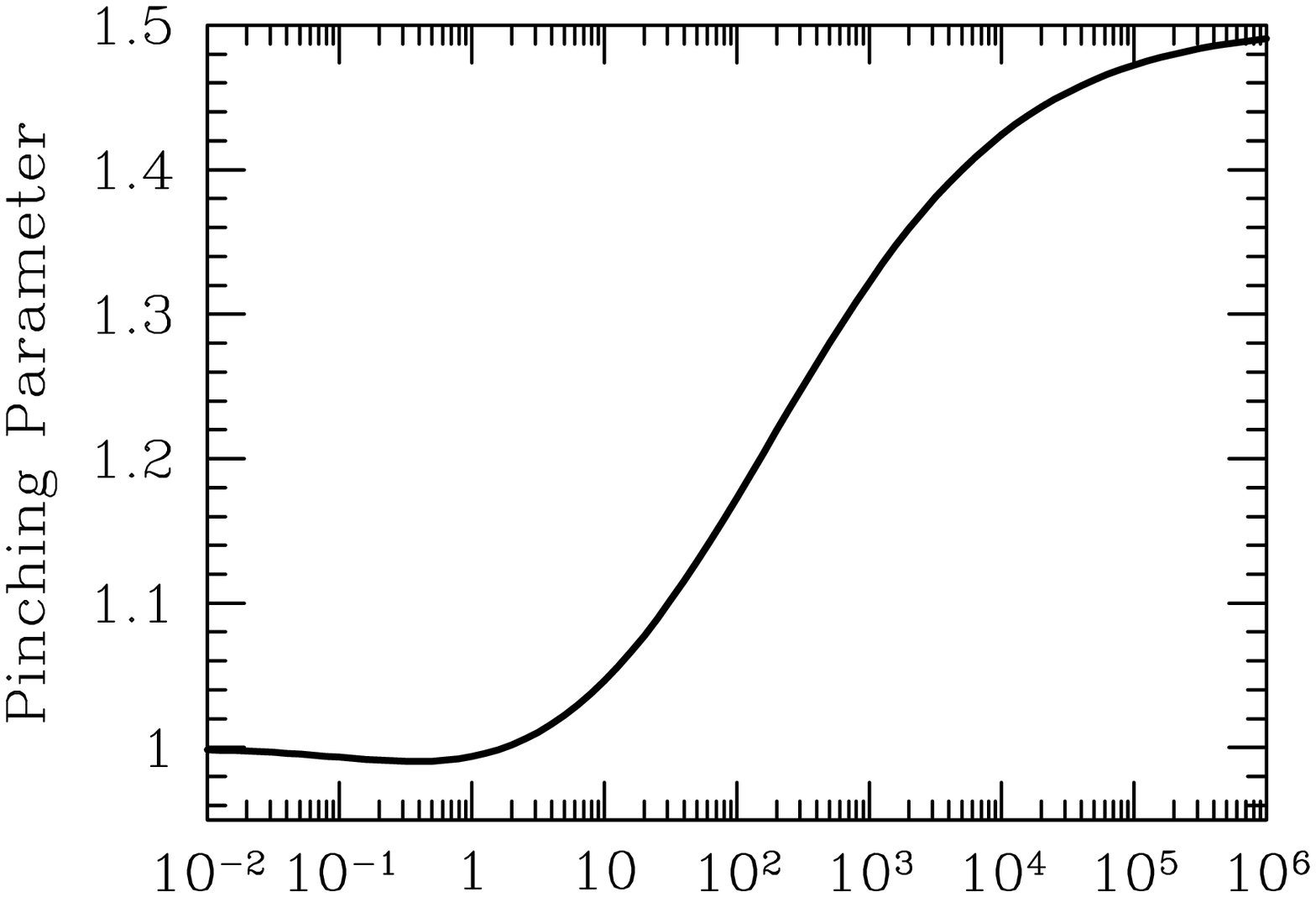}\\
\plotone{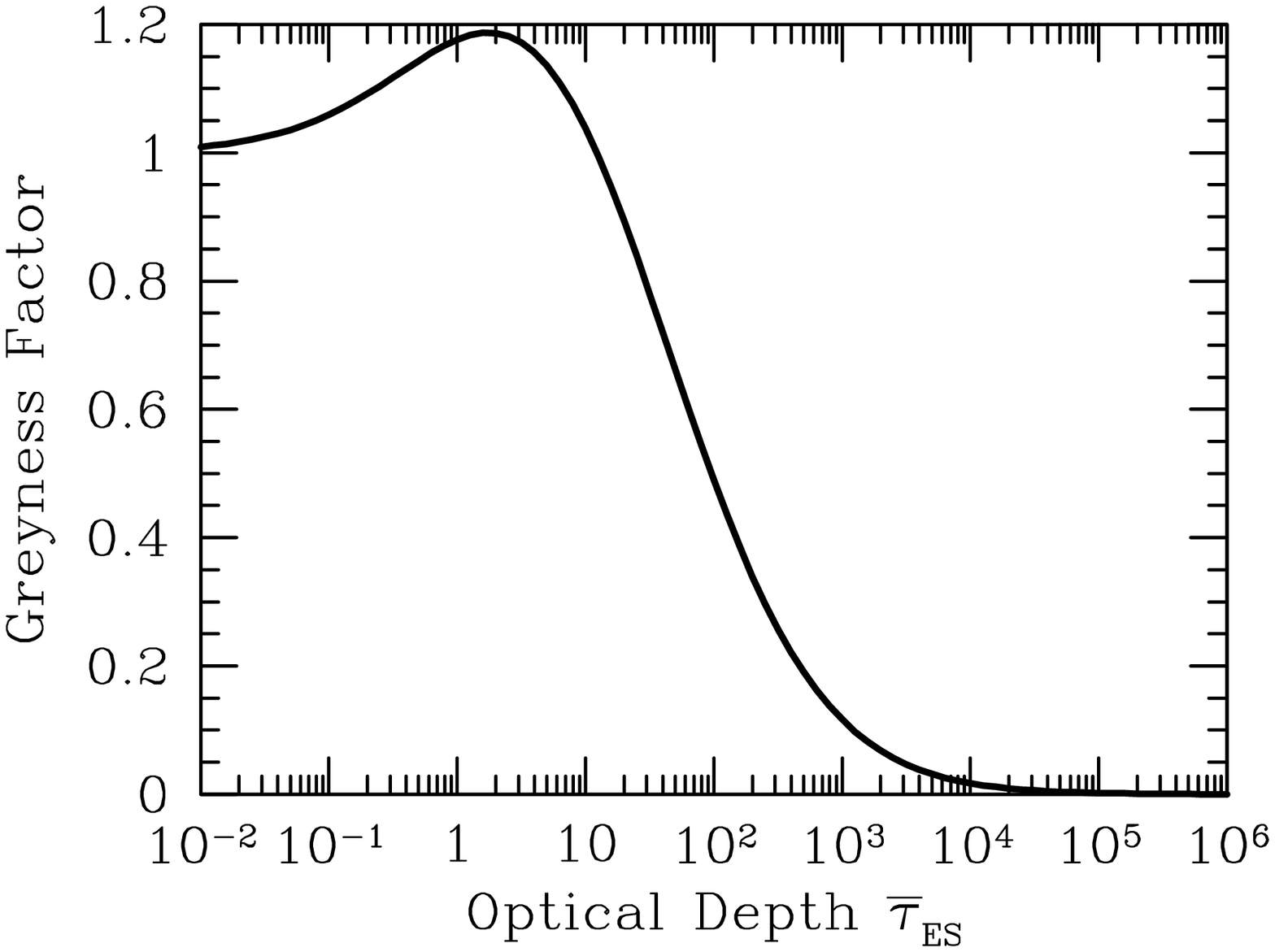}
\caption{\label{fig:tflux} Spectral characteristics of 
  the transmitted flux.}
\end{figure}

The flux temperature $T_{\rm flux}$ in units of $T_{\rm ES}$ as a
function of $\bar\tau_{\rm ES}$ is shown in the upper panel of
Fig.~\ref{fig:tflux}.  For $\bar\tau_{\rm ES}\ll1$ we have, of course,
$T_{\rm flux}=T_{\rm ES}$.  In the opposite limit $\bar\tau_{\rm
ES}\to\infty$ the transmission factor is proportional
to~$\epsilon^{-2}$ so that the transmitted spectrum is proportional to
$e^{-\epsilon/T_{\rm ES}}$ rather than the primary
$\epsilon^2e^{-\epsilon/T_{\rm ES}}$, implying $T_{\rm
flux}=\frac{1}{3}\,T_{\rm ES}$.  $T_{\rm flux}$ is significantly
smaller than $T_{\rm ES}$ even for relatively small values of
$\bar\tau_{\rm ES}$. An analytic approximation formula is
\begin{eqnarray}\label{eq:tfluxapproximation}
\frac{T_{\rm flux}}{T_{\rm ES}}
&=&\Biggl[\frac{1}{3}+\frac{2}{3\,(1+0.35\,
\bar\tau_{\rm ES}^{0.53})}\Biggr]\nonumber\\
&\times&\Biggl[1-0.03\,(1-0.75\,y^2)\,e^{-0.27\,y^2}\Biggr]
\end{eqnarray}
where $y=\log_{10}(\bar\tau_{\rm ES})-0.9$. The first bracket alone
approximates the true result to better than 4\% while the second
bracket improves the approximation to better than $\pm 0.3\%$
everywhere. To study power-law models it is also useful to have a
power-law representation. We find that
\begin{equation}\label{eq:powerlawflux}
\frac{T_{\rm flux}}{T_{\rm ES}}
=\frac{0.83}{\bar\tau_{\rm ES}^{0.128}}
\end{equation}
works well in the range $1\lesssim\bar\tau_{\rm ES}\lesssim100$ 
(dashed line in the upper panel of Fig.~\ref{fig:tflux}).

The deviation of the spectrum from a thermal shape may be
characterized by the ``pinching parameter''
\begin{equation}
p\equiv\frac{3}{4}\,\frac{\langle\epsilon^2\rangle}
{\langle\epsilon\rangle^2}
\end{equation}
where $p=1$ for a Maxwell-Boltzmann spectrum.  If the high- and
low-energy parts of the spectrum are relatively suppressed, the
spectrum is called ``pinched'' ($p<1$), in the opposite case
``anti-pinched'' ($p>1$). The limiting cases are $p=1$ for
$\bar\tau_{\rm ES}\to0$ and $\frac{3}{2}$ for $\bar\tau_{\rm
ES}\to\infty$.  Intermediate values are shown in Fig.~\ref{fig:tflux}
and Table~\ref{tab:spectrum}.

\begin{deluxetable}{rrrr}
\tablecaption{\label{tab:spectrum} Spectral modification of neutrino 
flux.}
\tablewidth{0pt}
\tablehead{\colhead{$\bar\tau_{\rm ES}$}&
\colhead{$T_{\rm flux}/T_{\rm ES}$}&
\colhead{$p$}&
\colhead{$\phi$}
}
\startdata
0       &   1&1&1\\
1       &   0.82&  0.99 & 1.18   \\
3       &   0.72&  1.01 & 1.17   \\
10      &   0.62&  1.05 & 1.04   \\
30      &   0.54&  1.10 & 0.79   \\
100     &   0.47&  1.17 & 0.49   \\
$\infty$&  1/3&3/2&0      \\
\enddata
\end{deluxetable}

The concept of spectral pinching refers to the width of the
energy distribution. Therefore, another choice for the
pinching parameter might have been something like
$(\langle\epsilon^2\rangle-\langle\epsilon\rangle^2)/
\langle\epsilon\rangle^2$. This definition is equivalent to $p$
aside from a shift of the zero point and the normalization. Either
way one uses the quadratic moment $\langle\epsilon^2\rangle$ in
addition to $\langle\epsilon\rangle$ to characterize the spectrum.

In the literature, non-thermal SN neutrino spectra are often
described by a degeneracy parameter $\eta$.  This representation
has the disadvantage that anti-pinched spectra are not covered because
the limit $\eta\to-\infty$ represents a Boltzmann distribution.

A further characteristic of the transmitted flux is its luminosity.
It may be larger or smaller than given by the Stefan-Boltzmann law in
terms of $T_{\rm flux}$.  We call the ratio of the true luminosity
relative to this Stefan-Boltzmann value the ``greyness factor''
$\phi$. When $\phi<1$ the transmitted flux is ``grey'' in that it
carries less energy than a blackbody flux of the same spectral shape.
In the bottom panel of Fig.~\ref{fig:tflux} we show $\phi$; some
values are given in Table~\ref{tab:spectrum}.

If the scattering atmosphere is very thick ($\bar\tau_{\rm ES}\gg1$)
the emerging spectrum approaches a limiting spectral temperature of
$\frac{1}{3}T_{\rm ES}$ as explained above
Eq.~(\ref{eq:tfluxapproximation}), while the flux is arbitrarily
diluted. Therefore, the greyness factor will become arbitrarily small.
It is remarkable, however, that $\phi=1$ within $\pm20\%$ for
$\bar\tau_{\rm ES}$ as large as 30.  Therefore, if in realistic
examples of density and temperature profiles the energy sphere is at
$\bar\tau_{\rm ES}<30$, then the emerging flux will be surprisingly
close to a Stefan-Boltzmann flux as calculated with the spectral
temperature $T_{\rm flux}$. Put another way, for $\bar\tau_{\rm
ES}<30$ the overall flux dilution and spectral shift conspire to mimic
the Stefan-Boltzmann law for the emerging flux. This unintuitive
behavior is a consequence of the $\epsilon^2$ dependence of the
transport cross section. If the cross section were energy independent,
then the flux would be diluted without a spectral shift, and even for
small $\bar\tau_{\rm ES}$ the emerging flux would appear diluted
(grey) relative to its spectral temperature.

The neutrino fluxes and spectra from SN cores are usually described as
showing equipartition of energy between different flavors, yet a
hierarchy of spectral temperatures. At late times when the star is
settled and the density and temperature profiles in the atmosphere are
steep, the radiating surfaces for different flavors are similar.
Therefore, as stressed by Janka (1995), similar luminosities with
different spectral temperatures must be explained by the flux dilution
caused by the scattering atmosphere. Janka (1995) did not worry about
the spectral shift caused by the $\epsilon^2$ variation of the
transport cross section so that his conclusions would have applied to
any optical depth of the energy sphere.  Our more refined treatment
suggests that large spectral differences and similar luminosities for
the different flavors at late times require average optical depths of
the $\nu_\mu$ energy sphere exceeding about $30$. However, it is
precisesly at late times when $Y_e$ is small and the density gradients
are steep that the $NN$ bremsstrahlung process is likely to dominate
over the leptonic processes. Once $NN$ bremsstrahlung is included in
future numerical simulations it remains to be seen if at late times
the energy sphere moves to such large optical depths as to cause
significantly ``grey'' $\nu_\mu$ or $\nu_\tau$ fluxes.

In summary, for values of $\bar\tau_{\rm ES}$ up to 30 the transmitted
flux resembles a thermal spectrum in both its absolute flux as well as
its spectral shape.  However, the relevant temperature is not $T_{\rm
ES}$, but rather a reduced temperature $T_{\rm flux}$.  For
$\bar\tau_{\rm ES}=10$--30 we find $T_{\rm flux}/T_{\rm
ES}=0.62$--0.54.  Therefore, we expect that the $\nu_\mu$ and
$\nu_\tau$ flux from a SN core is roughly thermal with a temperature
$T_{\rm flux}\approx 0.6\,T_{\rm ES}$ and a luminosity given
approximately by the Stefan-Boltzmann law, taking $T_{\rm flux}$ for
the temperature and the area of the energy sphere for the radiating
surface. Only for $\bar\tau_{\rm ES}<30$ is the emerging flux
significantly diluted below the Stefan-Boltzmann level corresponding
to its spectral temperature $T_{\rm flux}$.

\begin{figure}
\columnwidth=7.1cm
\plotone{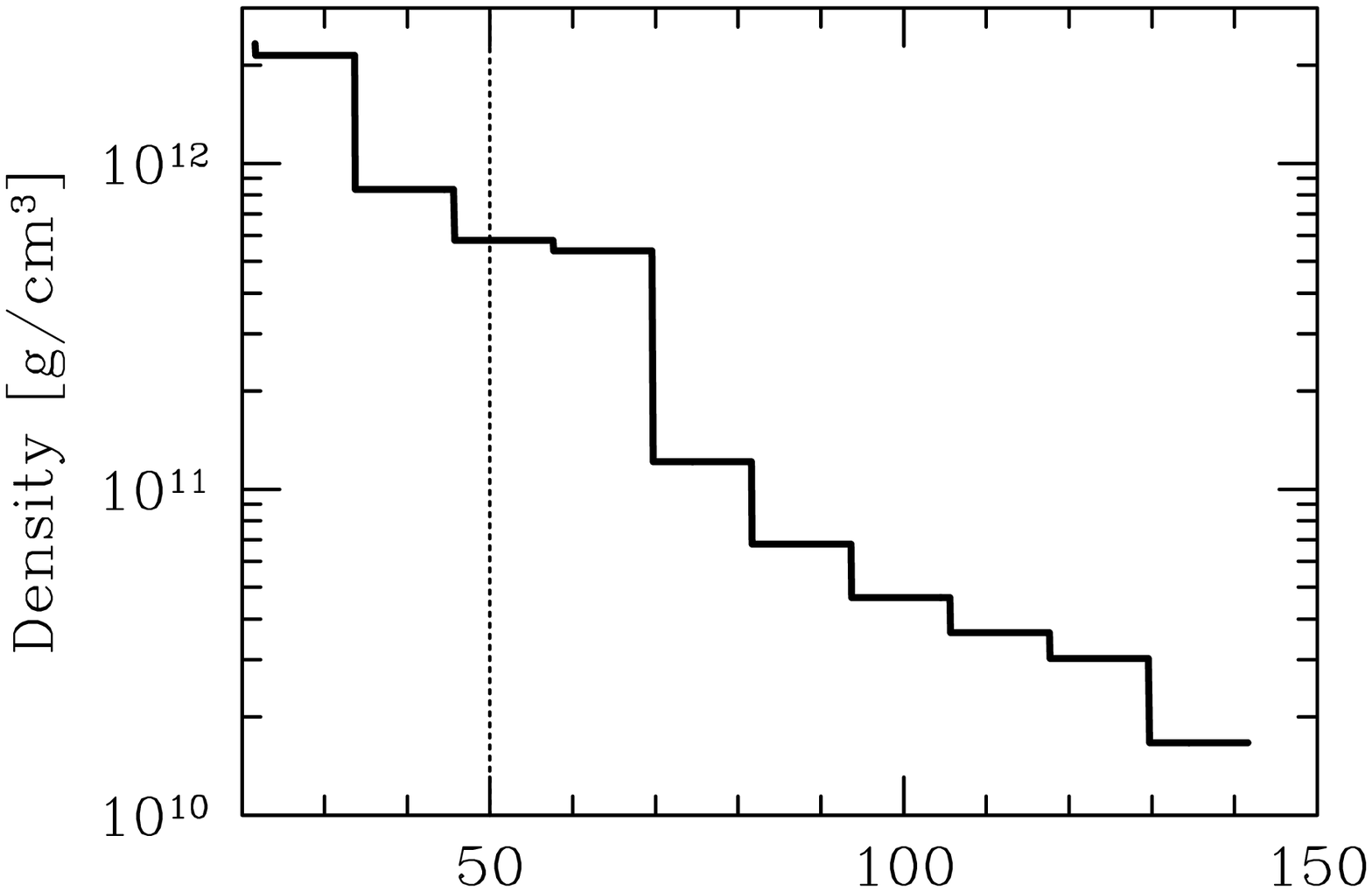}\\
\plotone{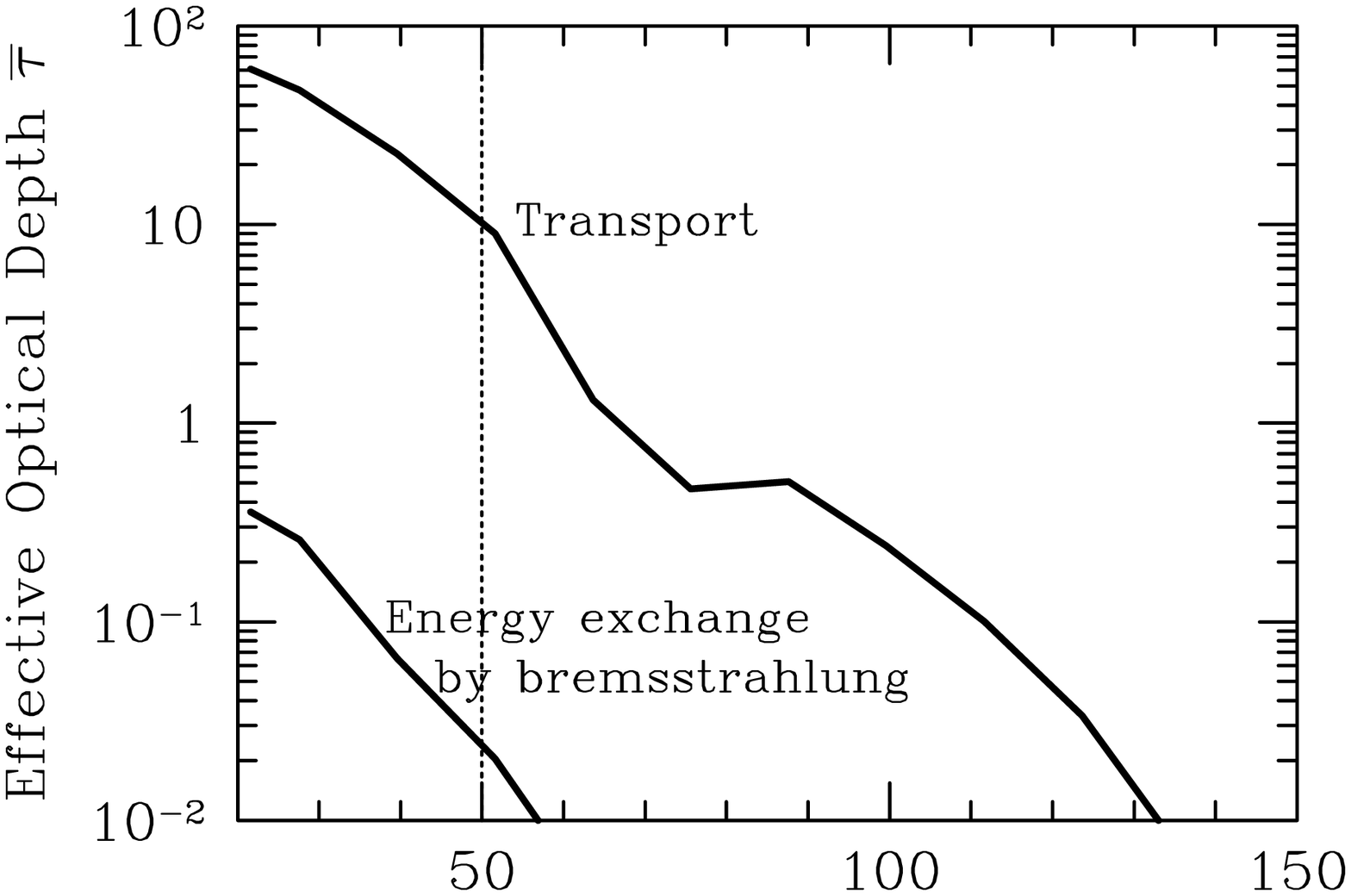}\\
\plotone{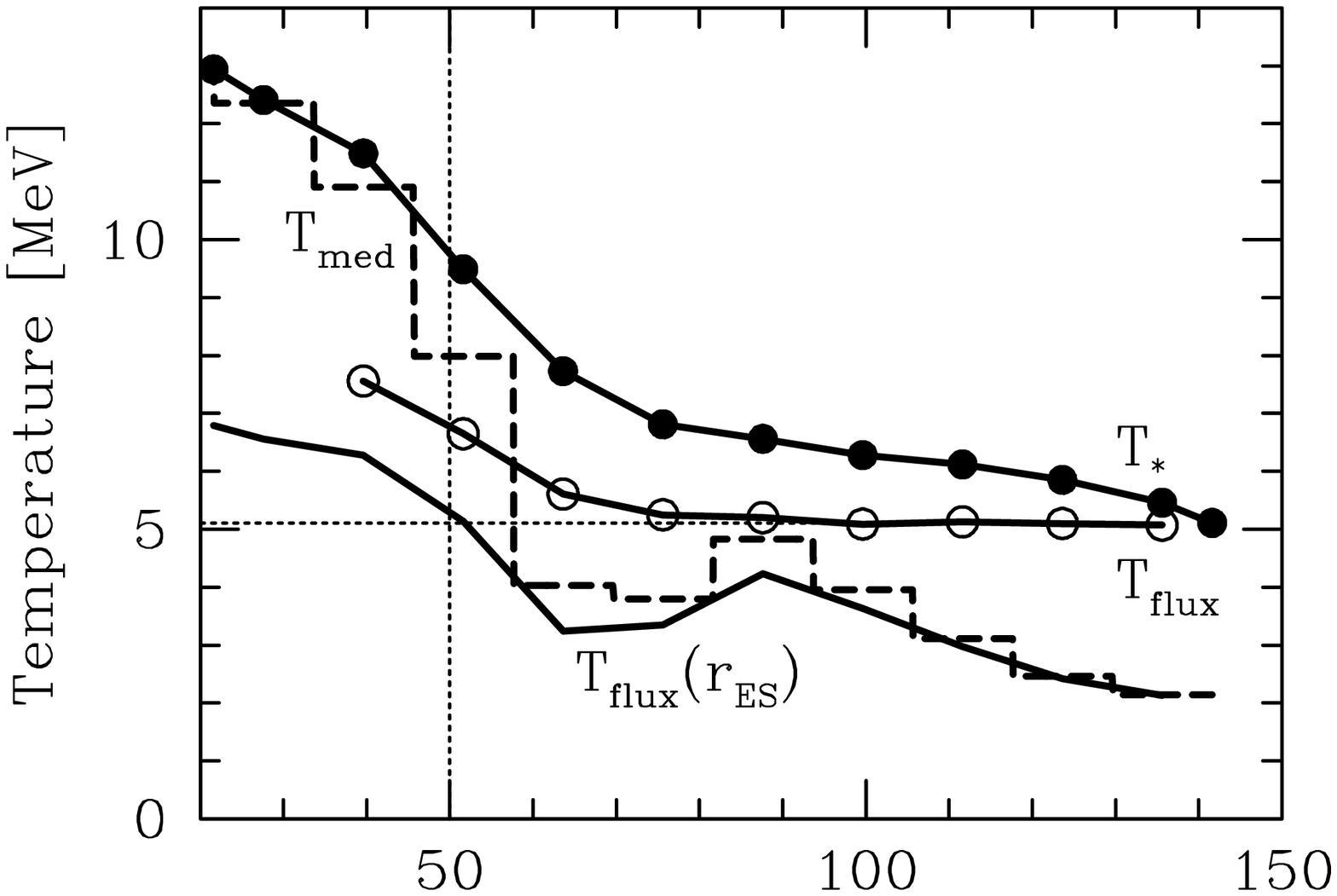}\\
\plotone{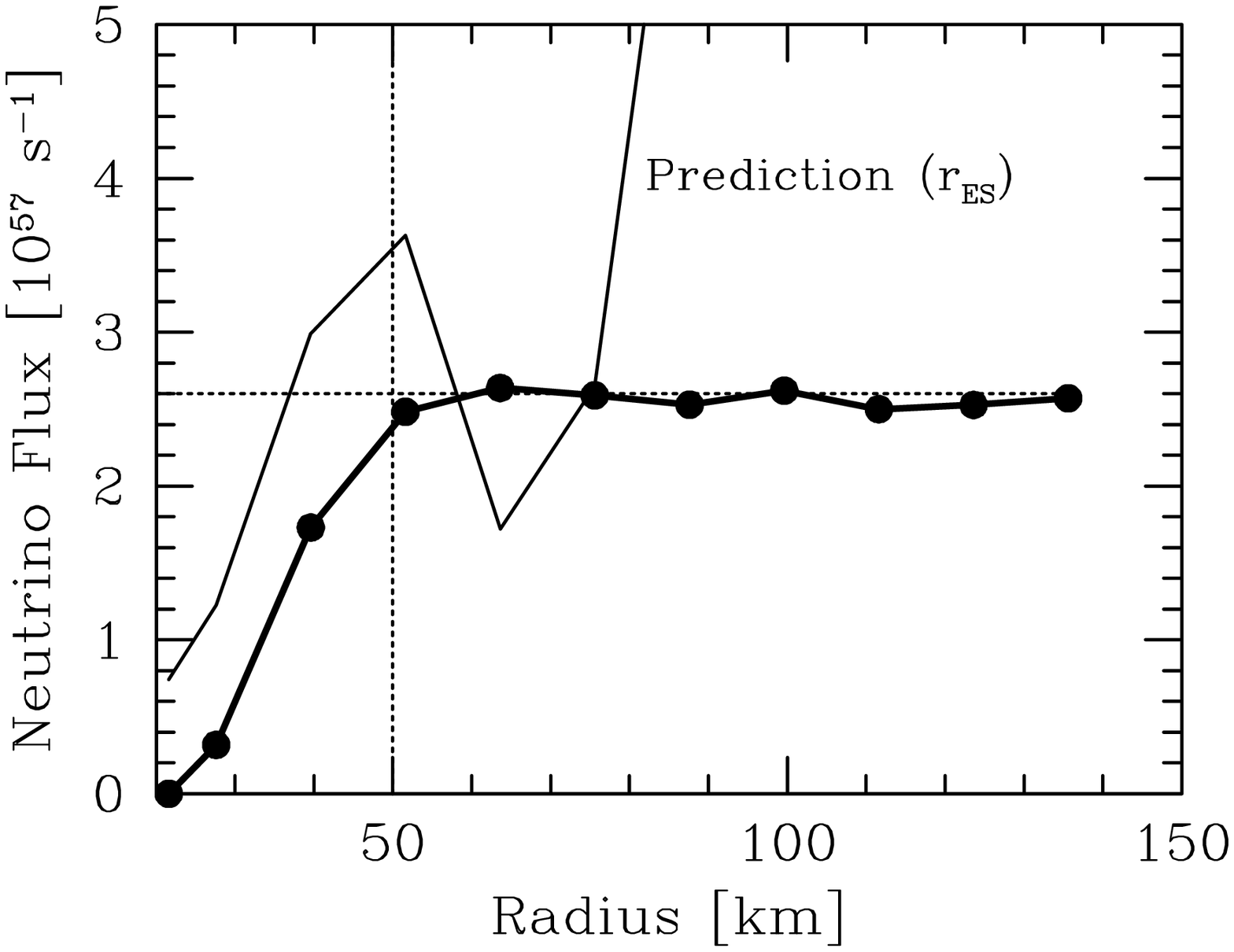}
\caption{\label{fig:jankamod}SN model for the Monte Carlo transport
  study of Janka \& Hillebrandt (1989b), Model~II.}
\end{figure}

\begin{figure}
\columnwidth=7.1cm
\plotone{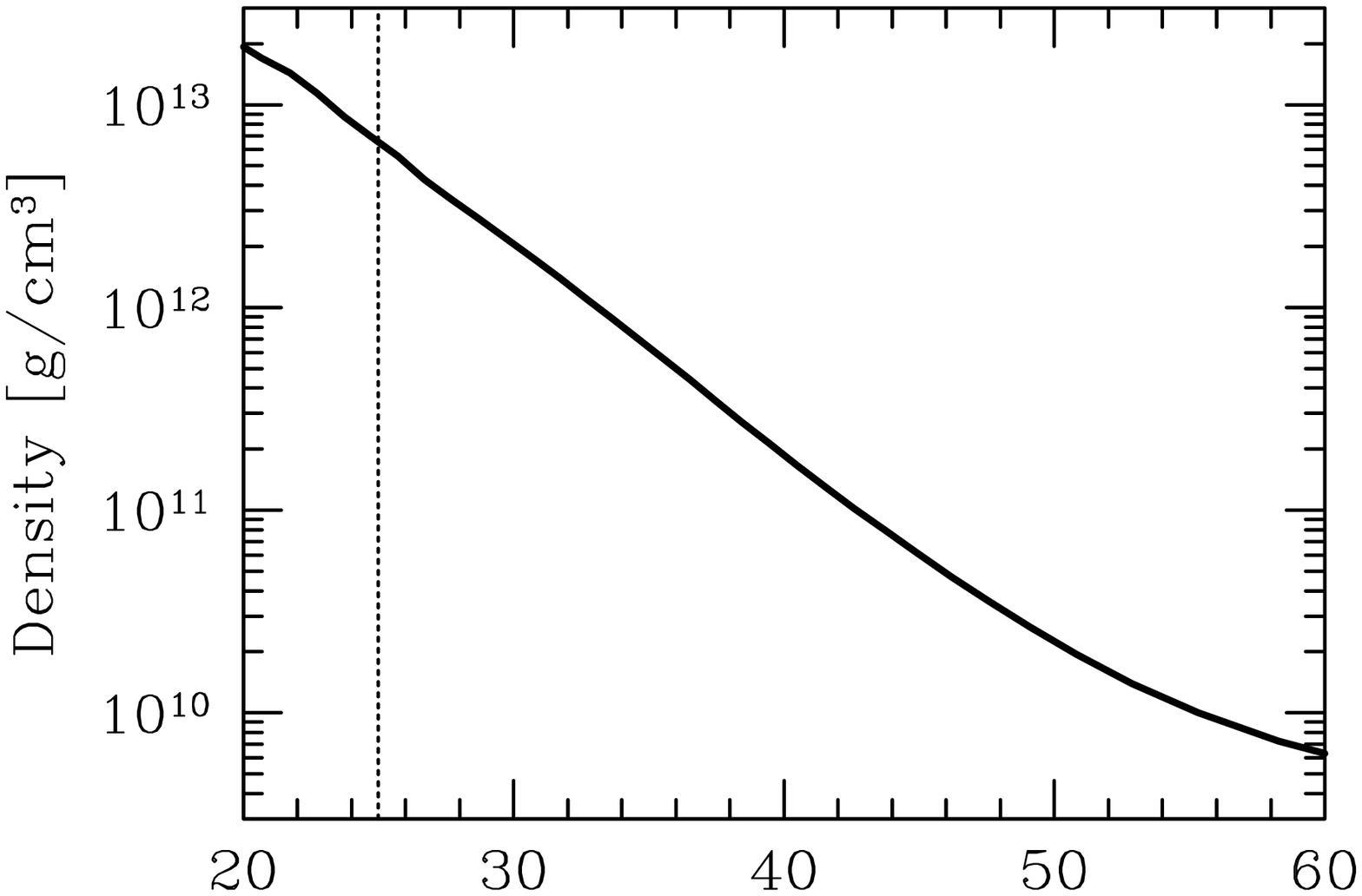}\\
\plotone{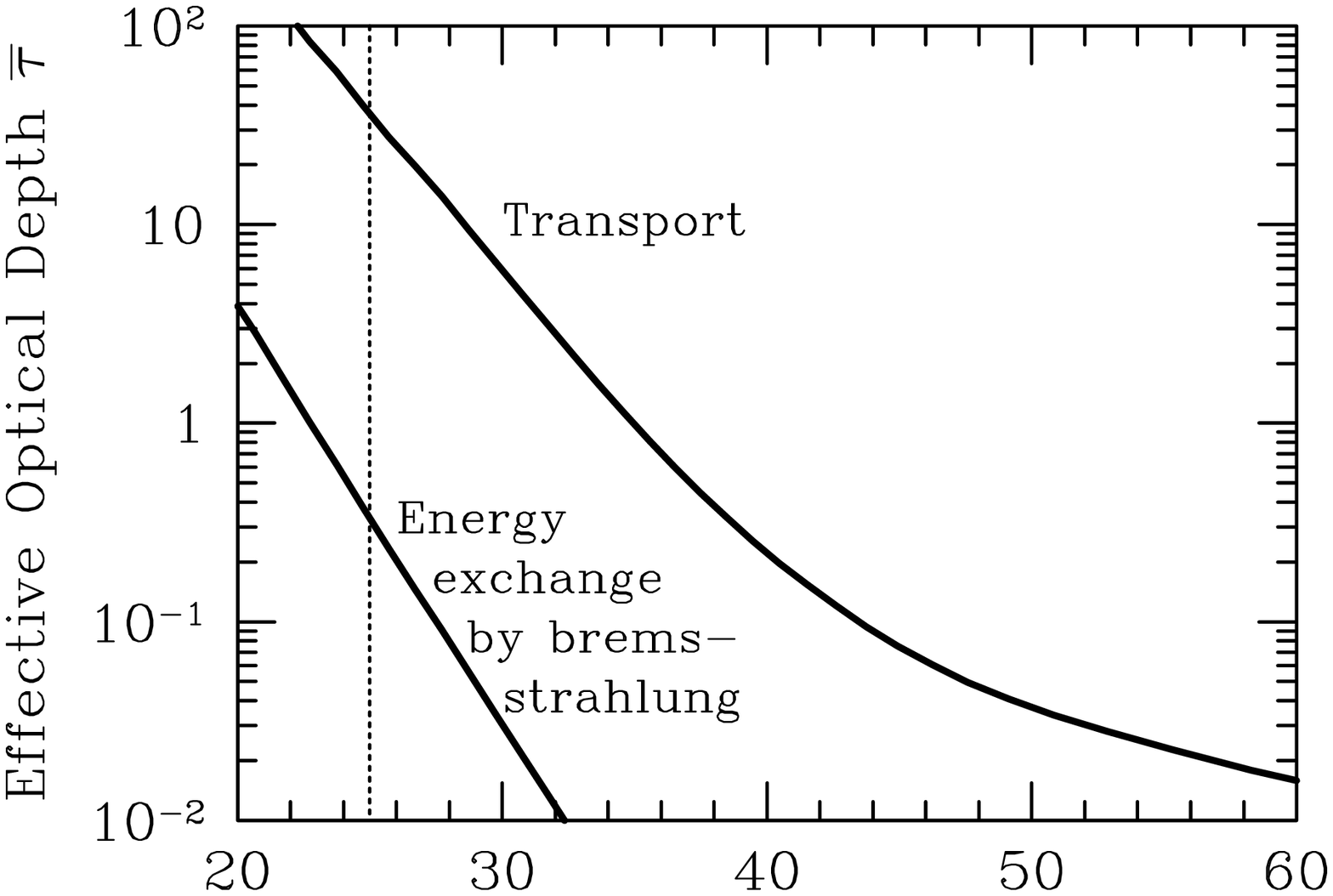}\\
\plotone{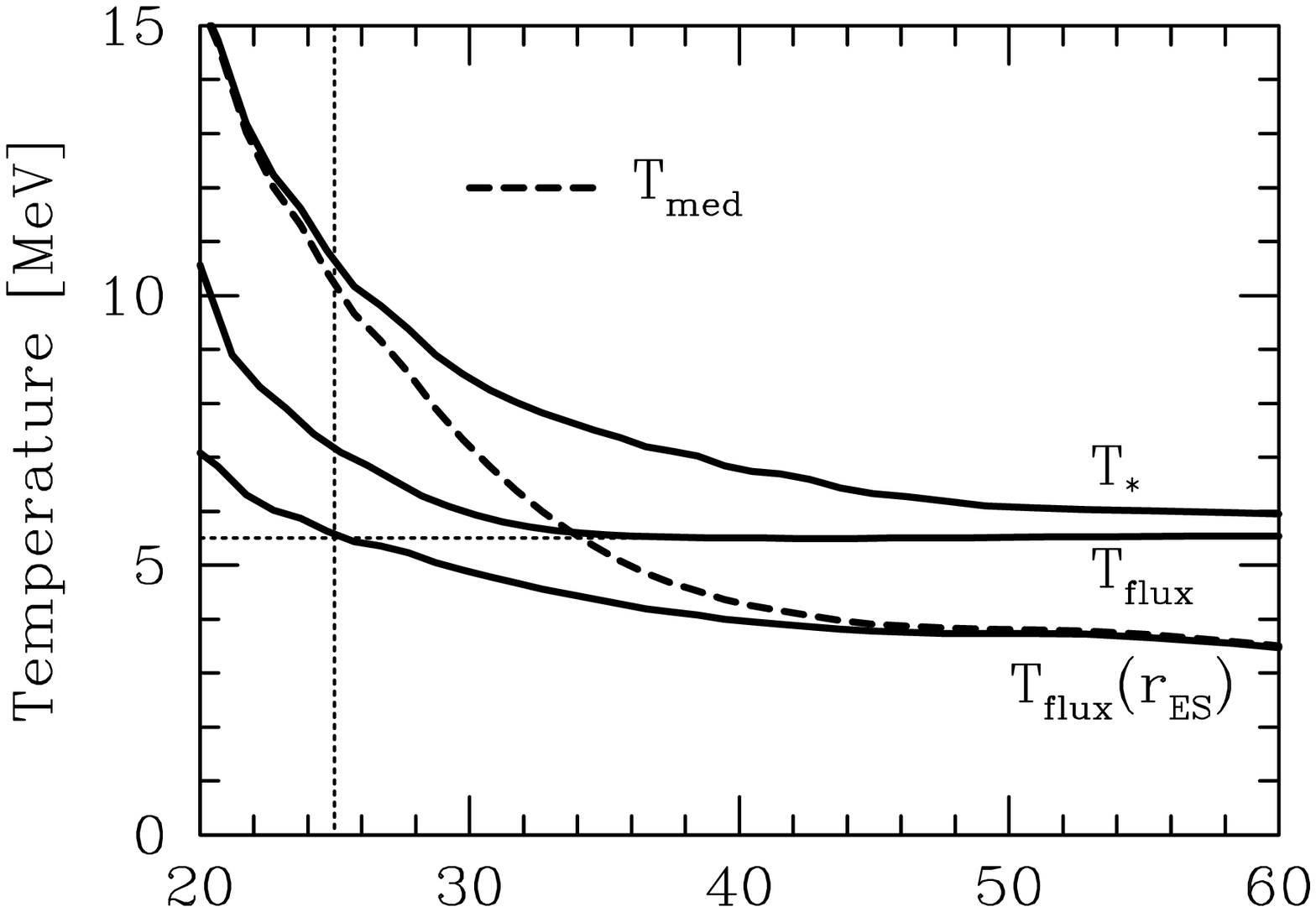}\\
\plotone{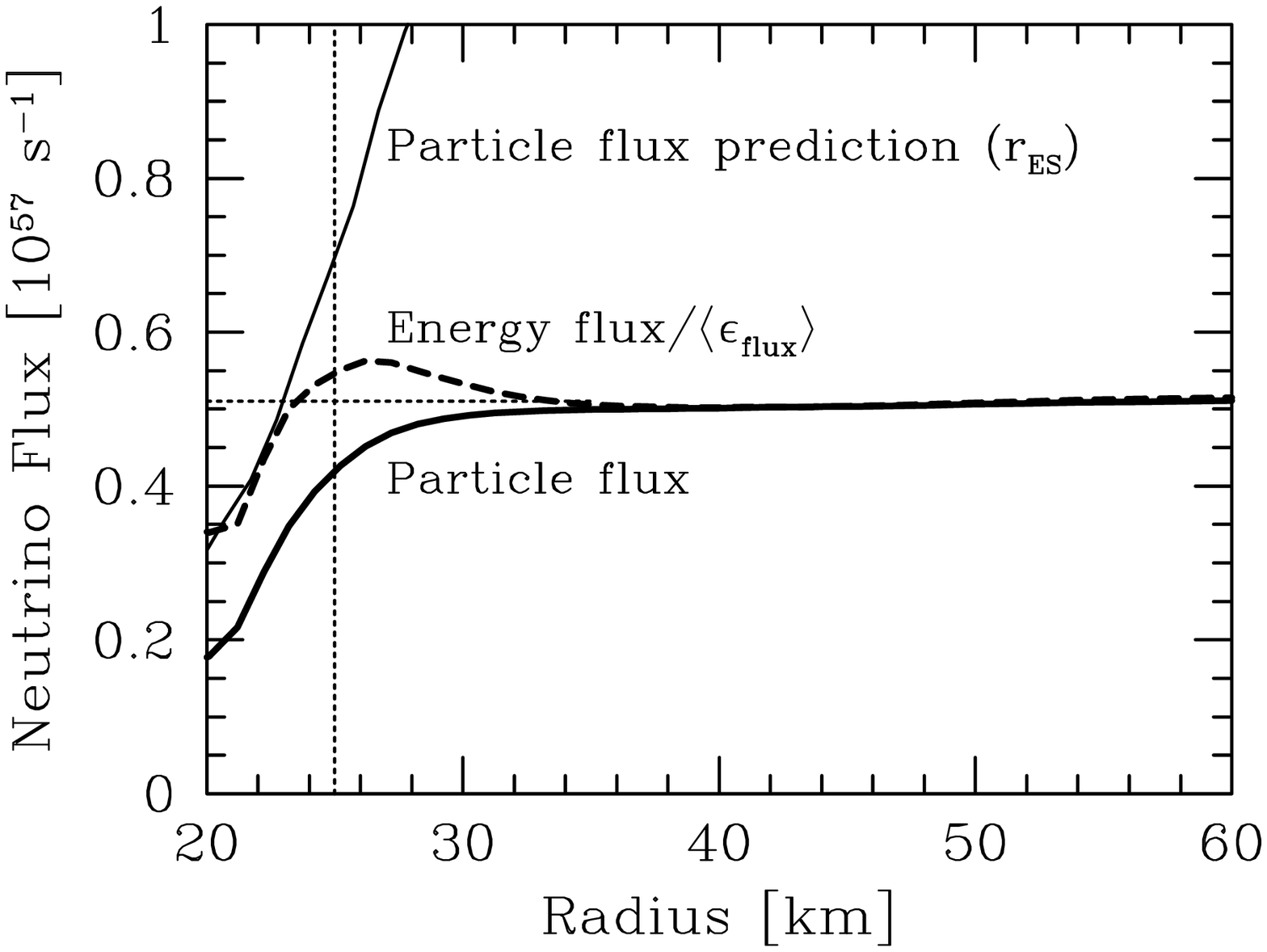}
\caption{\label{fig:messermod}SN model 324~ms after bounce
  from a Newtonian calculation (Messer et al.\ 2001).}
\end{figure}

\subsection{Comparing with Full-Scale Simulations}

\subsubsection{Janka \& Hillebrandt (1989b) Model}

As a ``reality check'' on our explanation of $\nu_\mu$ and $\nu_\tau$
spectra we turn to comparing our predictions with full-scale numerical
simulations in the literature. As a first case we consider the Monte
Carlo transport simulations of Janka \& Hillebrandt (1989a,b) where
the multi-flavor neutrino fluxes were calculated on the basis of a
fixed background model.  This model consisted of ten radial zones,
each 12~km wide; in Fig.~\ref{fig:jankamod} we show the temperature
and density profiles.  (I thank Thomas Janka for providing these
profiles.)  The local $\nu_\mu$ and $\nu_\tau$ temperature $T_*$ found
in this Monte Carlo simulation is indicated by filled circles in the
third panel of Fig.~\ref{fig:jankamod}; the flux temperature $T_{\rm
flux}$ by open circles.\footnote{The data are extracted from Table~IV
of Janka \& Hillebrandt (1989b). $T_*$ is obtained by dividing
$\langle\epsilon_\nu\rangle$ in the fifth column by 3.15 rather than
by 3 because Fermi-Dirac statistics were used here.  $T_{\rm flux}$ is
obtained by multiplying $T_*$ with $\xi_\epsilon/\xi_\nu$ which are
given in columns 11 and 9, respectively.}  As expected, $T_*$ and
$T_{\rm flux}$ coincide for large radii, but are vastly different in
regions where the neutrinos still scatter.

In the second panel of Fig.~\ref{fig:jankamod} we show the average
transport optical depth $\bar\tau$.  To this end we calculate the
transport optical depth, modified for the spherical case, and average
it thermally with the local medium temperature.  For the transport
opacity we use Eq.~(\ref{eq:scatteringopactiy}) which ignores the
vector-current contribution and also ignores nucleon degeneracy
effects, two approximations which work in opposite directions, and
anyway, are small corrections for the situation at hand.

On the basis of this $\bar\tau(r)$ and $T_{\rm med}(r)$ we calculate
for each radius the predicted flux temperature $T_{\rm flux}(r_{\rm
ES})$ of the escaping neutrino flux if the energy sphere were at that
radius.  This is the lowest curve in the third panel of
Fig.~\ref{fig:jankamod}. The actual $T_{\rm flux}$ at large radii
found in the Monte Carlo study is shown as a horizontal dotted line in
the third panel. It intersects the $T_{\rm flux}(r_{\rm ES})$ curve at
a radius of about 50~km, indicated by a vertical dotted line.  This
radius is $r_{\rm ES}$ if our picture is correct. We note that at
$r_{\rm ES}=50~{\rm km}$ we have $\bar\tau\approx 10$.

The neutrino temperature $T_*$ (top curve in the third panel)
separates from $T_{\rm med}$ at about the same radius so that it is
consistent to interpret 50~km as $r_{\rm ES}$.

Likewise, the numerical particle flux shown in the fourth panel
reaches its asymptotic value at about the same radius so that, indeed,
the energy sphere must be around 50~km. Within our simple picture we
can also predict the particle flux, assuming that the energy sphere is
at a given radius.  This prediction is shown as a thin line in the
fourth panel.  If $r_{\rm ES}=50~{\rm km}$ we would predict that a
neutrino flux of about $3.5\times10^{57}~{\rm s}^{-1}$ is emitted from
this SN model.  Comparing this with the Monte Carlo result of
$2.6\times10^{57}~{\rm s}^{-1}$, the agreement is not too bad.

In this numerical study the only number-changing reaction was
$e^+e^-\leftrightarrow \nu\bar\nu$, for energy exchange there was also
$\nu e\leftrightarrow e\nu$. As these processes may freeze out at
different radii, the simple picture of one energy sphere is certainly
not entirely adequate.  Therefore, it is quite surprising how well our
simple description can account for the main features of the Monte
Carlo results.

\subsubsection{Messer et~al.~(2001) Model}

As a second case we apply the same analysis to a model of Messer
et~al.\ (2001), provided to us by B.~Messer. It is based on a
full-scale Newtonian collapse simulation of the Woosley \& Weaver
$15\,M_\odot$ progenitor model labelled s15s7b, calculated with the SN
code described by Mezzacappa et~al.\ (2001).  We use a snapshot at
324~ms after bounce when the shock wave is at a radius of about
120~km, i.e.\ the SN core still accretes matter.  We plot in
Fig.~\ref{fig:messermod} the radial profile of the density, medium
temperature, $\nu_\mu$ temperatures $T_*$ and $T_{\rm flux}$, particle
flux, and luminosity (energy flux). Proceeding as before, we find that
the energy sphere should be at about 25~km (vertical dotted line) and
that $\bar\tau(r_{\rm ES})\approx 40$.  The main conclusions are the
same as in the previous example.

Our simple model for the $\nu_\mu$ and $\nu_\tau$ spectra apparently
works so well because the predicted flux temperature is rather
insensitive to the assumed location of the energy sphere.  From the
particle and energy fluxes in the bottom panel of
Fig.~\ref{fig:messermod} one might have located the energy
sphere, say, anywhere in the range $r_{\rm ES}=23$--29~km. One still
would have predicted $T_{\rm flux}$ within $\pm10\%$ of the numerical
value. Our approach of representing the energy sphere by a blackbody
surface looks like a robust approximation.

\subsection{Is {\boldmath$NN$} Bremsstrahlung Really Important?}

Within these specific models we can return to the question if
nucleonic bremsstrahlung $NN\leftrightarrow NN\nu\bar\nu$ is really
important relative to the leptonic processes that were actually
included in these simulations.  Based on the bremsstrahlung process
alone we have calculated the thermalization depth $\bar\tau_{\rm
therm}$ as defined in Eq.~(\ref{eq:thermalizationdepth}), averaged
over a thermal spectrum with the local medium temperature.  For
simplicity we have used the bremsstrahlung rate as in
Eq.~(\ref{eq:bremsmfp}) with $B(x)=1$.  We show $\bar\tau_{\rm therm}$
as the lower curve in the second panel of Figs.~\ref{fig:jankamod}
and~\ref{fig:messermod}.

In the Janka \& Hillebrandt (1989b) model $\bar\tau_{\rm therm}\ll1$
at the location of the energy sphere so that bremsstrahlung would have
been unimportant even if it had been included.  Bremsstrahlung depends
on $\rho^2$ and thus drops quickly in low-density regions.  This SN
atmosphere has rather low densities so that it is not surprising that
$NN$ bremsstrahlung is unimportant.

In the Messer et al.~(2001) model the actual energy sphere is
close to the bremsstrahlung thermalization sphere, i.e.\ the radius
where the lower curve in the second panel is 2/3.  Therefore, in this
model $NN$ bremsstrahlung would have been roughly competitive with the
leptonic processes that actually were included. Note that the
density at the energy sphere is about 10 times larger than in the
Janka \& Hillebrandt (1989b) case.  

In these models the atmospheres are still dominated by accretion.
After a successful explosion they will settle to more compact
dimensions, presumably enhancing the role of $NN$ bremsstrahlung. We
note that the model used by Hannestad \& Raffelt (1998) represented a
settled proto neutron star so that their finding of the importance of
bremsstrahlung is not inconsistent with the present discussion.
Apparently it depends on the detailed atmospheric structure which
processes predominate for the $\nu_\mu$ and $\nu_\tau$ thermalization.

\subsection{Sensitivity of 
{\boldmath$T_{\rm flux}$} to Thermalization Process}

We may explicitly test the sensitivity of the predicted flux
temperature $T_{\rm flux}$ to the strength of the assumed
thermalization process by means of a power-law model of the stellar
atmosphere as in Eq.~(\ref{eq:powerlaw}), i.e.\ $\rho\propto r^{-p}$
and $T\propto r^{-q}$.  For the scattering and energy-exchange rates
we assume
\begin{eqnarray}
\lambda_T^{-1}&\propto&\rho\,\epsilon^2\propto r^{-p}\,\epsilon^2,
\nonumber\\
\lambda_E^{-1}&\propto&\beta\,\rho^u\,T^v\,\epsilon^w
\propto \beta r^{-up-vq}\,\epsilon^w,
\end{eqnarray}
where for $NN$ bremsstrahlung approximately $u=2$, $v=3$ and $w=-1$.
Further, $\beta$ is a fudge factor which allows us to adjust
the strength of the energy-exchange process.
The thermalization rate then scales as
\begin{equation}
\left(\lambda_T^{-1}\lambda_E^{-1}\right)^{1/2}
\propto \beta^{1/2}\,r^{-x}\,\epsilon^y
\end{equation}
with $x=\frac{1}{2}[(u+1)p+vq]$ and $y=\frac{1}{2}(2+w)$.  Finally,
according to Eq.~(\ref{eq:powerlawflux}) we use
\begin{equation}\label{eq:tfluxpowerlaw}
T_{\rm flux}\propto T_{\rm ES}\,\bar\tau_{\rm ES}^{-t}
\end{equation}
with $t=0.128$.

It is now a matter of simple integrations to show that the transport
optical depth and thermalization depth scale as $\tau\propto
r^{-(p-1)}\,\epsilon^2$ and $\tau_{\rm therm}\propto
\beta^{1/2}\,r^{-(x-1)}\,\epsilon^y$.  The local thermal averages of
the energy expressions are $\langle\epsilon^2\rangle\propto T^2\propto
r^{-2q}$ and $\langle\epsilon^y\rangle\propto T^y\propto r^{-yq}$ so
that the thermally averaged optical depths are
\begin{eqnarray}
\bar\tau&\propto&r^{-(p+2q-1)},\nonumber\\
\bar\tau_{\rm therm}&\propto&\beta^{1/2}\,r^{-(x+yq-1)}.
\end{eqnarray}
The energy sphere is located where $\bar\tau_{\rm therm}=2/3$
so that 
\begin{equation}
r_{\rm ES}\propto\beta^{\textstyle\frac{1}{2(x+yq-1)}}
\end{equation}
Therefore, the optical depth and temperature at the
energy sphere scale as
\begin{eqnarray}
T_{\rm ES}&\propto&\beta^{-\textstyle\frac{q}{2(x+yq-1)}},\nonumber\\
\bar\tau_{\rm ES}&\propto&
\beta^{-\textstyle\frac{p+2q-1}{2(x+yq-1)}}.
\end{eqnarray}
Finally, with Eq.~(\ref{eq:tfluxpowerlaw}) we find
\begin{equation}
T_{\rm flux}\propto\beta^z
\end{equation}
with
\begin{equation}
z=\frac{t(p+2q-1)-q}{2(x+yq-1)}.
\end{equation}
If $NN$ bremsstrahlung is the thermalization process this
translates into
\begin{equation}
z_{\rm brems}=\frac{t(p+2q-1)-q}{3p+4q-2}.
\end{equation}
Since typically the density and temperature gradients are steep,
dropping the $-1$ and $-2$ terms will introduce only a small error so
that
\begin{equation}
z_{\rm brems}=\frac{t(1+2q/p)-q/p}{3+4q/p}.
\end{equation}
Noting that typically we will have $q/p=0.25$--0.35 we find that
$z_{\rm brems}$ is between $-0.015$ and $-0.105$. Taking this latter
number as typical, a factor of 3 uncertainty in the bremsstrahlung
rate translates roughly into a 10\% uncertainty of the predicted
$T_{\rm flux}$. Of course, this estimate depends on the temperature
and density profile remaining unchanged by the variation of $\beta$; a
self-consistent treatment is not possible with the present approach.

\subsection{Flavor Dependence of Spectral Temperature}

We may also address the question of the relative spectral flux
temperature between $\bar\nu_e$ or $\nu_e$ and the other flavors for
the case of a recoil-free scattering atmosphere. To this end we must
specify temperature and density profiles of the medium which we choose
as power laws of the form Eq.~(\ref{eq:powerlaw}) with $T_0$, $r_0$
and $\rho_0$ the energy-sphere values.  For a plane-parallel geometry,
the medium temperature profile as a function of the transport optical
depth $\tau(\epsilon)$ for a given neutrino energy $\epsilon$ is
easily found to be
\begin{equation}
\frac{T_{\rm med}}{T_{\rm ES}}=
\left(\frac{12\,T_{\rm ES}^2\,\tau(\epsilon)}
{\epsilon^2\,\bar\tau_{\rm ES}}\right)^u
\end{equation}
with $u=q/(p-1)$. If the density falls steeply so that $p\gg 1$ we
have $u\approx q/p$. Realistic values for this power-law index are in
the range, say, $u=0.25$--0.35.

For a given energy $\epsilon$ the $\nu_e$ or $\bar\nu_e$ decouple at
their respective energy spheres. The main energy-exchange process
relevant for these species are charged-current (CC) processes
involving electrons or positrons such as $\bar\nu_e+p\leftrightarrow
n+e^+$. In addition there are neutral-current (NC) elastic collisions
on nucleons.  Since both CC and NC processes have the same
$\epsilon^2$ behavior we may write $\lambda_{\rm
CC}^{-1}=\xi\lambda_{\rm NC}^{-1}$ with $\xi$ a factor which depends
on the medium's composition and the degree of electron and nucleon
degeneracy.  Moreover, $\xi$ is different for $\nu_e$ and
$\bar\nu_e$. In the absence of degeneracy effects we have
$\xi_{\nu_e}\approx 4\,(1-Y_e)$ and $\xi_{\bar\nu_e}\approx 4\,Y_e$
where $Y_e$ is the electron fraction per baryon.  Note that the CC
cross section is roughly 4 times the NC one and that only neutrons are
available as CC targets for $\nu_e$, and only protons for
$\bar\nu_e$. We now apply the Shapiro-Teukolsky criterion
Eq.~(\ref{eq:thermalizationdepth}) and find that the energy-dependent
mfp for thermalization is $\lambda_{\rm therm}^{-1}= (\lambda_{\rm
CC}^{-1}\lambda_{\rm NC}^{-1})^{1/2} =[\xi(1+\xi)]^{1/2}\,\lambda_{\rm
NC}^{-1}$.  Integrating over radius to obtain the optical depth then
yields $\tau_{\rm therm}(\epsilon)=[\xi(1+\xi)]^{1/2}\tau(\epsilon)$
where $\tau(\epsilon)$ is the NC transport optical depth. The
energy-dependent energy sphere is then located where $\tau_{\rm
therm}(\epsilon)=2/3$ so that the NC transport optical depth at this
location is $\tau(\epsilon)=(2/3)[\xi(1+\xi)]^{-1/2}$. The above
profile then informs us that
\begin{equation}
\frac{T_\epsilon}{T_{\rm ES}}=
\left(\frac{8}{\bar\tau_{\rm ES}\sqrt{\xi(1+\xi)}}\right)^u
\left(\frac{T_{\rm ES}}{\epsilon}\right)^{2u}
\end{equation}
is the medium temperature at the energy dependent energy spheres for
$\nu_e$ or $\bar\nu_e$.

The main assumption for calculating the $\nu_e$ or $\bar\nu_e$ spectra
is that every energy group is emitted with a thermal flux
corresponding to $T_\epsilon$, i.e.\ we ignore chemical-potential
effects. The escaping spectrum is then proportional to
$\epsilon^2\,e^{-\epsilon/T_\epsilon}$. Note that every energy group
is at the same transport optical depth because both NC and CC
processes have the same energy dependence. (For every energy group the
energy sphere is at $\tau_{\rm therm}(\epsilon)=2/3$ so that this
location also corresponds to a common value for the transport optical
depth.) Therefore, the flux dilution caused by NC scatterings does not
modulate the spectral shape.

It is now a simple matter of integration to determine the $\nu_e$ or
$\bar\nu_e$ flux temperature and pinching parameter,
\begin{eqnarray}
T_{{\rm flux},\nu_e}&=&
\left(\frac{8}{\bar\tau_{\rm ES}\sqrt{\xi(1+\xi)}}\right)^{\textstyle
\frac{u}{1+2u}}
\frac{\Gamma\left(\frac{5+2u}{1+2u}\right)}
{\Gamma\left(\frac{4+2u}{1+2u}\right)},
\nonumber\\
p_{\nu_e}&=&\frac{4\,\Gamma\left(\frac{4+2u}{1+2u}\right)
\Gamma\left(\frac{6+2u}{1+2u}\right)}
{5\left[\Gamma\left(\frac{5+2u}{1+2u}\right)\right]^2},
\end{eqnarray}
where $\Gamma$ is the Gamma function.  These spectra are always
pinched because $p<1$ for $u\geq 0$.  Around the realistic value
$u=0.3$ we find the expansion
\begin{eqnarray}
T_{{\rm flux},\nu_e}&\approx&
\left(\frac{8}{\bar\tau_{\rm ES}\sqrt{\xi(1+\xi)}}\right)^{0.19+\Delta u}
\nonumber\\
\noalign{\medskip}
&\times&
\bigl(0.465 - 0.65\,\Delta u\bigr),
\nonumber\\
\noalign{\smallskip}
p_{\nu_e}&\approx&
0.911 - 0.18\,\Delta u,
\end{eqnarray}
where $\Delta u\equiv u-0.3$.

As a first consequence we can estimate the relative spectral
temperature between $\nu_e$ and $\bar\nu_e$,
\begin{equation}
\frac{T_{{\rm flux},\nu_e}}{T_{{\rm flux},\bar\nu_e}}
=\left(\frac{4Y_e(4Y_e+1)}{4(1-Y_e)[4(1-Y_e)+1]}
\right)^{\frac{u}{2(1+2u)}}.
\end{equation}
For $u=0.3$ and to lowest order in $Y_e$ this is
$0.69\,(Y_e/0.1)^{0.094}$, in the ball park of what is found in
numerical simulations.

\eject

With the power-law representation of the $\nu_\mu$ flux temperature of
Eq.~(\ref{eq:powerlawflux}) we find
\begin{eqnarray}
\frac{T_{{\rm flux},\nu_\mu}}{T_{{\rm flux},\nu_e,\bar\nu_e}}
&\approx&\Bigl(1.39 + 1.9\,\Delta u\Bigr)\nonumber\\
&\times&[\xi(1+\xi)]^{0.094+0.2\,\Delta u}\nonumber\\
&\times&
\left(\frac{\bar\tau_{\rm ES}}{10}\right)^{0.06+0.4\,\Delta u}.
\end{eqnarray}
For $\bar\nu_e$ we have $\xi(1+\xi)=4Y_e(1+4Y_e)$. Thus for small
$Y_e$ we have to lowest order
\begin{eqnarray}
\frac{T_{{\rm flux},\bar\nu_\mu}}{T_{{\rm flux},\bar\nu_e}}
&\approx&\Bigl(1.32 + 1.8\,\Delta u\Bigr)\nonumber\\
&\times&(Y_e/0.1)^{0.094+0.2\,\Delta u}\nonumber\\
&\times&
\left(\frac{\bar\tau_{\rm ES}}{10}\right)^{0.06+0.4\,\Delta u}.
\end{eqnarray}
This ratio is almost independent of the optical depth of the $\nu_\mu$
energy sphere; changing $\bar\tau_{\rm ES}$ by a factor of 2 modifies
the temperature ratio by around 5\%.  A change $\Delta u=0.05$ implies
a 7\% modification of the temperature ratio.  Finally, changing $Y_e$
by a factor of 2 causes something like a 7\% modification.

In deriving this result we have made several simplifying assumptions
so that the final answer should not be overinterpreted.  Still, it is
gratifying that our simple model, without any fine-tuning, reproduces
relative neutrino flux temperatures which fully concur with what is
found in numerical simulations. This finding, again, encourages us to
put some faith in our explanation of the SN neutrino spectra.


\section{Nucleon Recoils}

\label{sec:NucleonRecoils}

\subsection{First Example}

\begin{figure}[t]
\columnwidth=6.5cm
\plotone{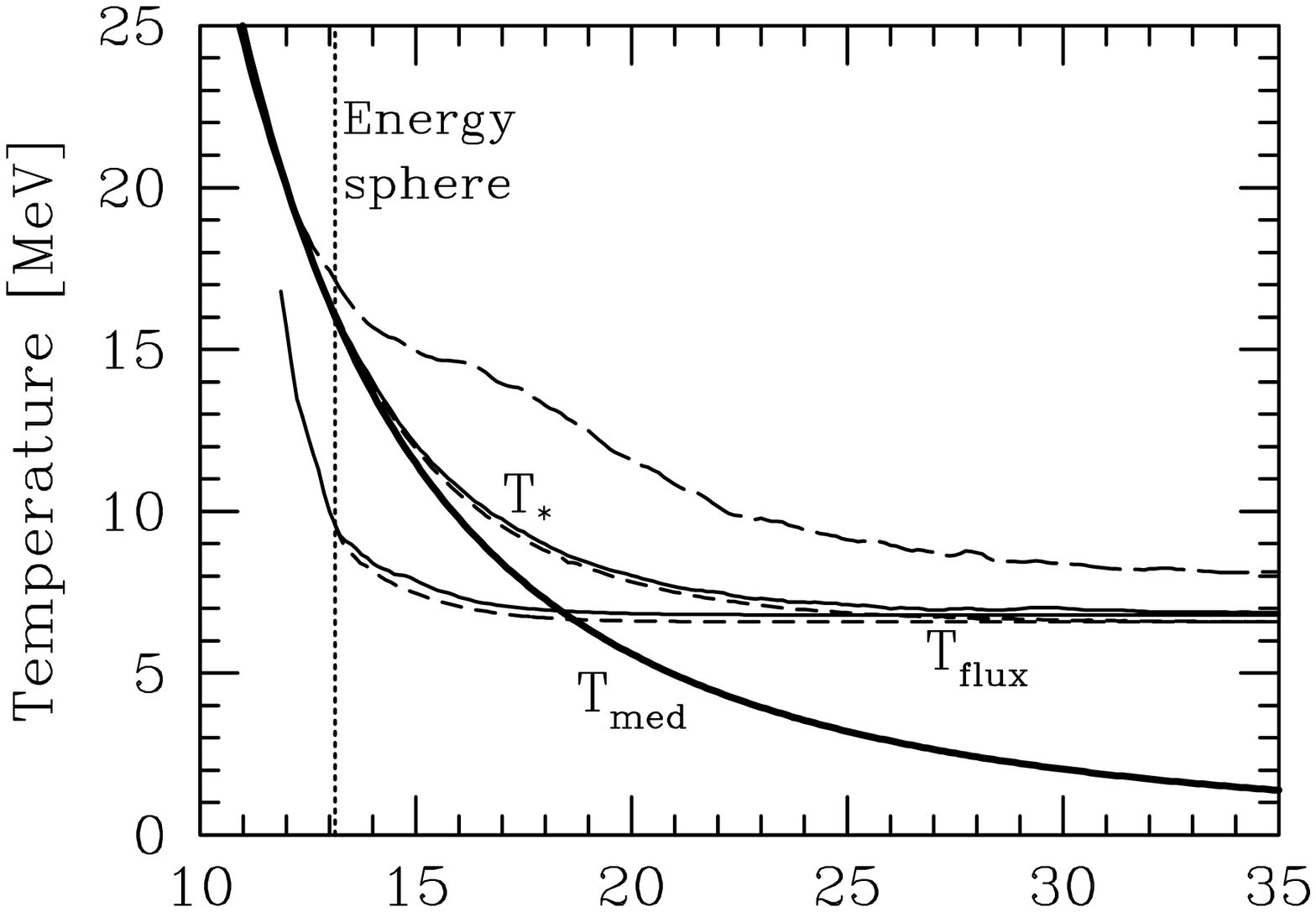}\\
\plotone{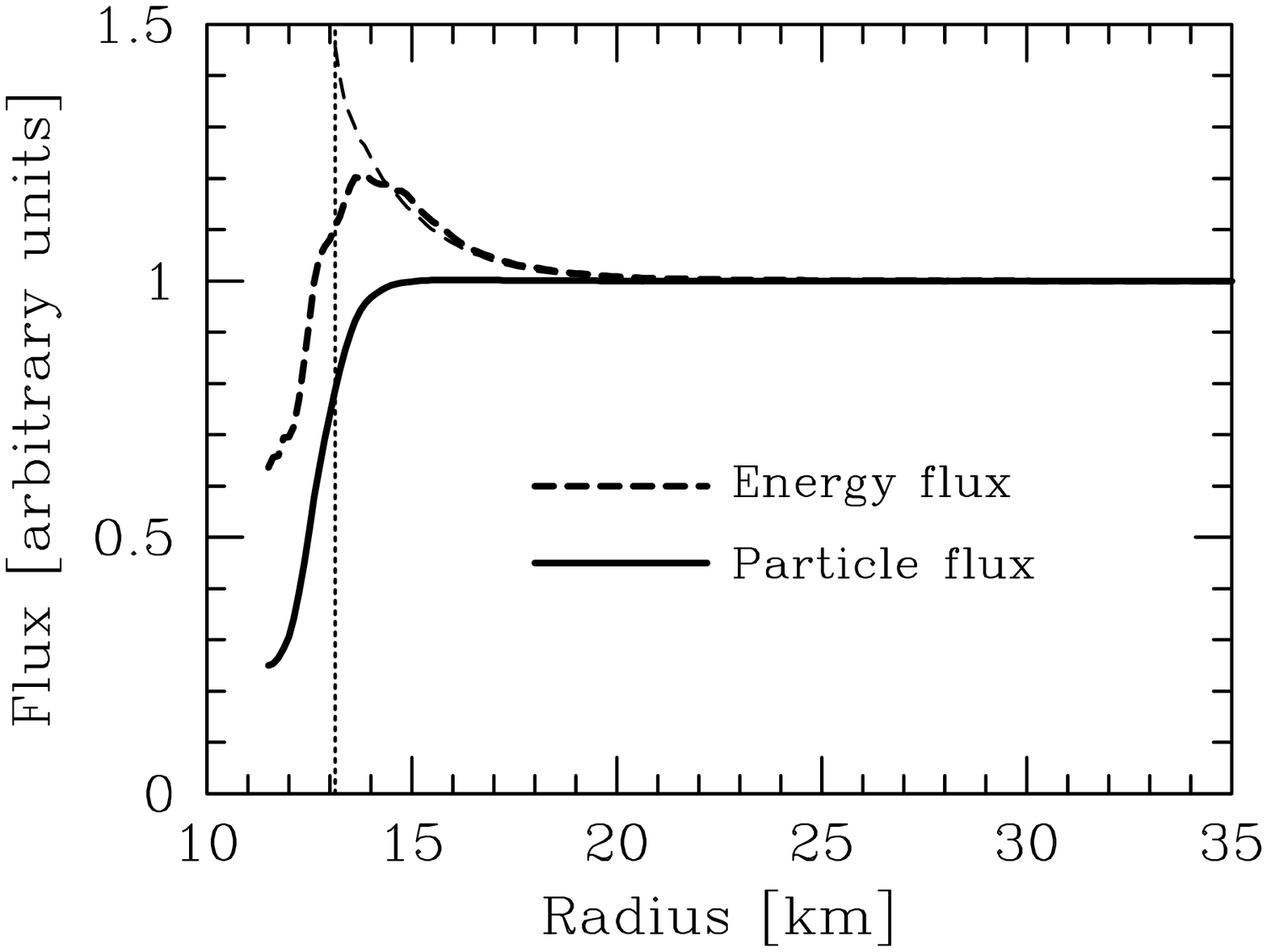}
\caption{\label{fig:recoilprofile}Example
  of Fig.~\protect\ref{fig:profile} with nucleon recoils included. The
  thin dashed lines refer to the case with a blackbody boundary
  condition $T_{\rm ES}=16~{\rm MeV}$ at the energy sphere.  The
  long-dashed line is $T_*$ for the no-recoil case
  of Fig.~\protect\ref{fig:profile}.}
\end{figure}

We now turn to the main goal of our investigation, the effect of
including nucleon recoils in the scattering atmosphere.  To illustrate
the type and magnitude of effect to be expected we begin with the same
plane-parallel power-law model that was shown in
Fig.~\ref{fig:profile}, except that now we allow for recoil energy
transfers in $\nu N$ collisions according to the formalism described
in Appendix~\ref{sec:NeutrinoNucleonScattering}.  The resulting radial
runs of parameters are shown in Fig.~\ref{fig:recoilprofile}

Nucleon recoils cause the neutrino temperature $T_*$ to follow more
closely the medium; the no-recoil case is shown as a long-dashed line.
The $T_{\rm flux}$ profile is shifted downward so that the neutrinos
emerge with a lower $T_{\rm flux}$ at the surface.  The luminosity
profile has a ``bump'' near the energy sphere, indicating that the
neutrinos transfer energy to the medium within the scattering
atmosphere.

The energy sphere shown here is the same as in
Fig.~\ref{fig:recoilprofile}, i.e.\ strictly speaking we mean the
thermalization depth of $NN$~bremsstrahlung when we say ``energy
sphere.''  Of course, in the present situation the concept of an
energy sphere no longer applies in the sense that nucleon recoils
allow for energy transfers throughout the scattering atmosphere, which
in turn no longer is strictly a scattering atmosphere.

Next, we switch off $NN$ bremsstrahlung entirely and use a
blackbody boundary condition with $T_{\rm ES}=16~{\rm MeV}$ at $r_{\rm
ES}$.  The resulting profiles are shown as short-dashed thin lines.
We practically obtain the same $T_*$ and $T_{\rm flux}$ profiles
whether we allow the neutrinos to thermalize by $NN$ bremsstrahlung or
if we enforce a blackbody surface at $r_{\rm ES}$, confirming our
previous arguments.

This approach has one more advantage. In general, $\Delta T_{\rm
  flux}$ depends on both the density and temperature profiles of
the scattering atmosphere.  However, if $\nu N$ scattering is the only
remaining process, then the only natural measure of distance is the
transport optical depth~$\tau$.  Put another way, the medium density
disappears from the Boltzmann collision equation if we use a quantity
proportional to $\int dr\, n_B(r)$ as a radial coordinate.  For us it
is most practical to use the transport optical depth, averaged over a
thermal neutrino spectrum at $T_{\rm ES}$.  Note that we use the same
thermal average everywhere so that the $T_{\rm med}$ profile does not
enter the transformation from $r$ to $\bar\tau$. 
Then the only relevant profile is $T_{\rm med}(\bar\tau)$.

\begin{figure}[b]
\columnwidth=6.8cm
\plotone{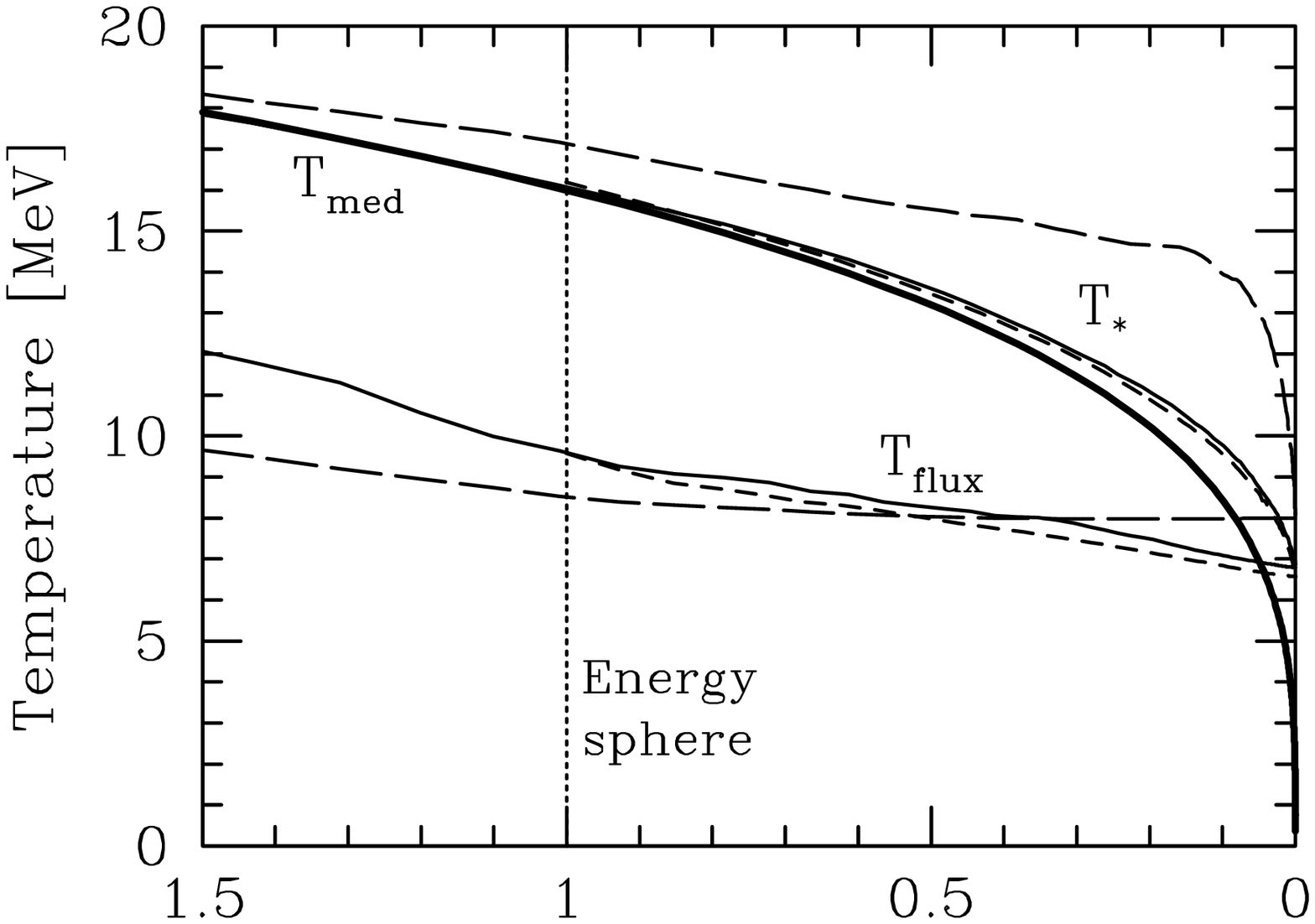}\\
\plotone{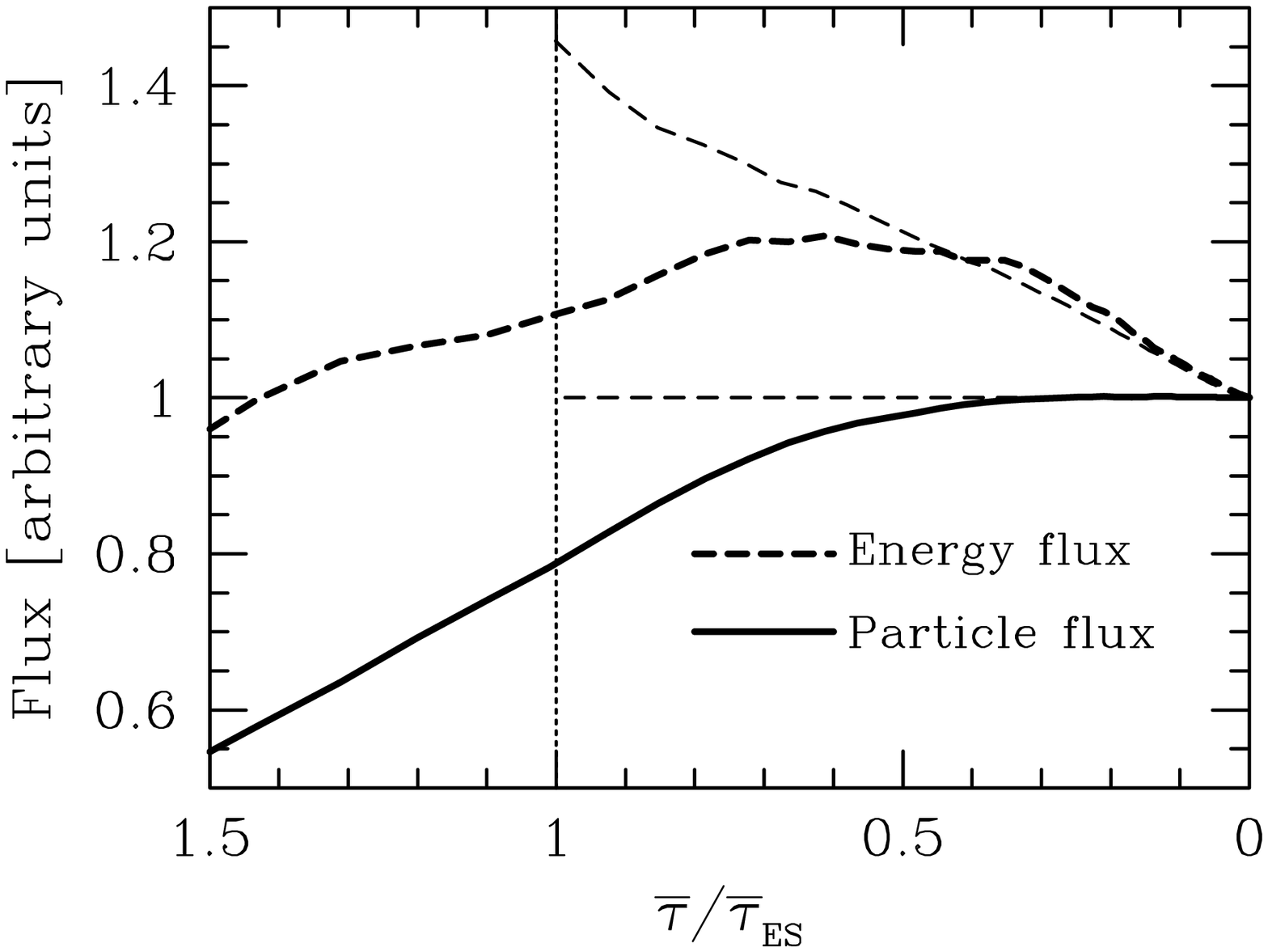}
\caption{\label{fig:recoilprofile2}
  Same as Fig.~\protect\ref{fig:recoilprofile}, except with the
  optical depth as the radial coordinate.}
\end{figure}

Therefore, radial profiles such as Fig.~\ref{fig:recoilprofile} are
actually plotted in the ``wrong'' coordinate for issues related to
neutrino transport. For illustration we show the same profiles in
Fig.~\ref{fig:recoilprofile2} with $\bar\tau$ as the radial
coordinate, scaled to $\bar\tau_{\rm ES}$. We recognize that all
temperature profiles vary slowly in the neighborhood of the energy
sphere.

The ``bump'' in the luminosity profile turns out to extend throughout
the scattering atmosphere.  For the blackbody boundary condition, the
luminosity decreases nearly linearly toward the surface so that the
neutrinos transfer energy to the medium throughout the scattering
atmosphere with approximately the same rate everywhere.  It is
interesting that the luminosity profiles are very different between
the $NN$-thermalization case and that with a blackbody boundary
condition.

\subsection{Constant Temperature}

Turning to a systematic study of the impact of nucleon recoils we
consider a scattering atmosphere with a fixed temperature $T_{\rm
  med}$, whereas the blackbody surface at the bottom is at the
temperature $T_{\rm ES}\geq T_{\rm med}$. What is the spectrum of the
emerging neutrinos?

Actually not even the sign of $\Delta T_{\rm flux}$ caused by nucleon
recoils is a priori obvious. Since $T^0_{\rm flux}<T_{\rm ES}$ it is
conceivable that energy exchange will {\it increase\/} $T_{\rm flux}$
in a situation when $T_{\rm med}>T^0_{\rm flux}$.  (The superscript 0
refers to the no-recoil case.)  Conversely, one expects the energies
of the emerging neutrinos to be lowered when $T_{\rm med}<T^0_{\rm
flux}$.

\begin{figure}[t]
\columnwidth=6.5cm
\plotone{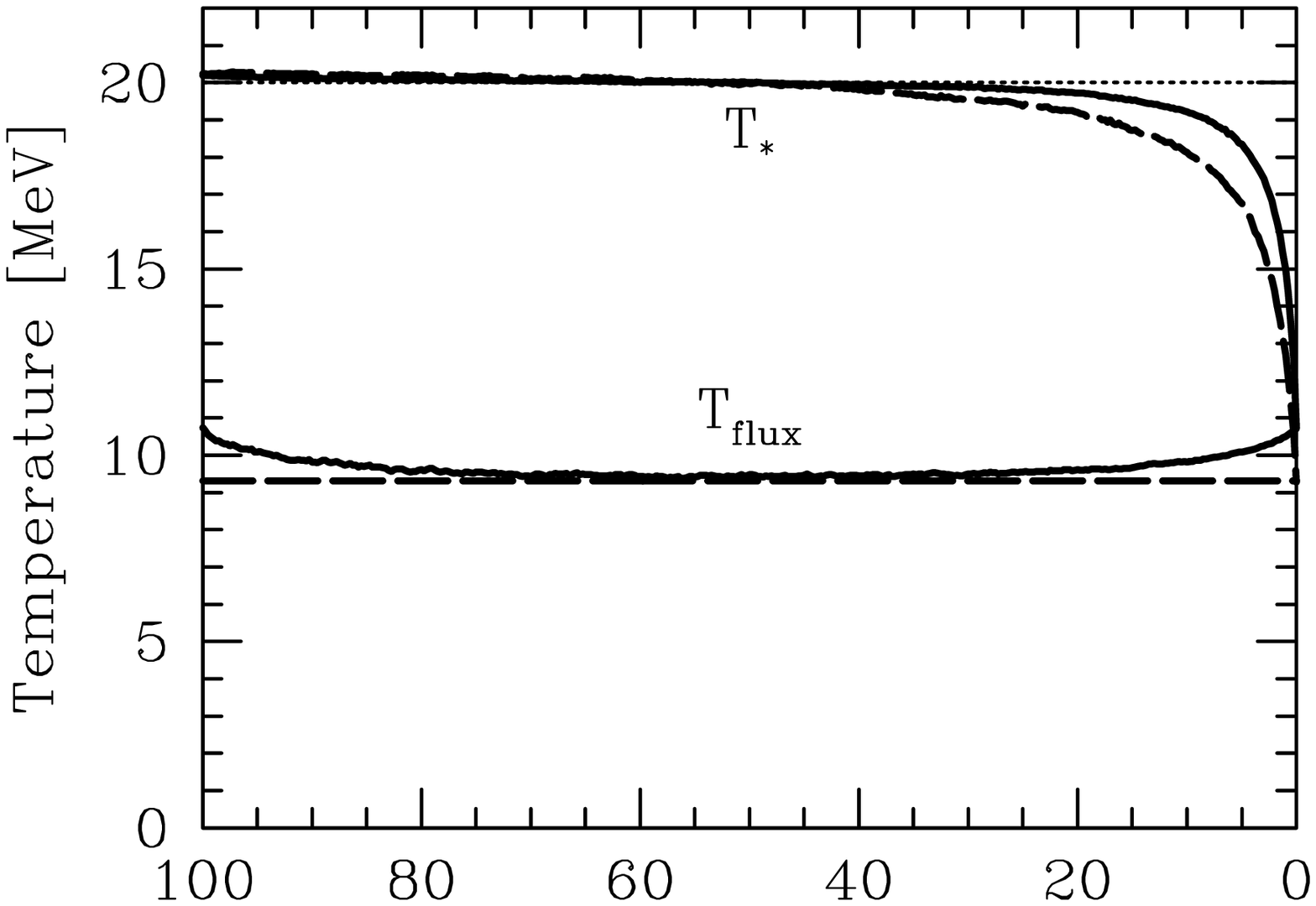}\\
\plotone{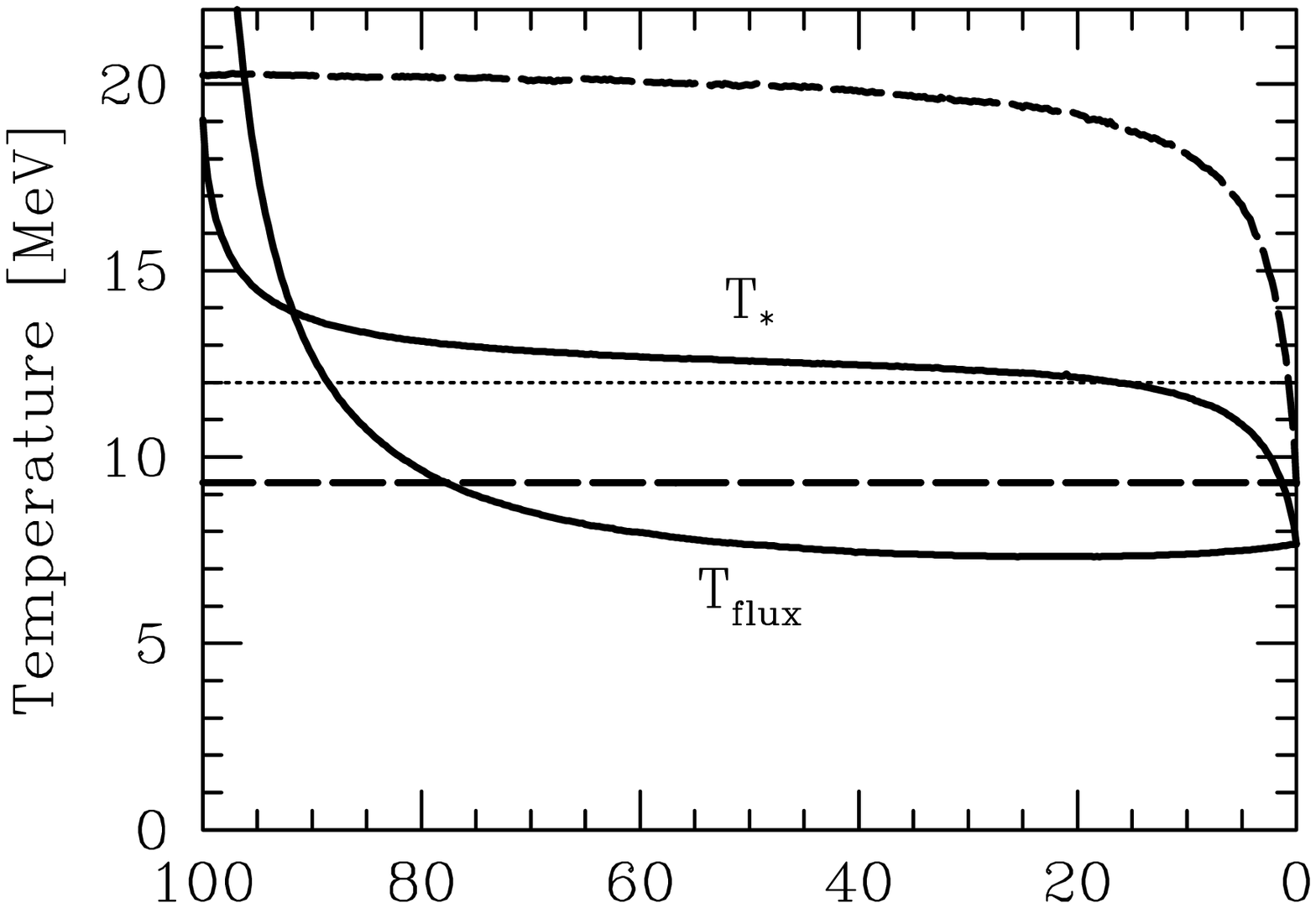}\\
\plotone{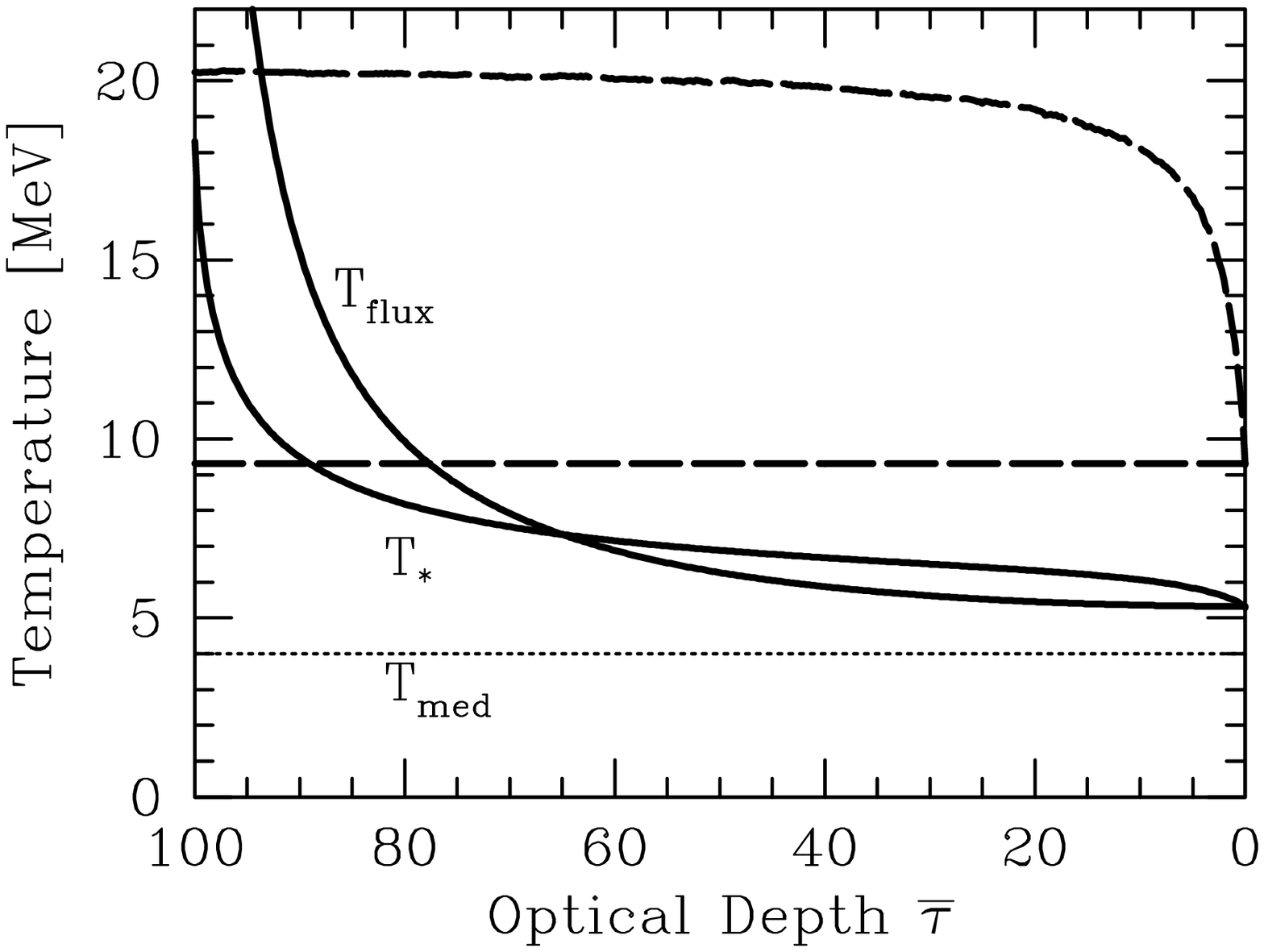}
\caption{\label{fig:fixprofile} Neutrino temperature profiles $T_*$
and $T_{\rm flux}$ for $T_{\rm ES}=20~{\rm MeV}$ and
$T_{\rm med}=20$, 12 and $4~{\rm MeV}$ (top to bottom).  Dashed lines
for the no-recoil case, dotted lines for $T_{\rm med}$.}
\end{figure}

The real answer is more complicated. In Fig.~\ref{fig:fixprofile} we
show the $T_*$ and $T_{\rm flux}$ profiles for a scattering atmosphere
with $\bar\tau_{\rm ES}=100$ and $T_{\rm ES}=20~{\rm MeV}$.  The
dashed lines are for the no-recoil case.  The three panels are for the
different medium temperatures 20, 12, and 4~MeV as indicated by dotted
lines.

\begin{figure}[t]
\columnwidth=6.5cm
\plotone{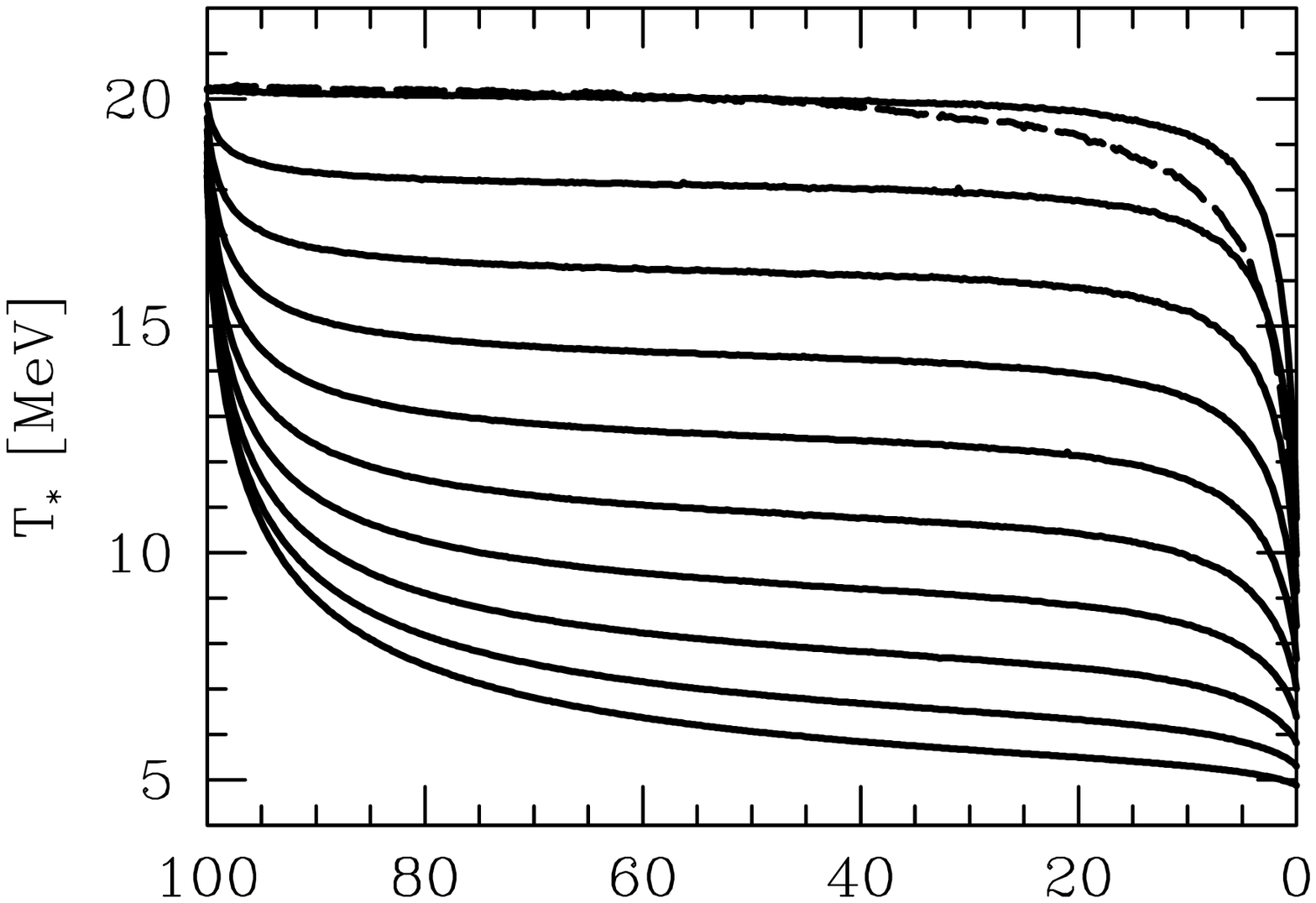}\\
\plotone{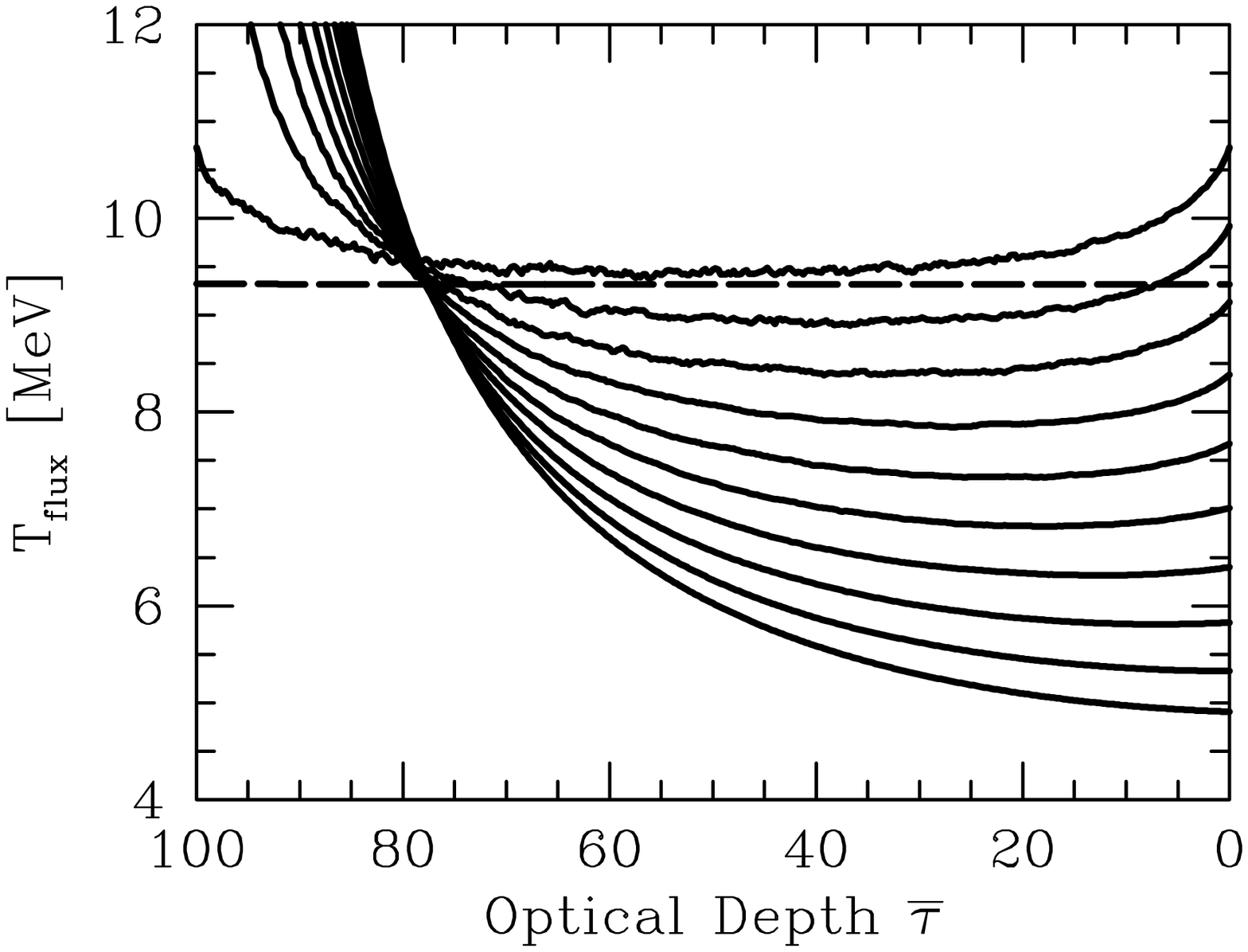}
\caption{\label{fig:fixprofile2} Neutrino temperature profiles $T_*$
and $T_{\rm flux}$ for $T_{\rm ES}=20~{\rm MeV}$.  From top to bottom
the curves are for $T_{\rm med}=20$ to 2~MeV in steps of
2~MeV.  Dashed lines are for the no-recoil case.}
\end{figure}

\begin{figure}[t]
\columnwidth=7cm
\plotone{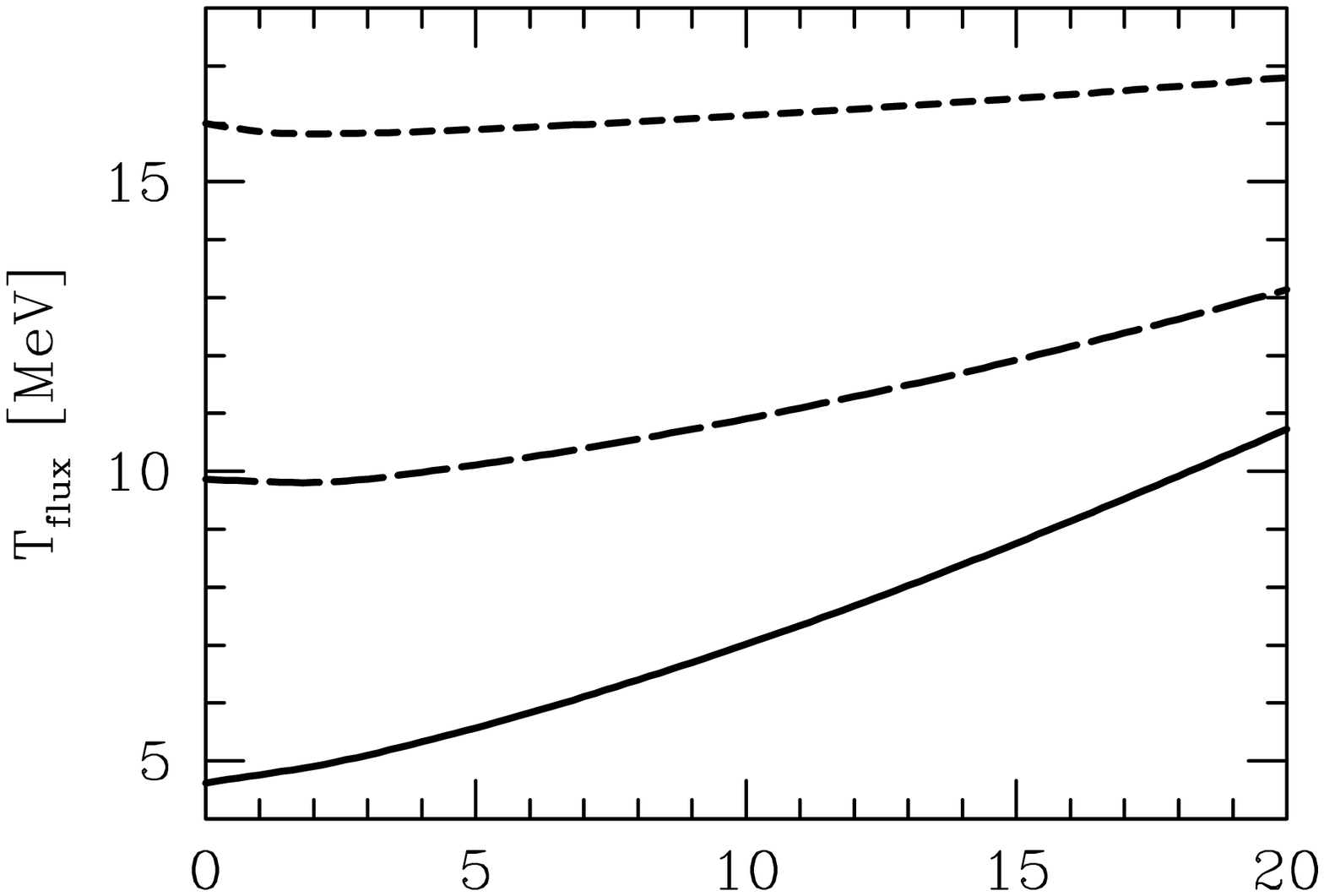}\\
\plotone{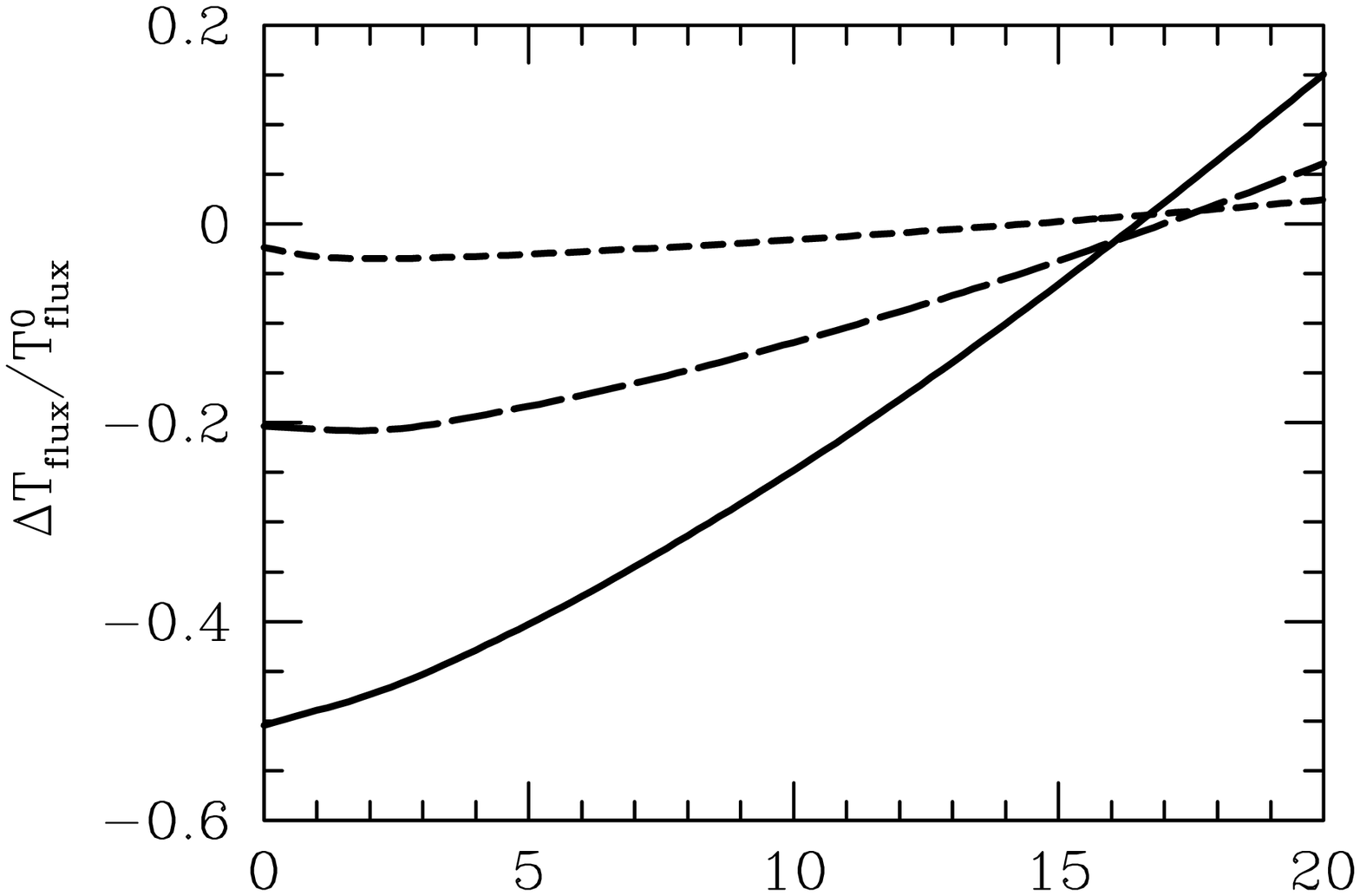}\\
\plotone{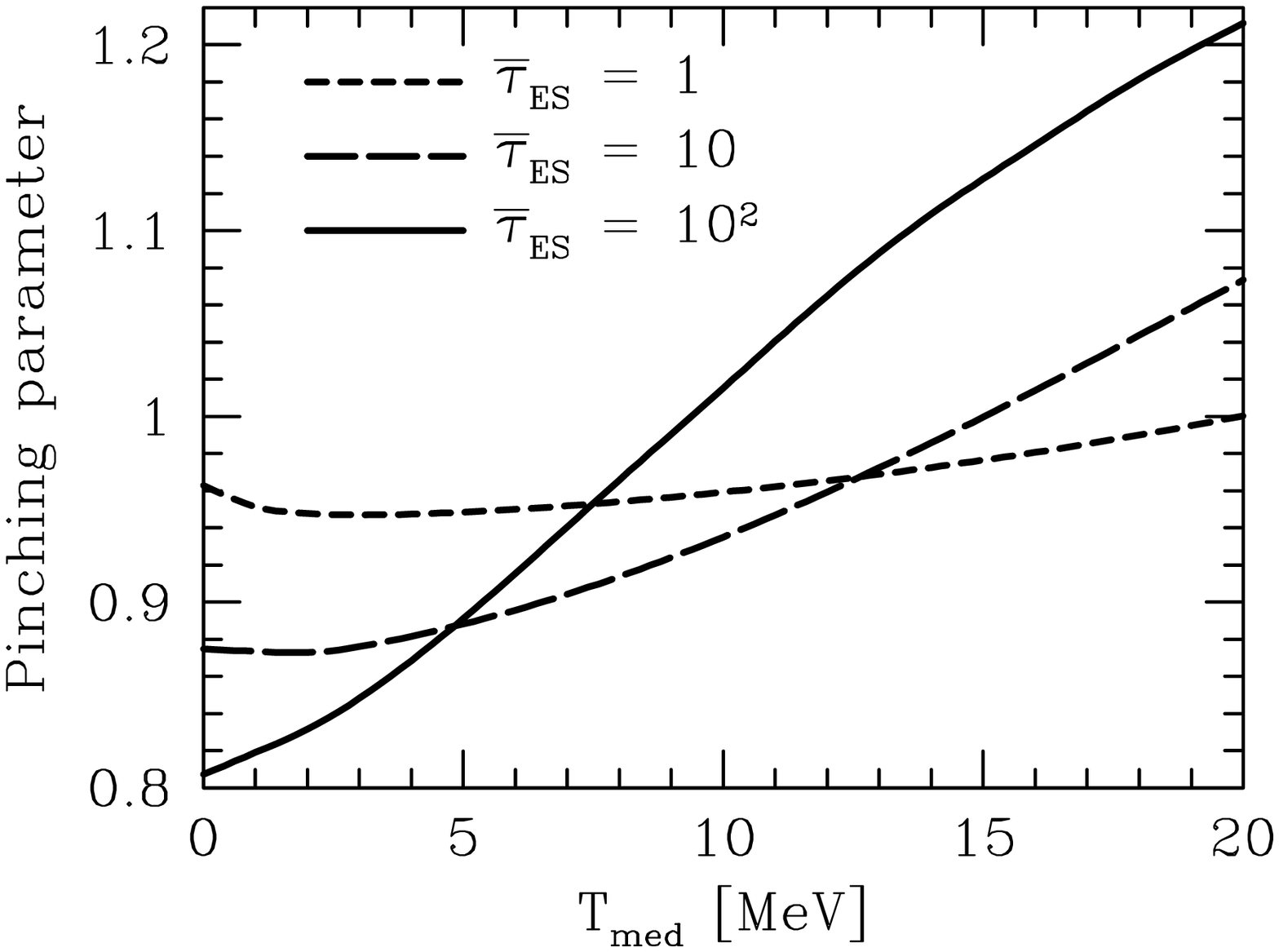}
\caption{\label{fig:delt}Variation of neutrino flux parameters
  at the surface with medium temperature for several values of
  $\bar\tau_{\rm ES}$.}
\end{figure}

In the first panel where $T_{\rm med}=T_{\rm ES}$ the trapped
neutrinos approach more closely $T_{\rm med}$, while $T_{\rm flux}$ at
the surface actually {\it increases!}  In the second panel we use
$T_{\rm flux}^0<T_{\rm med}<T_{\rm ES}$.  The $T_*$ profile is close
to $T_{\rm med}$ while $T_{\rm flux}$ shoots up near the energy
sphere.  However, in spite of $T_{\rm flux}^0<T_{\rm med}$, at the
surface $T_{\rm flux}$ reaches a value below the no-recoil case.  The
overall behavior is similar in the third panel where $T_{\rm
med}<T_{\rm flux}^0<T_{\rm ES}$.

In Fig.~\ref{fig:fixprofile2} we show the $T_*$ and $T_{\rm flux}$
profiles for $T_{\rm med}$ varying from $T_{\rm ES}=20~{\rm MeV}$ to
2~MeV in steps of 2~MeV. While the $T_*$ curves certainly look vaguely
as expected, the qualitative behavior of the $T_{\rm flux}$ profiles
would not have been intuitively obvious.  Interestingly, the $T_{\rm
  flux}$ curves all intersect essentially in one point.

Finally we show in Fig.~\ref{fig:delt} for several values of
$\bar\tau_{\rm ES}$ how the parameters of the emerging neutrino flux
vary with $T_{\rm med}$. In the first panel we show $T_{\rm flux}$, in
the second panel the relative change $\Delta T_{\rm flux}/T^0_{\rm
  flux}$ where the superscript 0 again refers to the no-recoil case.
These latter curves intersect essentially in one point, suggesting
that the medium temperature for which the neutrino flux spectrum
remains unchanged is nearly independent of the optical depth of the
scattering atmosphere.

In the third panel we show the pinching parameter. The effect of
nucleon recoils is not only to shift $T_{\rm flux}$, but also to
modify the relative shape of the spectrum. If the difference between
$T_{\rm ES}$ and $T_{\rm med}$ is large enough the spectra become
pinched.

In summary, we have found that nucleon recoils can have the effect of
shifting $T_{\rm flux}$ of the emerging neutrinos both up or down,
depending on how much lower $T_{\rm flux}$ is relative to $T_{\rm
  med}$. Likewise, the spectral shape can become more or less pinched.

In a SN core the temperature varies slowly as a function of
$\bar\tau$, i.e.\ for most of the scattering atmosphere $T_{\rm med}$
is close to $T_{\rm ES}$ (Fig.~\ref{fig:recoilprofile2}).  Therefore,
it is by no means obvious if nucleon recoils will shift the energies
of the emerging neutrinos up or down, even though naively one would
have expected that an energy-transfer channel between neutrinos and
nucleons must always shift them downward.  A simple inspection of the
temperature profiles in the no-recoil case does not allow one to
predict the direction of the energy shift.  This unintuitive behavior
is a consequence of the huge difference between the neutrino
temperature and their flux temperature, which in turn is a consequence
of the energy dependence of the $\nu N$ scattering cross section.  If
this cross section were energy independent, these complications would
not arise.

\begin{figure}[b]
\columnwidth=6.5cm
\plotone{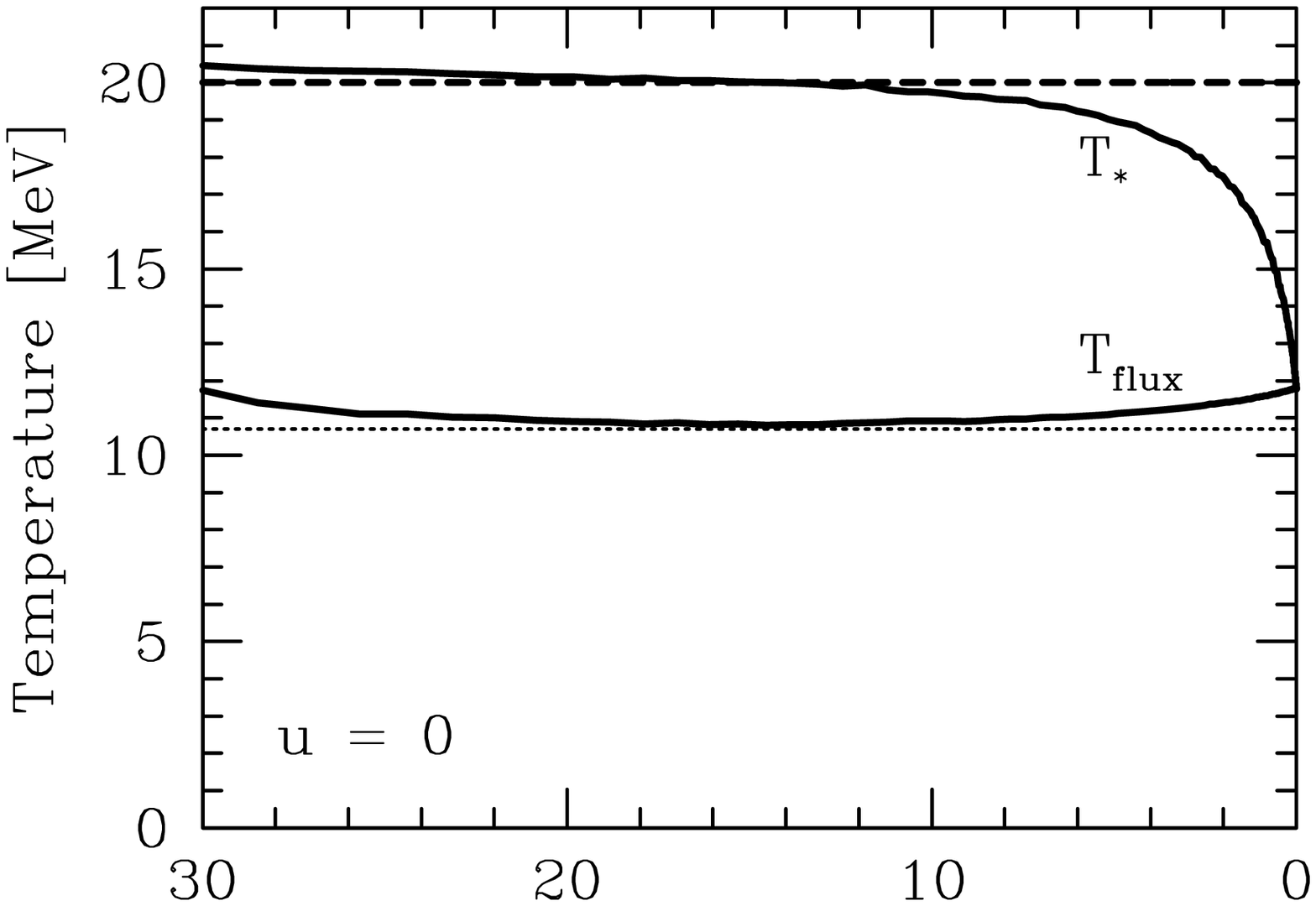}\\
\plotone{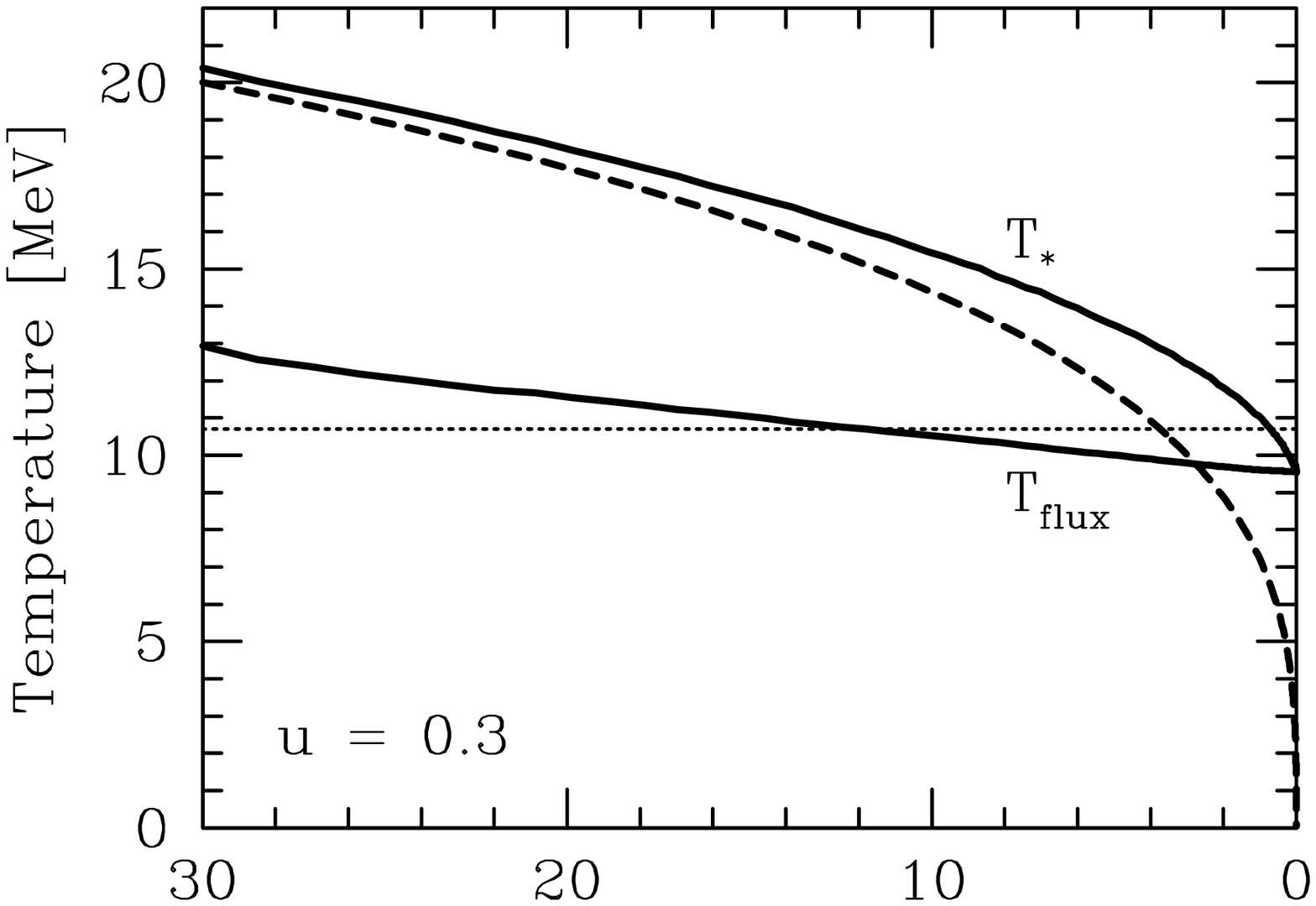}\\
\plotone{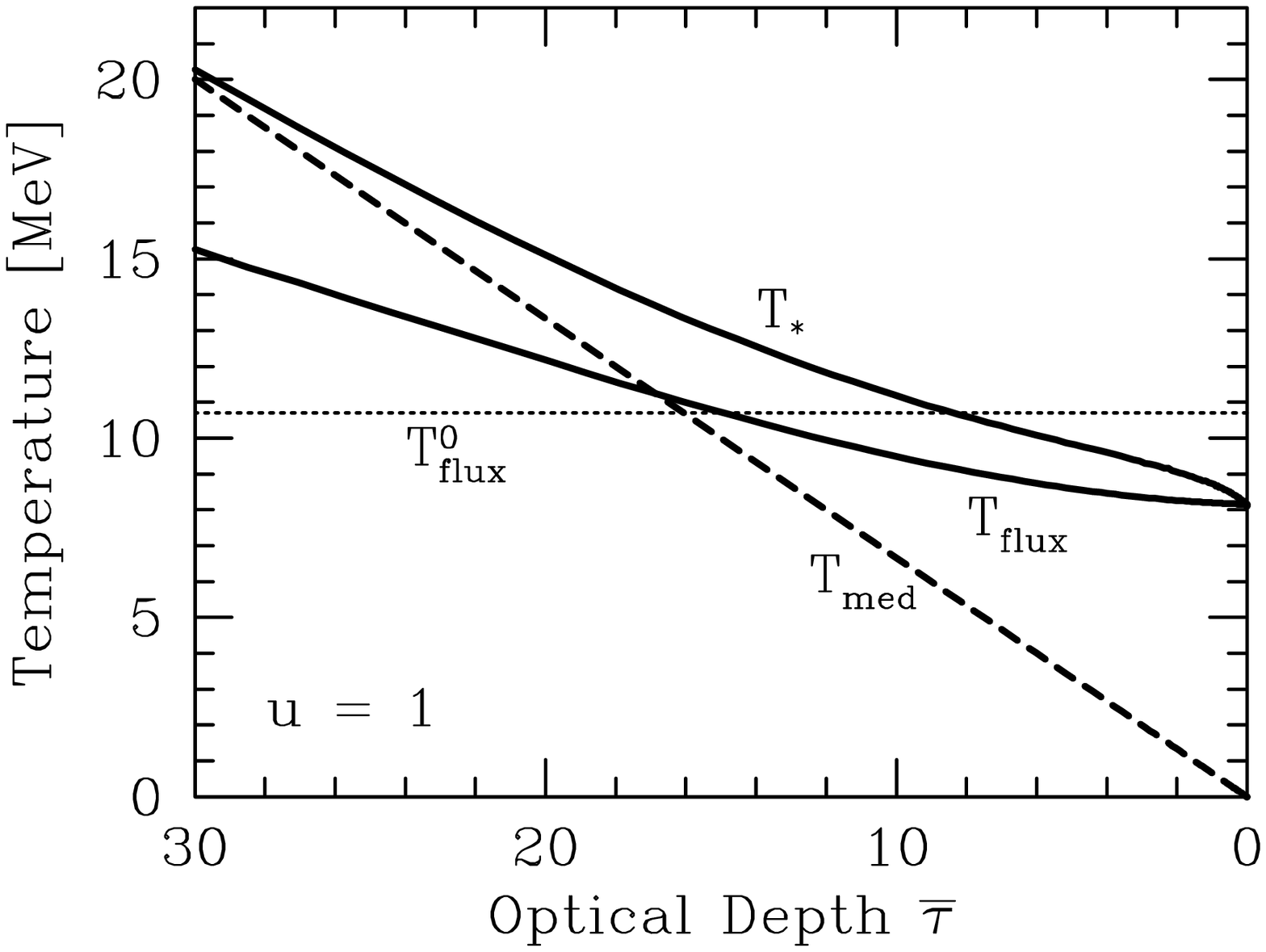}
\caption{\label{fig:powerlawprofile}
  Temperature profiles $T_{\rm med}$, $T_*$ and $T_{\rm flux}$ for
  power-law indices $u=0$, 0.3 and~1 (top to bottom panel).  Dotted
  line for the no-recoil case, thick dashed line for $T_{\rm med}$.}
\end{figure}

\subsection{Power-Law Profiles}

As a next step we consider temperature profiles which resemble a 
realistic SN core, i.e.\ power-laws of the form
\begin{equation}
T_{\rm med}=T_{\rm ES}\,
\left(\frac{\bar\tau}{\bar\tau_{\rm ES}}\right)^u.
\end{equation}
If the density varies with radius as $\rho\propto r^{-p}$ and
$T\propto r^{-q}$, then $u=q/(p-1)$. The density is a steeply
decreasing function of $r$, i.e.\ $p\gg 1$ so that $u\approx q/p$. A
realistic range is $q/p=\frac{1}{4}$ -- $\frac{1}{3}$, i.e.\ the
relative temperature profile does not depend sensitively on $p$.  In
the following we will consider the range $0\leq u\leq 1$ where the
extreme case $u=0$ corresponds to a constant $T_{\rm med}=T_{\rm ES}$
while the other extreme $u=1$ corresponds to $T_{\rm med}$ decreasing
linearly from the energy sphere to the surface.

\begin{figure}
\columnwidth=6.5cm
\plotone{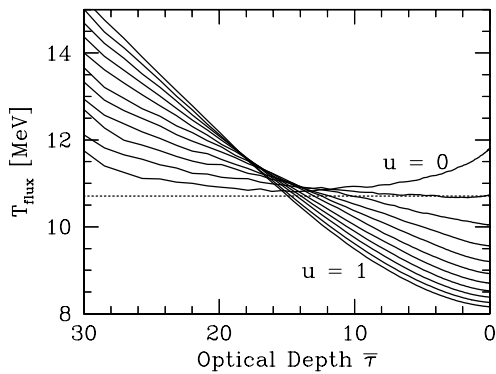}
\caption{\label{fig:powerlawprofile2} 
  $T_{\rm flux}$ profiles for $u=0$ to~1 in steps of 0.1.}
\end{figure}

For $\bar\tau_{\rm ES}=30$ we show in Fig.~\ref{fig:powerlawprofile}
profiles for $T_{\rm med}$, $T_*$ and $T_{\rm flux}$, from top to
bottom for $u=0$, $0.3$ and 1.  We also show as dotted horizontal
lines the no-recoil value $T_{\rm flux}^0$.  In
Fig.~\ref{fig:powerlawprofile2} we show the $T_{\rm flux}$ profiles
for $u=0$ to 1 in steps of 0.1.  For $u\gtrsim 0.1$ the neutrinos
emerging from the surface have a lower temperature than in the
no-recoil case.  For a realistic power-law index $u\approx 0.3$,
$T_{\rm flux}$ decreases nearly linearly from the energy sphere to the
surface. It deserves mention that relative to the no-recoil case
$T_{\rm flux}$ near the energy sphere has increased, at the surface it
has decreased.  For power-law indices which are not too small, the
$T_{\rm flux}$ profiles roughly intersect in one point, roughly
halfway between the energy sphere and the surface.

\begin{figure}[t]
\columnwidth=6.5cm
\plotone{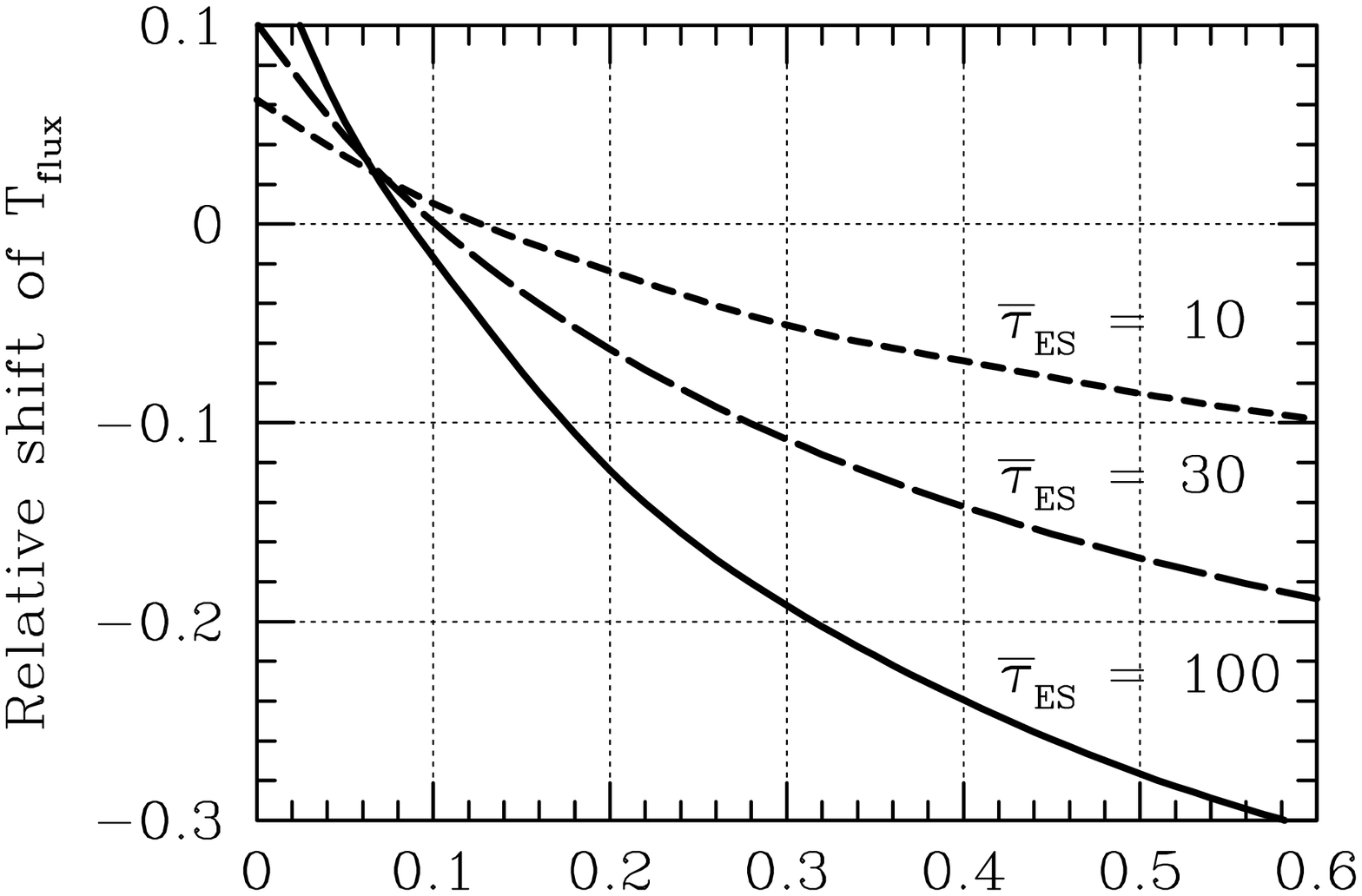}\\
\plotone{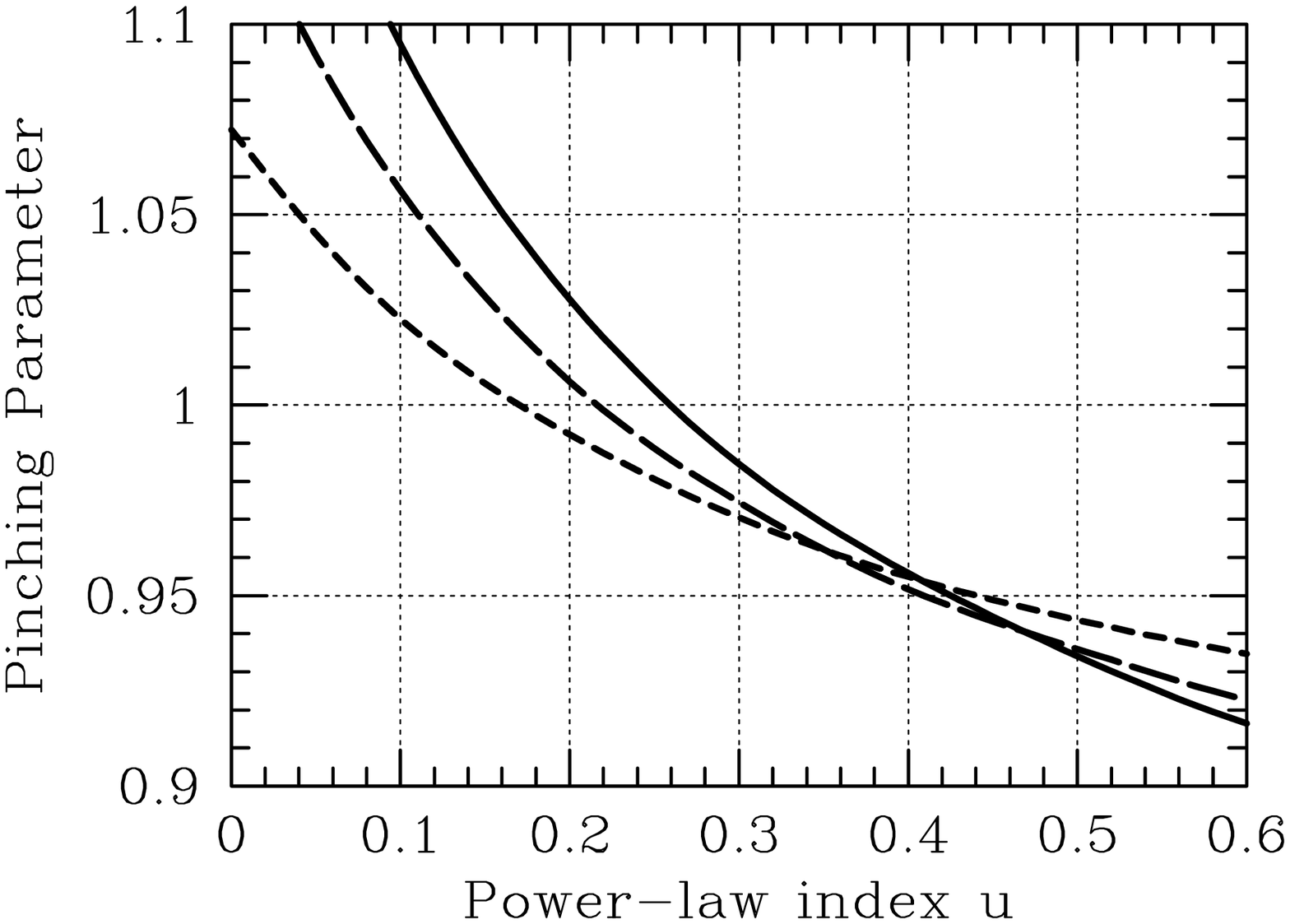}
\caption{\label{fig:powershift} 
  Spectral properties at the surface for power-law models.}
\end{figure}

In Fig.~\ref{fig:powershift} we show the relative shift $\Delta T_{\rm
  flux}/T_{\rm flux}^0$ of the surface flux temperature caused by
nucleon recoils as a function of $u$ for $\bar\tau_{\rm ES}=10$, 30
and~100.  The shift is upward for $u\lesssim 0.1$, the exact value
depending on~$\bar\tau_{\rm ES}$. For realistic values around $u=0.3$
the shift is between about $-5\%$ and $-20\%$, depending on
$\bar\tau_{\rm ES}$. We also show the pinching parameter which is
larger than 1 for small $u$, indicating anti-pinched spectra.  For
large $u$ the spectra are pinched. In the realistic range around
$u=0.3$ the pinching parameter is slightly below 1 so that the spectra
are pinched, but rather close to a thermal shape.

\subsection{Variation with Nucleon Mass}

Next we address the question of how sensitive $\Delta T_{\rm flux}$
depends on the efficiency of recoil energy transfer.  The only
dimensionless parameter governing recoil is $T/m$ with $m$ the nucleon
mass.  Therefore, we may either change the absolute scale of the
medium's temperature profile, or we may artificially change the
nucleon mass.

\begin{figure}[t]
\columnwidth=6.5cm
\plotone{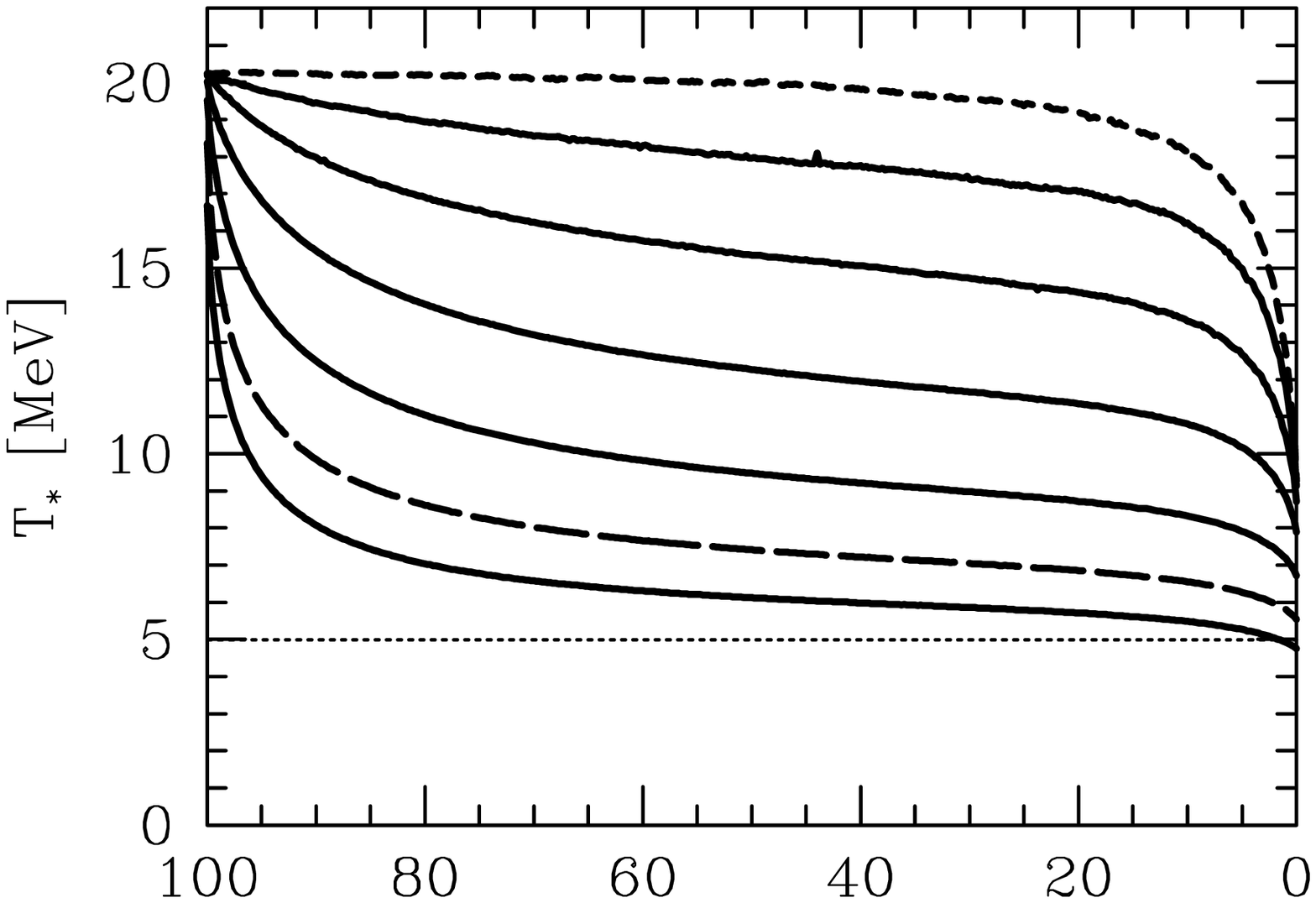}\\
\plotone{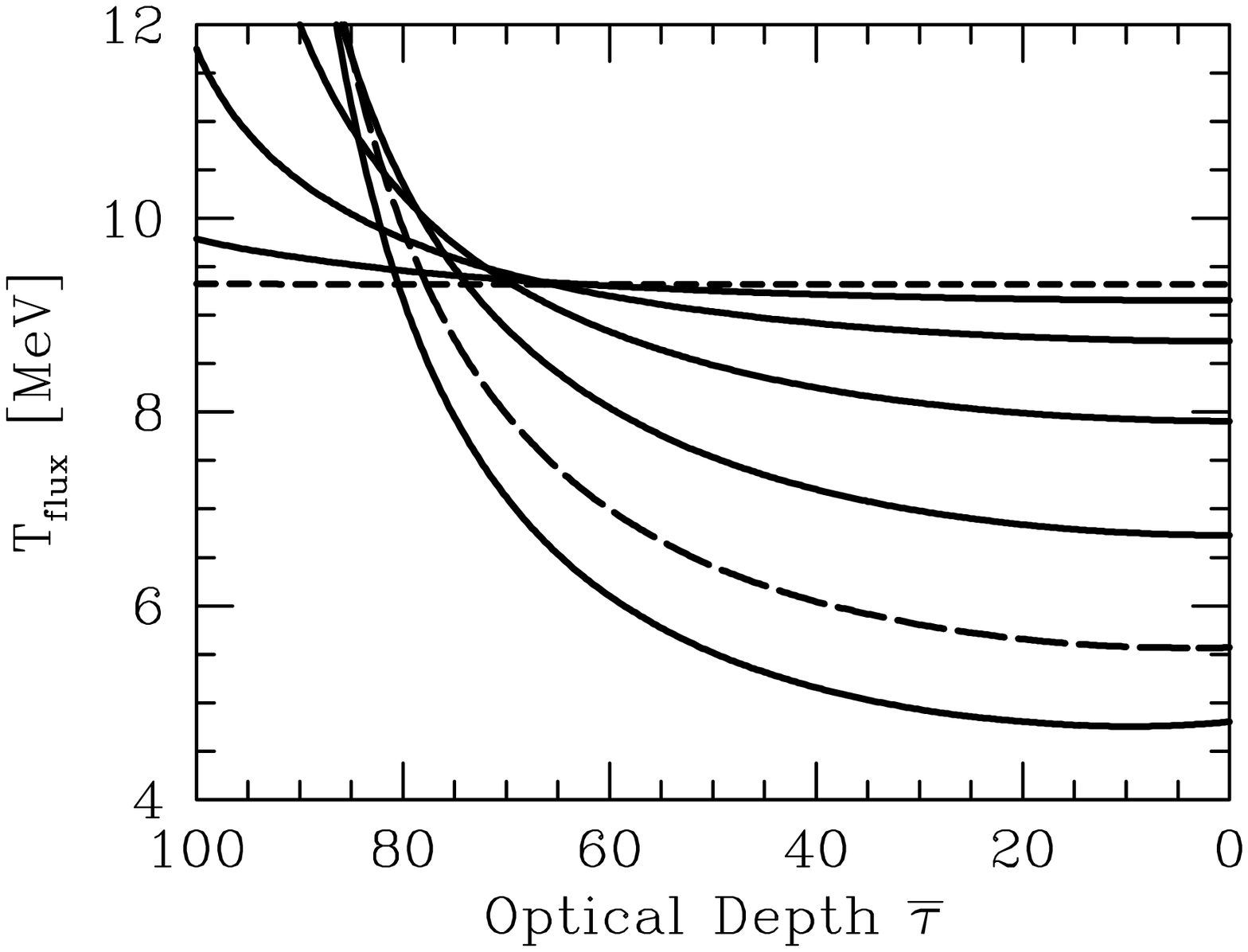}
\caption{\label{fig:massprofile} Neutrino
  temperature profiles $T_*$ and $T_{\rm flux}$ for $T_{\rm
    ES}=20~{\rm MeV}$ and $T_{\rm med}=5~{\rm MeV}$ (thin dotted
  line).  From top to bottom the curves are for nucleon masses
  $m=\infty$ (no-recoil, short dashes), $m/m_0=10^4$,
  $10^3$, $10^2$, $10$, $1$ (standard case, long
  dashes), and $10^{-1}$.}
\end{figure}

We have opted for varying $m$ and show in Fig.~\ref{fig:massprofile}
profiles for $T_*$ and $T_{\rm flux}$ for a constant $T_{\rm
  med}=5~{\rm MeV}$ and for $\bar\tau_{\rm ES}=100$. Short-dashed
lines show the no-recoil case, corresponding to $m=\infty$, while
long-dashed lines refer to the standard recoil case with $m=m_0$ where
$m_0=938~{\rm MeV}$ is the physical nucleon mass.  The solid curves
are from top to bottom for nucleon masses $m/m_0=10^4$, $10^3$,
$10^2$, $10$ and $10^{-1}$.

Surprisingly, even a huge $m$ allows for a large modification of the
local neutrino temperature.  Neutrinos scatter frequently so that even
small energy transfers compound to a large effect. In
Fig.~\ref{fig:srecoil} we showed the spread in final-state energies
for a typical neutrino interacting with thermal nucleons. It is worth
noting that the width of this curve scales with $(T/m)^{1/2}$ so that
the analogous curve for $m=10^4\,m_0$ would have 1\% of the original
width.

\begin{figure}[t]
\columnwidth=6.5cm
\plotone{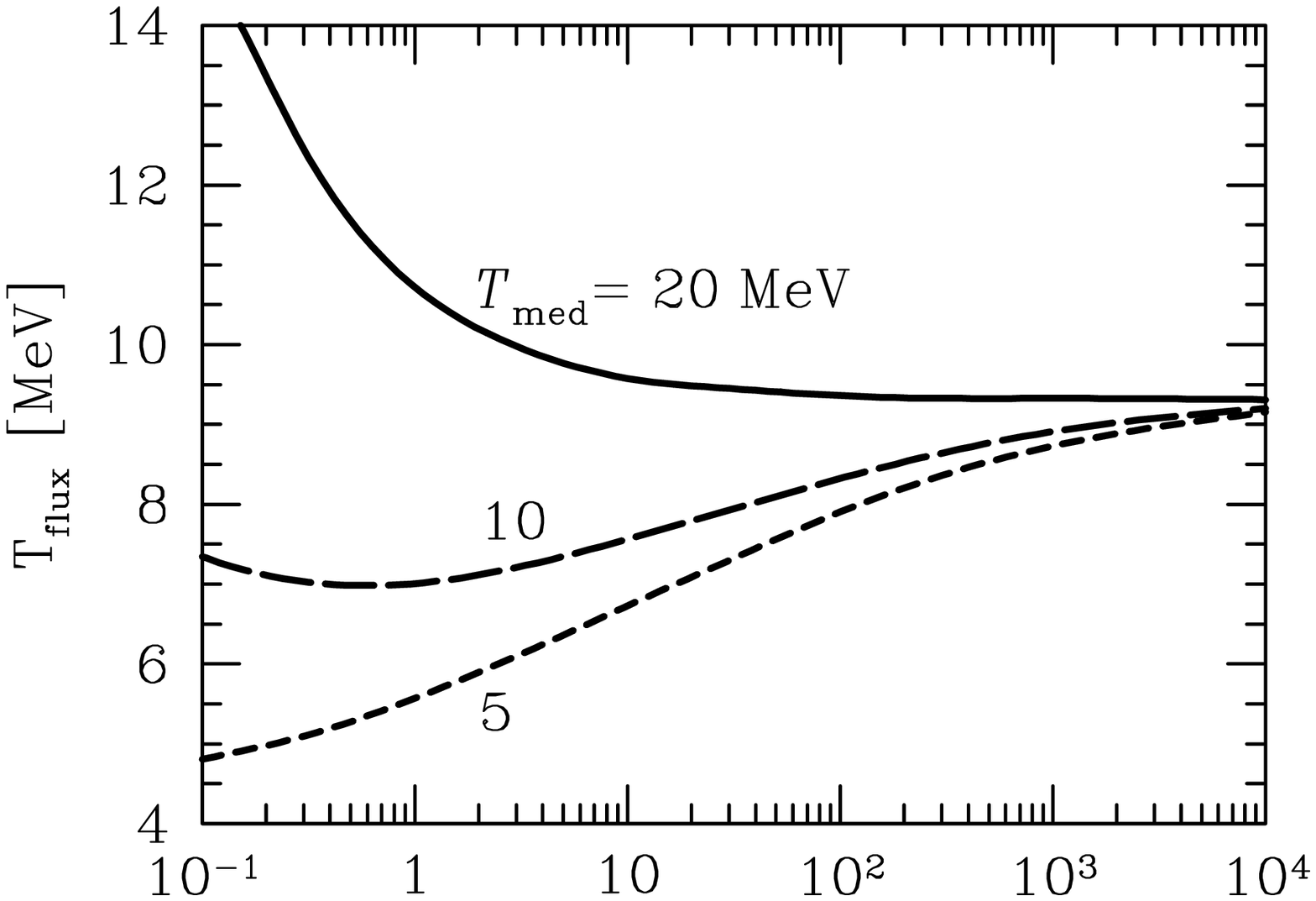}\\
\plotone{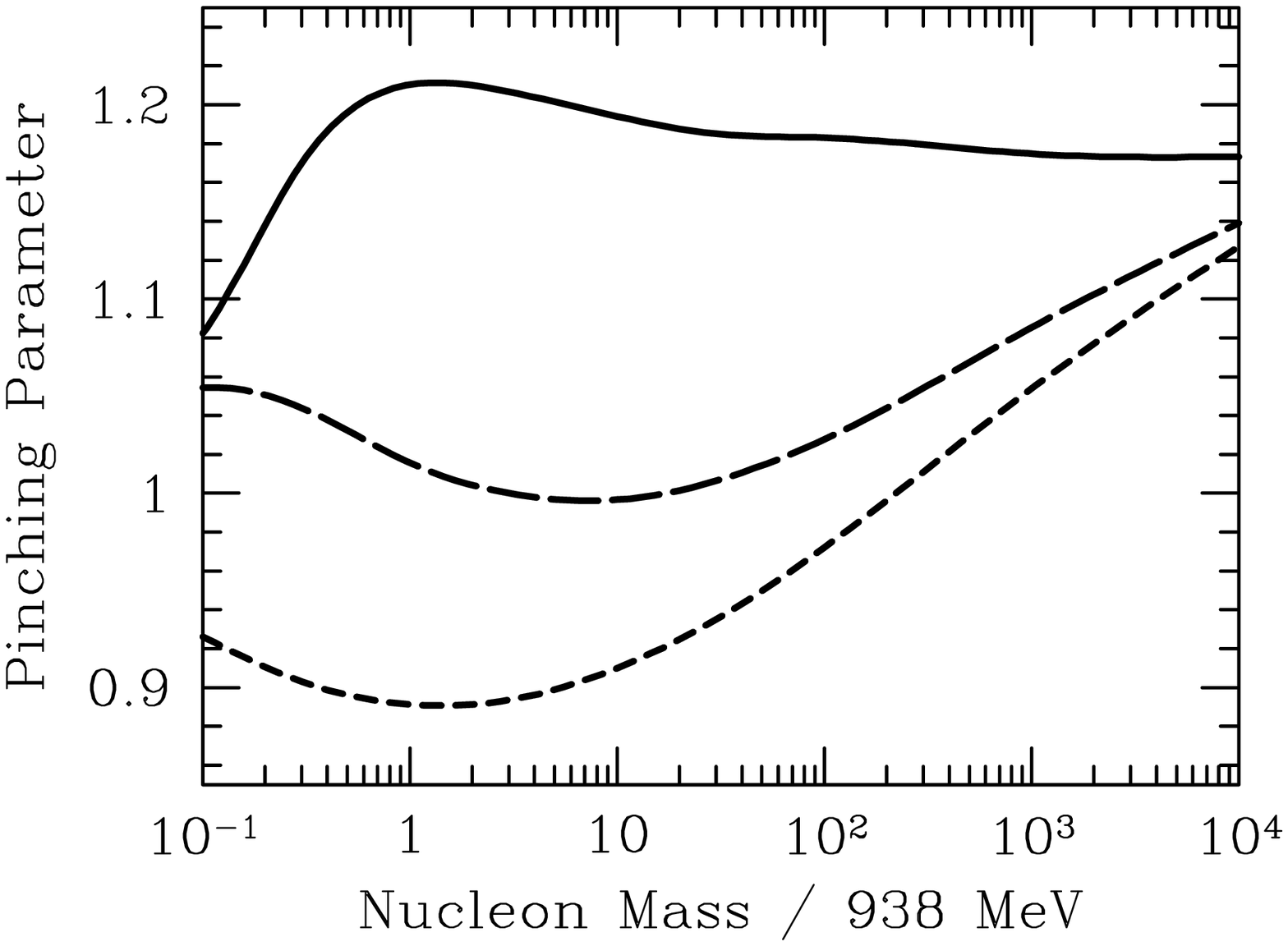}
\caption{\label{fig:tempmass} Spectral properties of escaping 
  neutrinos for constant $T_{\rm med}$ as indicated, and with $T_{\rm
    ES}=20~{\rm MeV}$.}
\end{figure}

This is further illustrated in Fig.~\ref{fig:tempmass} where we show
the $m$-dependent $T_{\rm flux}$ of the escaping neutrinos for
$\bar\tau_{\rm ES}=100$, $T_{\rm ES}=20~{\rm MeV}$ and $T_{\rm
med}=5$, 10 and 20~MeV. For $m\to\infty$ (no-recoil) the asymptotic
value is 9.30~MeV.  When $T_{\rm med}=20~{\rm MeV}$ recoils shift
$T_{\rm flux}$ upward, for $T_{\rm med}=5~{\rm MeV}$ downward, both
effects being monotonic with $m$.  For $T_{\rm med}=10~{\rm MeV}$ the
shift is downward, but $T_{\rm flux}$ is not a monotonic function
of~$m$. In the lower panel we show the variation of the pinching
parameter with $m$.

\begin{figure}[t]
\columnwidth=6.5cm
\plotone{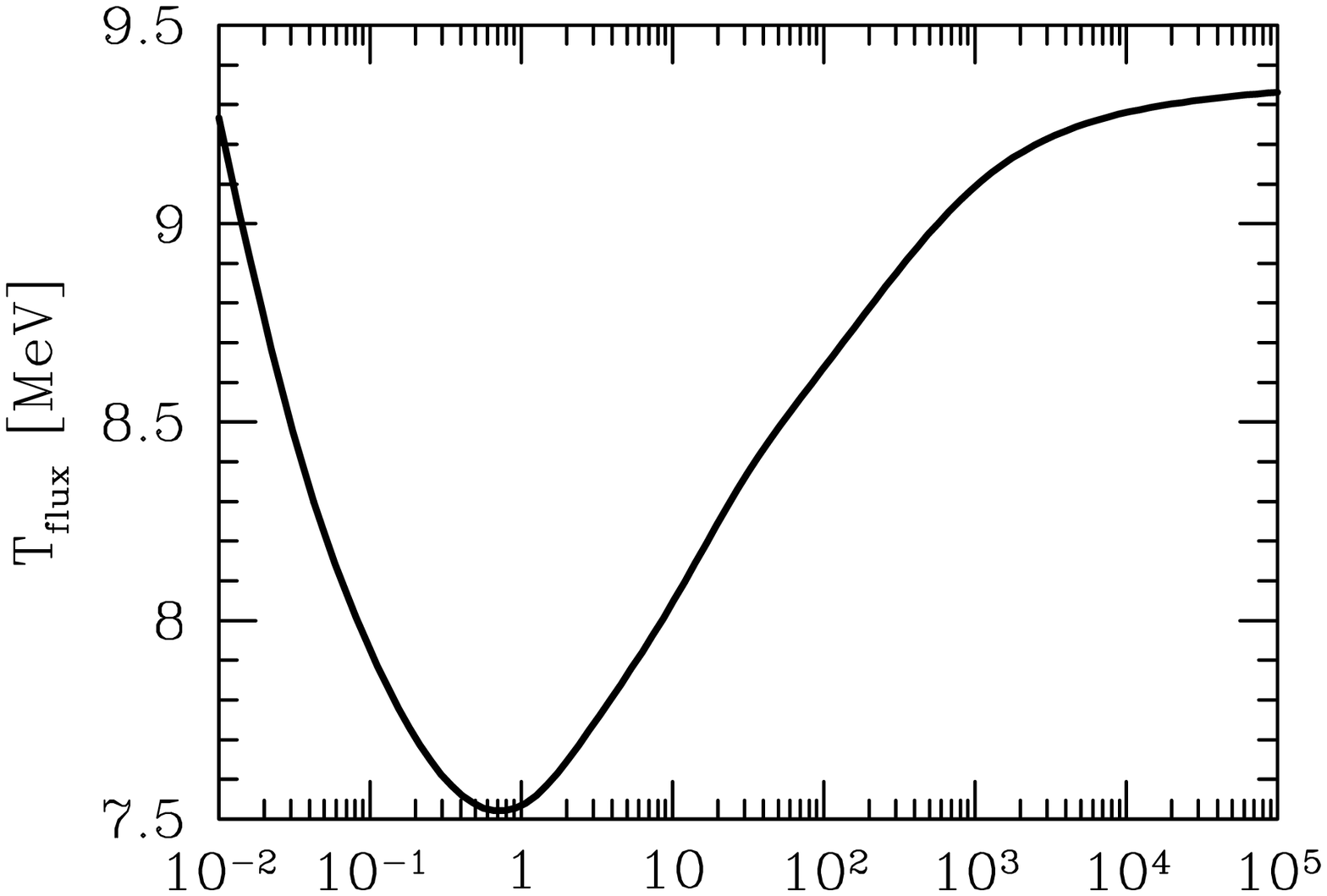}\\
\plotone{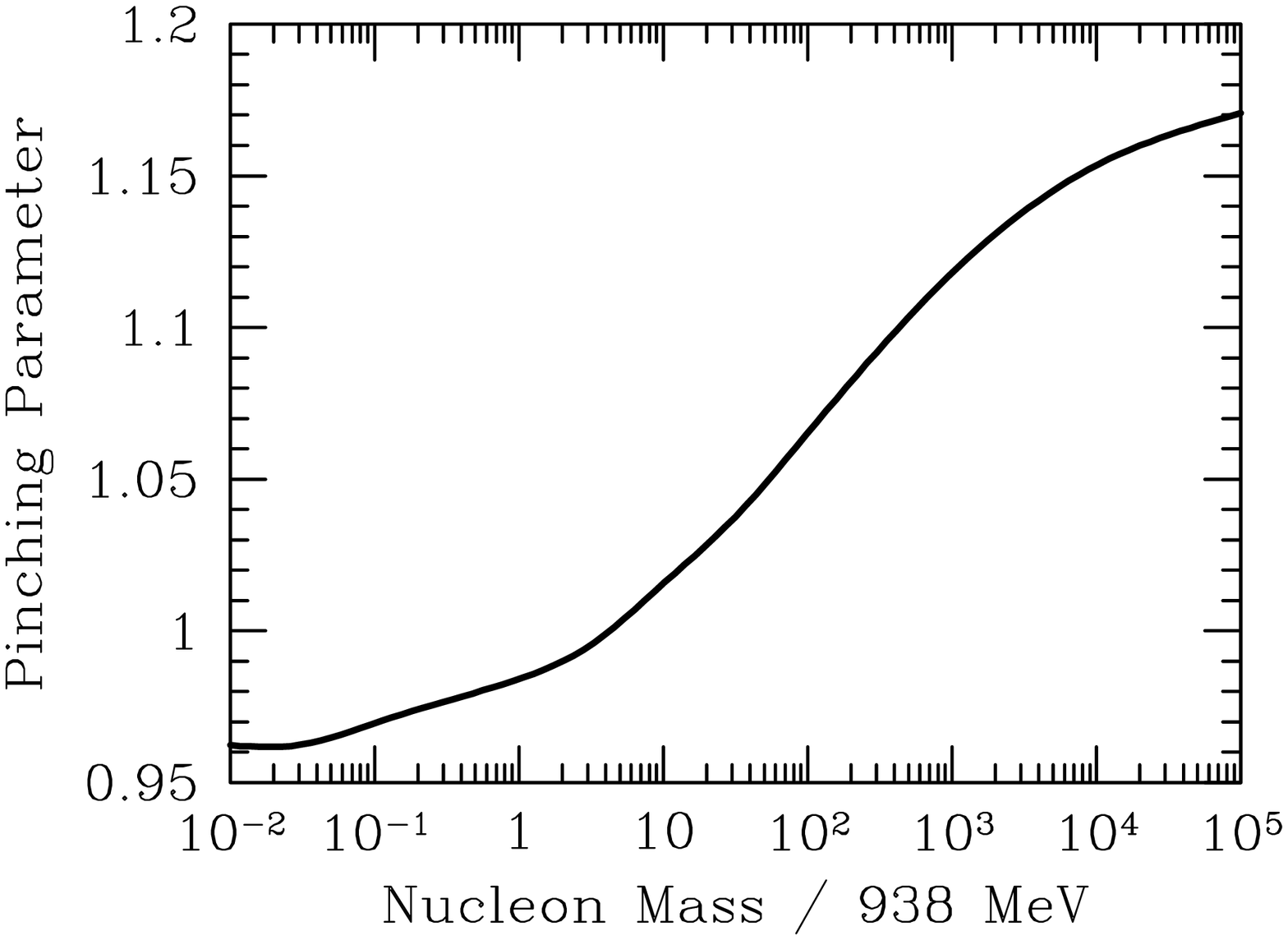}
\caption{\label{fig:umass} Spectral properties of escaping neutrinos 
  for a medium with a power-law $T_{\rm med}$ profile with $u=0.3$,
  $T_{\rm ES}=20~{\rm MeV}$, and $\bar\tau_{\rm ES}=100$.}
\end{figure}

In Fig.~\ref{fig:umass} we perform the same analysis for a power-law
medium with $u=0.3$, $T_{\rm ES}=20~{\rm MeV}$, and $\bar\tau_{\rm
ES}=100$.  Again, $T_{\rm flux}$ of the escaping neutrinos is not a
monotonic function of $m$; the shift $\Delta T_{\rm flux}$ relative to
the no-recoil case is most extreme for $m$ around the physical
value. Moreover, the pinching parameter shown in the lower panel is
quite close to~1 for $m$ near its physical value so that the spectrum
takes on a nearly thermal shape. We believe it is coincidental that a
nucleon mass near its physical value maximizes the energy-transfer
effect for the conditions of interest.

In summary, the energy shift of the escaping neutrinos and their
spectral shape are both extremely insensitive to the nucleon mass,
i.e.\ to the exact width of the medium's dynamical structure function
for energy exchange.  Changing the nucleon mass by a factor of a few
up or down has no significant effect whatsoever on the predicted
neutrino flux spectrum.

\subsection{Is Inelastic {\boldmath $\nu N$} scattering important?}

Many-body effects modify the nuclear medium's dynamical structure
function. However, even when one ignores nucleon-nucleon correlations,
neutrinos can transfer energy by ``inelastic scattering'' $\nu NN\to
NN\nu$, the crossed version of bremsstrahlung $NN\to NN\nu\bar\nu$.
This channel allows for energy transfer even if one ignores recoil.
For this effect one may use the same dynamical structure function
as for bremsstrahlung---see Appendices~\ref{sec:Bremsstrahlung}
and~\ref{sec:RecoilScatteringCombined} for technical matters.
Bremsstrahlung varies as $\rho^2$ so that it depends on the density if
recoil alone or inelastic scattering alone is more~important.

In view of the insensitivity of $T_{\rm flux}$ to the exact width of
the structure function it is clear that inelastic scattering can not
have a large impact in addition to recoil.  We have actually performed
a variety of runs where we include the combined effect of recoil and
inelastic scattering by means of the combined structure function
described in Appendix~\ref{sec:RecoilScatteringCombined}.  For
example, in a power-law model like the one shown in
Figs.~\ref{fig:profile} and~\ref{fig:recoilprofile}, including only
the inelastic $\nu N$ channel gives a significant but weaker shift of
$T_{\rm flux}$ than recoil alone.  Combining both effects, $T_{\rm
flux}$ is virtually indistinguishable from the recoil-only case.

Again, we conclude that the detailed implementation of energy transfer
does not matter as long as its efficiency is crudely comparable to
that of recoil.


\section{Discussion and Summary}

\label{sec:Summary}

We have studied how nucleon recoils in a SN core's scattering
atmosphere modify the escaping $\nu_\mu$ and $\nu_\tau$ spectra.  We
justified and used a model where the energy sphere at the bottom of
the scattering atmosphere is represented by a blackbody boundary
condition for the neutrino distribution function (temperature $T_{\rm
  ES}$).  The only remaining interaction process is $\nu N$
scattering.  If these collisions are taken to be iso-energetic, the
properties of the emerging neutrino flux are determined by exactly two
parameters, $T_{\rm ES}$ and the transport optical depth at the energy
sphere $\bar\tau_{\rm ES}$, averaged over a thermal spectrum.  For
spherically symmetric rather than plane-parallel geometry, the
analogous role of $\bar\tau_{\rm ES}$ is played by a quantity where
the column density is calculated by including a factor $(r_{\rm
  ES}/r)^2$ under the integral (energy-sphere radius $r_{\rm ES}$).
However, for the steep density gradients typical for SN cores the
difference between a spherical and plane-parallel treatment is small.
Realistic SN models have $\bar\tau_{\rm ES}\approx 10$--30.  In this
range the spectral flux temperature $T_{\rm flux}$ of the escaping
neutrinos is 62--54\% of $T_{\rm ES}$.
Equation~(\ref{eq:tfluxapproximation}) is an analytic approximation
formula for $T_{\rm flux}/T_{\rm ES}$ as a function of $\bar\tau_{\rm
  ES}$.  In summary, the spectral temperature of the $\nu_\mu$ and
$\nu_\tau$ flux emitted from a SN core is roughly 60\% of the medium
temperature at the $\nu_\mu$ and $\nu_\tau$ thermalization depth.
This is confirmed by SN simulations from the literature.

The effect of nucleon recoils on $T_{\rm flux}$ is in several ways
counter-intuitive.  We always assume monotonically decreasing
temperature profiles so that throughout the scattering atmosphere
$T_{\rm med} \leq T_{\rm ES}$. Still, the average emerging neutrino
energies can be both larger or smaller compared with the no-recoil
case. This behavior is explained by the large difference between the
local temperature $T_*$ of trapped neutrinos and their spectral flux
temperature $T_{\rm flux}$, a difference which is caused by the
$\epsilon^2$ dependence of the $\nu N$ scattering cross section.  If
we assume a power-law model $T_{\rm med}=T_{\rm ES}
(\bar\tau/\bar\tau_{\rm ES})^u$, the spectral shift is upward for
$u\lesssim 0.1$ (Fig.~\ref{fig:powershift}).  Realistic power-law
indices are around $u=0.3$ where the shift is downward by anywhere
between $-5\%$ and $-20\%$, depending on $u$ and $\bar\tau_{\rm ES}$.

The spectral shift is extraordinarily insensitive to the assumed value
of the nucleon mass $m$, i.e.\ to the details of the energy-transfer
mechanism. Therefore, for the spectra formation problem we do not need
to know the details of the nuclear medium's dynamical structure
function. As a consequence, the relative shift of $T_{\rm flux}$ is
also insensitive to the absolute scale of $T_{\rm med}$, it is only
sensitive to its radial profile, in our case the power-law index $u$.
For power-law profiles, $T_{\rm flux}/T_{\rm ES}$ depends on
$\bar\tau_{\rm ES}$ and $u$, but only extremely weakly on $T_{\rm ES}$
or~$m$.

The magnitude of the spectral shift found in our investigation is too
large to ignore in full-scale numerical simulations, but too small to
be wildly critical.  If the relative flux spectra, say, between
$\bar\nu_e$ and $\bar\nu_\mu$ differ by only 10--30\% as in some
simulations, then a reduction of the $\bar\nu_\mu$ energies by
10--20\% would be quite important. If, on the other hand, the spectra
are much more different as found in other simulations, then the recoil
reduction may not make much of a practical difference. The importance
of recoil effects may well depend on the exact phase within the SN
evolution.  There may be big differences between the post-bounce
pre-explosion accretion phase, and the later Kelvin-Helmholtz cooling
phase.

We also warn that our discussion, being based on a static background
model without self-consistent adjustment, may be more relevant as an
upper limit to the possible magnitude of the effect than a realistic
estimate. The energy transfer permitted by nucleon recoils allows the
neutrinos to heat the medium, modifying $T_{\rm med}$ such as to
reduce the rate of energy transfer. Of course, such a modified
temperature profile will affect the electron-flavored neutrinos,
perhaps increasing their energies. We expect that nucleon recoils
cause the $\bar\nu_e$ and $\bar\nu_\mu$ spectra to be more similar
than they would otherwise be, but it would not be correct to estimate
the multi-flavor spectra by taking the spectra from a simulation and
reducing the $\bar\nu_\mu$ energies by our recoil factors.

Another problem with the interpretation of our result is that we have
always neglected $\nu e$ scattering. This process provides a large
amount of energy exchange in a collision and thus should freeze out at
its thermalization sphere.  Nucleon recoils, on the other hand,
provide for a small amount of energy transfer in each collision, a
process which does not freeze out, except when neutrinos stream freely
beyond the transport sphere. Therefore, one would expect the role of
$\nu N$ and $\nu e$ scattering to be rather different.  We expect our
treatment to apply beyond the $\nu e$ freeze-out sphere.  However,
given the extreme insensitivity of the spectral shift to the exact
mode of energy transfer, we feel not entirely sure about what happens
when $\nu e$ scattering and $\nu N$ recoils are both present, even if
the $\nu e$ freeze-out sphere is in the neighborhood of the $NN$
bremsstrahlung sphere. The effect of nucleon recoils has turned out to
be rather counter-intuitive in several ways, cautioning us not to rush
into predictions about the behavior of this coupled system without
having investigated it.  Again, it appears safe to take our estimates
as upper limits to the differential change expected when nucleon
recoils are included in addition to the $\nu e$ process.

In a nutshell, the main conclusions of our investigation are that the
$\nu_\mu$ and $\nu_\tau$ spectra are well accounted for by the picture
of a scattering atmosphere with a blackbody boundary condition at the
bottom, that nucleon recoils lower the flux temperature for typical
conditions by as much as $-20\%$ if all else is kept equal, and that
this effect is astonishingly insensitive to the detailed treatment of
the energy-transfer mechanism.


\section*{Acknowledgments}

Long and fruitful discussions with Thomas Janka are gratefully
acknowledged as well as comments on the manuscript.  Bronson Messer
made the unpublished results of a collapse simulation available, and
also made helpful comments on the manuscript. Likewise, Mathias Keil
and Steen Hannestad read the manuscript and suggested useful
improvements.  This work was begun during an extended visit at the
TECHNION (Haifa, Israel) where support was granted by the Lady Davis
Trust.  I thank the Institute for Nuclear Theory (University of
Washington, Seattle) for support during a visit when this work was
finalized.  In Munich, this work was partly supported by the
Deut\-sche For\-schungs\-ge\-mein\-schaft under grant No.\ SFB 375 and
by the ESF network Neutrino Astrophysics.


\appendix

\section{Neutrino Transport in the Diffusion Limit}

\label{sec:DiffusionLimit}

\subsection{Plane Parallel Geometry}

In order to calculate the flux suppression, or ``flux dilution,'' by a
scattering atmosphere in the diffusion limit we assume that the
neutrinos interact with ``infinitely heavy'' (recoil free) scattering
centers. The cross section is taken to be
\begin{equation}\label{eq:dipolecrosssection}
\frac{d\sigma}{d\cos\theta}=\sigma_0\,\frac{1+b\cos\theta}{2},
\end{equation}
where $-1\leq b\leq+1$, $\theta$ is the scattering angle, and
$\sigma_0$ the total scattering cross section. The neutrino energy
$\epsilon$ is assumed to be fixed so that the energy dependence of
$\sigma_0$ can be ignored.

We ask for the neutrino flux penetrating a certain medium layer.  To
this end we need to solve the Boltzmann Collision Equation for the
distribution function $f$ of the neutrinos,
\begin{equation}
\frac{\partial f}{\partial t}+{\bf v}\cdot\nabla f
={\cal C}[f],
\end{equation}
where ${\bf v}$ is the neutrino velocity and ${\cal C}[f]$ the
collision term. When the neutrinos do not exchange energy with the
medium this equation applies to each neutrino energy $\epsilon$
separately so that indeed we may assume $\epsilon$ to be fixed.  

As a first case we assume that the medium is layered plane-parallel
transverse to some direction ${\bf r}$.  In a stationary state and
taking massless neutrinos ($|{\bf v}|=1$) the collision equation is
\begin{eqnarray}\label{eq:BCE1}
\mu\,\partial_r f(r,\mu)&=&-\sigma_0 n_B(r) f(r,\mu)\nonumber\\
&+&n_B(r)\int_{-1}^{+1}d\mu'\,\langle\sigma\rangle_{\mu\mu'} 
f(r,\mu'),\nonumber\\
\end{eqnarray}
where $n_B(r)$ is the number density of scattering centers
(``baryons''). The distribution function depends only on the one
spatial coordinate $r=|{\bf r}|$ and on $\mu$, the cosine of the
propagation angle relative to ${\bf r}$.  The azimuthally averaged
differential scattering cross section $\langle\sigma\rangle_{\mu\mu'}$
represents the probability for a neutrino to scatter from the
direction $\mu'$ to $\mu$. If the cross section is given by the
``dipole formula'' Eq.~(\ref{eq:dipolecrosssection}) we find
$\langle\sigma\rangle_{\mu\mu'}=\sigma_0(1+b\mu\mu')/2$ so that the
r.h.s.\ of the collision equation becomes
\begin{equation}
\sigma_0 n_B(r)\biggl[-f(r,\mu)
+\int_{-1}^{+1}d\mu'\,\frac{1+b\mu\mu'}{2}\,f(r,\mu')\biggr].
\end{equation}
We may further use the spatial coordinate
\begin{equation}
\rho\equiv\sigma_0\int_{r_0}^r dr'\,n_B(r'),
\end{equation}
corresponding to the optical depth at location $r$ as measured from
the radiating surface at the bottom of the medium layer ($r=r_0$).
The collision equation then simplifies to
\begin{eqnarray}\label{eq:BCE2}
\mu\,\partial_\rho f(\rho,\mu)&=& -f(\rho,\mu)\nonumber\\
&+&\int_{-1}^{+1}d\mu'\,\frac{1+b\mu\mu'}{2}\,f(\rho,\mu').\nonumber\\
\end{eqnarray}
To solve it, we must supplement it with the boundary conditions
$f(0,\mu)=\Phi_0$ for $0\leq\mu\leq 1$ and $f(\tau,\mu)=0$ for
$-1\leq\mu\leq 0$ at the medium's surface which is characterized by
$\tau\equiv\rho(\infty)$, the total optical depth of the scattering
atmosphere.

An exact analytic solution of Eq.~(\ref{eq:BCE2}) is found by an
ansatz corresponding to the usual moment expansion
\begin{equation}
f(\rho,\mu)=\Phi_0\left[1-\frac{\rho}{\tau}+a(1+\mu)\right]\,,
\end{equation}
where
\begin{equation}\label{eq:taubthird}
a^{-1}=\tau\,(1-b/3).
\end{equation}
If the medium is optically thick, $\tau\gg 1$, the boundary conditions
are approximately fulfilled.

The first moment of the distribution function, which is proportional
to the particle flux, is $2\pi\int_{-1}^{+1} d\mu\,\mu\,f(\rho,\mu)$
and is found to be $(4\pi/3)\,a\,\Phi_0$, independently of $\rho$,
i.e.\ the flux is conserved.  If the medium were absent, the solution
would be $f(\rho,\mu)=\Phi_0\Theta(\mu)$ with the flux $\pi\Phi_0$.
Hence in the diffusion limit we find
\begin{equation}
s(\tau)=\frac{4}{(3-b)\,\tau}
\end{equation}
for the neutrino flux suppression relative to the free-streaming case.

We observe that the r.h.s.\ of Eq.~(\ref{eq:taubthird}) corresponds to
the usual transport cross section, 
\begin{equation}
\sigma_{\rm T}=\int_{-1}^{+1} d\cos\theta\;(1-\cos\theta)\; 
\frac{d\sigma}{d\cos\theta}.
\end{equation}
With Eq.~(\ref{eq:dipolecrosssection}) this is $\sigma_{\rm
  T}=\sigma_0(1-b/3)$. Therefore, the relevant variable is the
transport cross section, not the scattering cross section. In the main
text we usually mean $\sigma_{\rm T}$ when we speak of the cross
section, and we mean $\tau (1-b/3)$, the ``transport optical depth,''
when we speak of the optical depth.

\subsection{Spherical Symmetry}

Next, we turn to the case of spherical symmetry instead of the
previous plane-parallel geometry. If one writes the neutrino
distribution function in terms of the variables $r$ and $\mu$, where
$\mu$ is the cosine of the propagation angle relative to the radial
direction, the l.h.s.\ of the stationary collision
equation~(\ref{eq:BCE1}) becomes
\begin{equation}
\left(\mu\,\partial_r+\frac{1-\mu^2}{r}\,\partial_\mu\right) 
f(r,\mu).
\end{equation}
With the moment expansion $f(r,\mu)=f_0(r)+\mu\,f_1(r)$ the collision
equation turns into
\begin{equation}
\mu\,\partial_r f_0+\mu^2\partial_r f_1+
\frac{1-\mu^2}{r}\,f_1
=-\sigma_T n_B\mu f_1.
\end{equation}
The neutrino flux is $(4\pi/3)\,f_1(r)$ so that flux conservation
through a given spherical shell implies $f_1(r)\propto r^{-2}$ or
\begin{equation}
\mu\,\partial_r f_0(r)=
-\left[\frac{1-3\mu^2}{r}+\sigma_T n_B(r)\mu\right]\, f_1(r).
\end{equation}
In the diffusion limit, the second term in square brackets is much
larger than the first for all $r$ which significantly exceed the
scattering mean free path. Neglecting the first term we find
\begin{equation}
\partial_r f_0(r)=-\sigma_T n_B(r)f_1(r).
\end{equation}
The only remaining difference to the plane-parallel case is that flux
conservation now dictates that $f_1(r)=\Phi_0\,a\,(r_0/r)^2$ with
$r_0$ the bottom of the scattering atmosphere. In the plane-parallel
case we had $f_1(r)=\Phi_0\,a$.  

Therefore, the spherically symmetric solution is obtained by the substitution
\begin{equation}
\sigma_{\rm T} n_B(r) \to \sigma_{\rm T} n_B(r) (r_0/r)^2.
\end{equation}
Put another way, in the diffusion limit the scattering atmosphere
dilutes the neutrino flux by the factor 
\begin{equation}
s(\tau_*)=\frac{4}{3\tau_*}\,,
\end{equation}
where
\begin{equation}
\tau_*\equiv\sigma_{\rm T}
\int_{r_0}^\infty dr\,
 n_B(r) \left(\frac{r_0}{r}\right)^2.
\end{equation}
Of course, even in the spherical case the proper optical depth is
defined without the $(r_0/r)^2$ factor under the integral.  However,
for the purpose of neutrino flux transmission the parameter $\tau_*$
plays the same role as the proper optical depth $\tau$ in the
plane-parallel case.

We may compare the two cases for an example where
the density drops as a power law, $n_B(r)\propto r^{-p}$, implying
\begin{equation}
\frac{\tau_*}{\tau}=\frac{p-1}{p+1}.
\end{equation}
If $p\gg 1$ this ratio is close to unity.  Put another way, if the
density drops quickly for $r>r_0$ the integral is dominated by
$r$-values near $r_0$ and we recover the transport optical depth of
the plane-parallel case.


\section{Neutrino-Nucleon Scattering}

\label{sec:NeutrinoNucleonScattering}

\subsection{Scattering Cross Section}

\label{sec:ScatteringCrossSection}

We need to derive an expression for the dynamical structure function
$S(\omega,k)$ which plays the role of a scattering kernel for neutrino
processes. In terms of the structure function the differential
scattering cross section of a neutrino with initial energy
$\epsilon_1$ is
\begin{equation}\label{eq:diffsigma}
\frac{d\sigma}{d\epsilon_2\,d\cos\theta}
=\frac{C_A^2(3-\cos\theta)}{2\pi}\,G_F^2\,\epsilon_2^2\,
\frac{S(\omega,k)}{2\pi}
\end{equation}
where $\epsilon_2$ is the final-state neutrino energy,
$\omega=\epsilon_1-\epsilon_2$ the energy transfer, $k=|{\bf k}_1-{\bf
k}_2|$ the modulus of the momentum transfer, and $\theta$ the
scattering angle. $S(\omega,k)$ is a function of $k$, and not of ${\bf
k}$, because the medium is assumed to be isotropic.

In this paper we include only one species of nucleons for which the
axial weak coupling constant is assumed to be $|C_A| = 1.26/2$.  In
general, there is also a vector-current interaction; the total
scattering cross section on non-relativistic nucleons is proportional
to $C_V^2+3C_A^2$. For protons, $C_V$ very nearly vanishes while even
for neutrons with $C_V=-1/2$ the vector-current contribution to the
cross section is a rather small correction. Ignoring the vector
current simplifies our discussion as we need only one structure
function.  In general, there is a different structure function for the
vector-current, the axial vector current, and the mixed term.

When the nucleons are ``infinitely heavy'' and thus unable to recoil,
the structure function is simply $S(\omega,k)=2\pi\,\delta(\omega)$,
i.e.~the collisions are iso-energetic (no energy exchange), and the
nucleon can absorb any amount of momentum. In this case the total
axial-current scattering cross section and the transport cross section
for a neutrino of energy $\epsilon$ are
\begin{eqnarray}
\sigma_0&=&\frac{3 C_A^2}{\pi}\,G_{\rm F}^2 \epsilon^2,
\nonumber\\
\sigma_{\rm T}&=&\frac{10 C_A^2}{3\pi}\,G_{\rm F}^2 \epsilon^2.
\end{eqnarray}
This corresponds to an inverse mean free path of
\begin{eqnarray}\label{eq:scatteringopactiy}
\lambda_0^{-1}&=&12.08~{\rm km}^{-1}\,\rho_{14}\,
\epsilon_{10}^2,
\nonumber\\
\lambda_{\rm T}^{-1}&=&
13.42~{\rm km}^{-1}\,\rho_{14}\,\epsilon_{10}^2,
\end{eqnarray}
where $\rho_{14}=\rho/10^{14}~\rm g~cm^{-3}$ and
$\epsilon_{10}=\epsilon/10~{\rm MeV}$.

\subsection{General Properties of the Structure Function}

The dynamical structure function can be expressed as a current-current
correlator, in our case of the nucleon axial vector current (see, for
example, Janka et~al.\ 1996). This representation allows one to derive
a number of general properties.  First, there is detailed balance
\begin{equation}\label{eq:detailedbalance}
S(-\omega,k)=e^{-\omega/T}\,S(\omega,k).
\end{equation}
Second, if there are no spin-spin interactions, we have the
normalization
\begin{equation}\label{eq:normalization}
\int_{-\infty}^{+\infty}\frac{d\omega}{2\pi}\,S(\omega,k)=1,
\end{equation}
and the $f$-sum rule
\begin{equation}\label{eq:fsumrule}
\int_{-\infty}^{+\infty} \frac{d\omega}{2\pi}\,\omega\,
S(\omega,k)=\frac{k^2}{2m}.
\end{equation}
Recent calculations of the dynamical structure functions in the
presence of nucleon-nucleon interactions, i.e.\ in the presence of
spin-spin correlations, have been performed by Burrows and Sawyer
(1998) as well as Reddy, Prakash and Lattimer (1998) 
and Reddy et~al.\ 
(1999). In our present study we ignore all nucleon-nucleon interaction
effects except for the calculation of nucleon-nucleon bremsstrahlung
(see below).

It is sometimes useful to construct a symmetric version of the
structure function,
\begin{eqnarray}
\bar S(\omega,k)&\equiv&\frac{S(\omega,k)+S(-\omega,k)}{2}
\nonumber\\
&=&\frac{1+e^{-\omega/T}}{2}\,S(\omega,k), 
\end{eqnarray}
and conversely
\begin{equation}\label{eq:unsymmSfromS}
S(\omega,k)=\frac{2}{1+e^{-\omega/T}}\,\bar S(\omega,k). 
\end{equation}
By construction $\bar S(\omega,k)$ is a symmetric function of
$\omega$; it obeys the same normalization as $S(\omega,k)$.

\subsection{Nucleon Recoil}

For a finite nucleon mass and ignoring nucleon degeneracy effects, the
structure function takes on the well-known form
\begin{equation}\label{eq:srecoil1}
S_{\rm recoil}(\omega,k)=
\sqrt{\frac{\pi}{\omega_k T}}\,
\exp\left\{-\frac{(\omega-\omega_k)^2}{4 T \omega_k}\right\},
\end{equation}
where $\omega_k\equiv k^2/2m$.  One easily checks that this
``displaced Gaussian'' fulfills detailed balancing
Eq.~(\ref{eq:detailedbalance}) as well as the sum rules
Eqs.~(\ref{eq:normalization}) and~(\ref{eq:fsumrule}).

The distribution $d\sigma/d\epsilon_2$ of final-state neutrino
energies is obtained by integrating Eq.\ (\ref{eq:diffsigma}) over
$\cos\theta$ with Eq.~(\ref{eq:srecoil1}) for $S(\omega,k)$ and
observing that $\omega=\epsilon_1-\epsilon_2$ and
$k=(\epsilon_1^2+\epsilon_2^2-2\epsilon_1\epsilon_2\cos\theta)^{1/2}$.
As an example we show in Fig.~\ref{fig:srecoil} this distribution for
$T=10~{\rm MeV}$ and an initial-state neutrino energy
$\epsilon_1=30~{\rm MeV}$. Evidently the distribution is rather broad.
For other values of $\epsilon_1$ it looks similar when the horizontal
axis is scaled accordingly, i.e.~the fractional width relative to
$\epsilon_1$ does not depend much on $\epsilon_1$.


\section{Bremsstrahlung}

\label{sec:Bremsstrahlung}

\subsection{Nucleon Spin Relaxation Rate}

In order to calculate the bremsstrahlung emission of neutrino pairs
$NN\to NN\nu\bar\nu$ and related processes
we need to include nucleon-nucleon interactions.
The neutrino energy-loss rate of a single-species non-relativistic
non-degenerate thermal nucleon gas can be expressed as
(see, e.g., Raffelt 1996)
\begin{eqnarray}\label{eq:generalbrems}
Q_{\nu\bar\nu}&=&\left(\frac{C_A G_{\rm F}}{\sqrt{2}}\right)^2 n_B
\nonumber\\
&\times&
\int\frac{d^3{\bf k}_1}{2\epsilon_1(2\pi)^3}
\frac{d^3{\bf k}_2}{2\epsilon_2(2\pi)^3}\,(\epsilon_1+\epsilon_2)
\nonumber\\
&\times&
8\epsilon_1\epsilon_2\,(3-\cos\theta)\,
S\left(-\epsilon_1{-}\epsilon_2,|{\bf k}_1{+}{\bf k}_2|\right),
\nonumber\\
\end{eqnarray}
where $\epsilon_{1,2}$ are the energies of the emitted neutrinos,
${\bf k}_{1,2}$ their momenta, and $S(\omega,k)$ is the same dynamical
structure function that describes axial-current neutrino-nucleon
scattering. In the non-relativistic limit the vector current
does not contribute to bremsstrahlung.

The phase-space integration in Eq.~(\ref{eq:generalbrems}) covers
energy-momentum transfers $(\omega,k)=(-\epsilon_1{-}\epsilon_2,
\break
|{\bf
  k}_1{+}{\bf k}_2|)$ which are time-like ($\omega^2\geq k^2$) whereas
for neutrino scattering they are space-like ($\omega^2\leq k^2$).
Therefore, even though both scattering and bremsstrahlung are
characterized by the same $S(\omega,k)$, it is different regions in
the $(\omega,k)$ plane that contribute to these processes. The recoil
structure function Eq.~(\ref{eq:srecoil1}) has only power for
space-like $(\omega,k)$ and thus cannot account for bremsstrahlung, in
keeping with the obvious insight that free nucleons cannot radiate.
Nucleon-nucleon interactions were included in various calculations of
$S(\omega,k)$, most recently by Burrows and Sawyer (1998) as well as
Reddy, Prakash and Lattimer (1998) and Reddy et~al.\ (1999). However,
these and previous works limited their calculations to the
random-phase approximation where nucleon spin-spin correlations are
included, but not nucleon spin fluctuations. Bremsstrahlung requires
nucleon spins to fluctuate. Graphically speaking, a nucleon spin needs
to be kicked to be able to emit radiation which couples to the spin
such as neutrino pairs. The $S(\omega,k)$ calculated in these works
have no power for time-like $(\omega,k)$ and thus do not account for
bremsstrahlung.

One finds essentially two different approaches in the literature to
calculating the bremsstrahlung rate. One may model the nucleon-nucleon
interaction potential, typically by a one-pion exchange potential, and
then evaluate the relevant Feynman amplitudes for $NN\to
NN\nu\bar\nu$.  Recently, Hanhart, Phillips and Reddy (2001) have
followed a different approach where they calculate the bremsstrahlung
matrix element from nucleon scattering data in the soft limit
($\omega\to0$) and then extend the results ``by hand'' to
non-vanishing energy transfers.

If one follows this latter approach one needs to guess a functional
form for $S(\omega,k)$. To this end one may use the ``long wavelength
approximation'' where the momentum transfer $k$ to the nucleons is
assumed to be negligible so that one approximates the structure
function by $S(\omega)=\lim_{k\to 0} S(\omega,k)$.  We further recall
that the structure function can be expressed as a nucleon spin-spin
correlator. The nucleon spin autocorrelation function has a nontrivial
time dependence only if there are spin-nonconserving forces between
the nucleons such as the nuclear tensor force. It is plausible (but
not necessary) that a nucleon spin, being kicked by other nucleons,
loses memory of its original orientation as $e^{-\Gamma t}$ where
$\Gamma$ is the spin relaxation rate. An exponential decline
corresponds to the assumption that the spin relaxation is a Markovian
process, i.e.\ a chain of uncorrelated random kicks.  The Fourier
transform of an exponentially declining autocorrelation function is
\begin{equation}\label{eq:lorentzian}
\bar S(\omega)=\frac{2\Gamma}{\omega^2+\Gamma^2},
\end{equation}
leading to a Lorentzian ansatz for the symmetric structure
function.\footnote{In previous papers (Janka et al.\ 1996 and
  references) we have used the notation $\Gamma_\sigma=2\Gamma$ and
  called $\Gamma_\sigma$ the spin fluctuation rate. However, the
  relationship between $\Gamma$ and an exponentially declining spin
  autocorrelation function (Raffelt \& Sigl 1999) implies that
  $\Gamma$, and not $\Gamma_\sigma$, has the intuitive interpretation
  of a spin relaxation or spin fluctuation rate.}

Calculating the bremsstrahlung rate in the one-pion exchange model in
Born approximation with uncorrelated, non-degenerate, single-species
nucleons, Raffelt \& Seckel (1995) find in the soft limit
($\omega\to0$) and ignoring the pion mass
\begin{equation}\label{eq:Gamma1}
\Gamma=2\sqrt\pi\,\alpha_\pi^2\,n_B T^{1/2} m^{-5/2}
\end{equation}
with $n_B$ the baryon density, $m$ the nucleon mass, and
$\alpha_\pi\equiv(f 2m/m_\pi)^2/4\pi\approx15$ with $f\approx1$ the
pion-nucleon ``fine-structure constant.''

In the one-pion approximation one can calculate the behavior of the
structure function for non-vanishing energy transfers and one may also
include the pion mass. Therefore, Eq.~(\ref{eq:lorentzian}) can be
supplemented with a dimensionless factor $\bar s(x,y)$ where
$x=\omega/T$ and
\begin{equation}
y\equiv \frac{m_\pi^2}{m T}=1.94\,T_{10}^{-1}
\end{equation}
with $T_{10}=T/10~{\rm MeV}$. One finds (Raffelt \& Seckel 1995;
Hannestad \& Raffelt 1998)
\begin{eqnarray}\label{eq:snd}
\bar s(x,y)&=&\frac{e^{-x/2}+e^{x/2}}{16}\nonumber\\
&\times&\int_{|x|}^{\infty}
dt\,\,e^{-t/2}\,\,\frac{3x^2+6ty+5y^2}{3}\nonumber\\
&\times&
\Biggl[\frac{2\sqrt{t^2-x^2}}{x^2+2ty+y^2}
\nonumber\\
&-&\frac{1}{t+y}\,
\log\left(\frac{t+y+\sqrt{t^2-x^2}}{t+y-\sqrt{t^2-x^2}}\right)\Biggr].
\nonumber\\
\end{eqnarray}
This expression is even in $x$ and also $\bar s(0,0)=1$.

\begin{figure}[t]
\columnwidth=6.5cm
\plotone{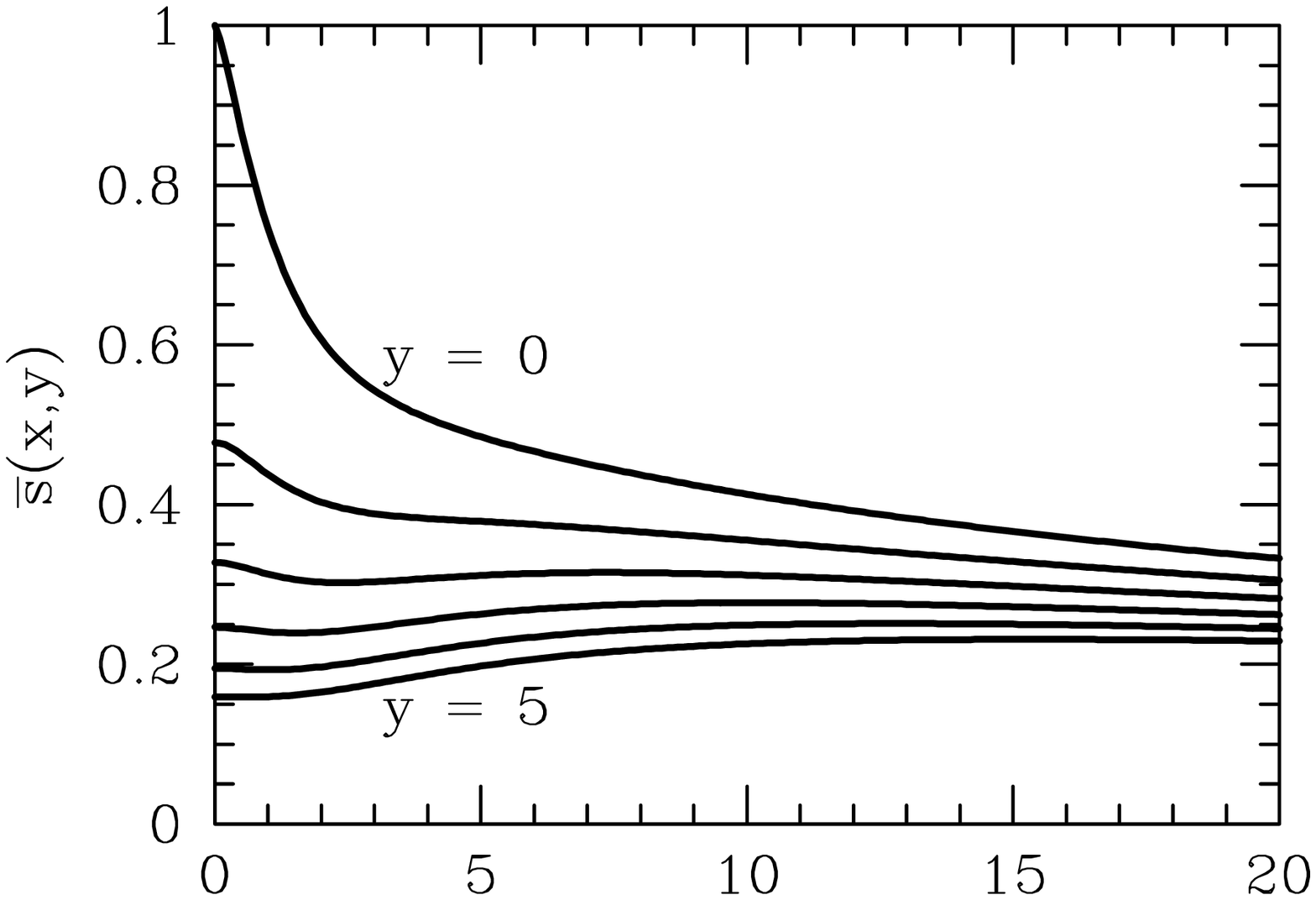}\\
\plotone{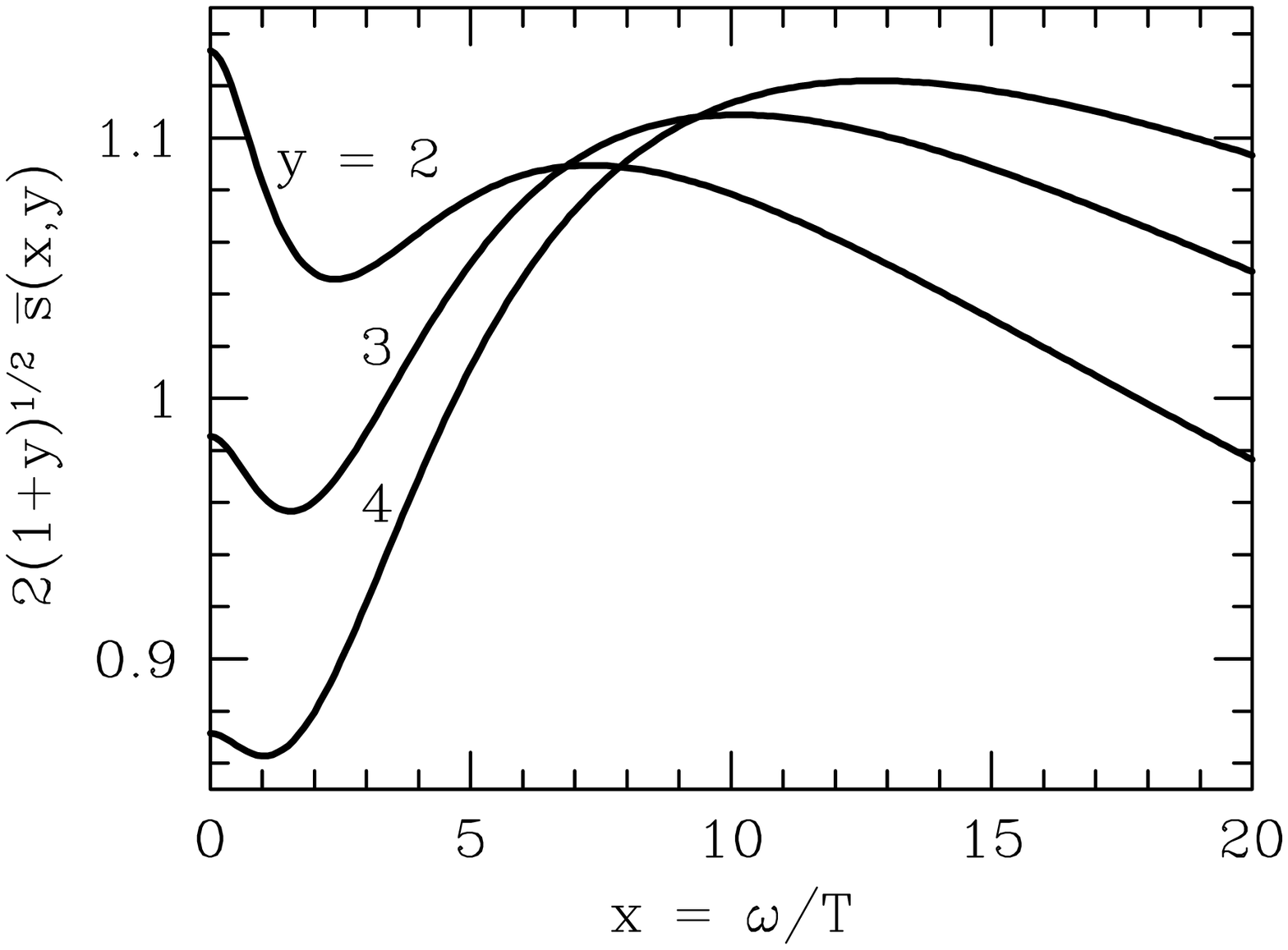}
\caption{\label{fig:snd}
  Dimensionless structure function according to
  Eq.~(\protect\ref{eq:snd}) for the indicated values of $y$.}
\end{figure}

In Fig.~\ref{fig:snd} we show $\bar s(x,y)$ as a function of $x$ for
several values of $y$.  In the outer parts of the SN core, we
have $y=2$--4, values for which $\bar s(x,y)$ varies slowly as a
function of $x$. The lower panel of Fig.~\ref{fig:snd} shows that in
this $y$-range we have $2\sqrt{1+y}\,\bar s(x,y)=1$ within about $\pm
10\%$. Therefore, at the relatively crude level of approximation that
we content ourselves with it is justified to use the Lorentzian
Eq.~(\ref{eq:lorentzian}) as a structure function with the spin
relaxation rate Eq.~(\ref{eq:Gamma1}), divided by $2\sqrt{1+y}$, i.e.\
\begin{equation}
\Gamma=\frac{\sqrt\pi\,\alpha_\pi^2 n_B\,T}{m^2\sqrt{mT+m_\pi^2}}\,.
\end{equation}
Within a few percent this is numerically
\begin{equation}\label{eq:Gammanumerical}
\gamma\equiv\frac{\Gamma}{T}=1.25\,\rho_{14}\,
\sqrt{\frac{3}{2+T_{10}}}\,,
\end{equation}
where $\rho_{14}=\rho/10^{14}~{\rm g}~{\rm cm}^{-3}$ and
$T_{10}=T/10~{\rm MeV}$.

It is not obvious how well or poorly this result represents the true
spin relaxation rate for the physical conditions at hand.  Hanhart,
Phillips and Reddy (2001) have calculated the equivalent of $\Gamma$
on the basis of nuclear scattering data.  They claim that at nuclear
density the true $\Gamma$ is about a factor of 4 smaller than the one
obtained from the OPE calculation.\footnote{These authors compare with
the OPE calculations of Friman and Maxwell (1979). However, Friman and
Maxwell have missed a term in the squared matrix element of $NN\to
NN\nu\bar\nu$; the correct result was derived by Brinkmann and Turner
(1988) as well as Raffelt and Seckel (1995). Therefore, the
discrepancy between the OPE result and that of Hanhart, Phillips and
Reddy (2001) is even larger, perhaps by as much as a factor of~1.5.}
In a one-species medium at nuclear density ($3\times10^{14}~{\rm
g~cm^{-3}}$) the nucleon Fermi momentum is about 350~MeV, much larger
than the pion mass, so that nucleon-nucleon collisions probe deeply
into the core of the nucleon-nucleon potential; it is no surprise that
the OPE model, which only mimics the peripheral part of the potential,
would yield poor results. However, we are here interested in a
non-degenerate medium at, say, $T=10~{\rm MeV}$ where a typical
nucleon momentum is $(3mT)^{1/2}\approx 170~{\rm MeV}$ so that the OPE
method would be expected to be much more justified.  Unfortunately,
Hanhart et~al.\ do not provide an estimate for such conditions.
Hannestad \& Raffelt (1998) have argued that the OPE result should not
underestimate the true answer by much more than about 30\% for such
conditions.

An additional complication arises in a mixed medium of protons and
neutrons.  Within the OPE model, the bremsstrahlung rate of such a
mixed medium far exceeds that of a single-species medium at the same
density because the $np$ matrix element is much larger. In the present
paper the medium is always modeled by a single-species composition,
leading to an underestimate of the bremsstrahlung rate of a realistic
SN core.  It is not quantitatively known how well these two effects
compensate each other, but we would have to be very unlucky if our
adopted value for the spin relaxation rate were off by more than a
factor of 2 in either direction.

\subsection{Energy Loss Rate}

We may now calculate the energy-loss rate of a medium due to
nucleon-nucleon bremsstrahlung emission of neutrino pairs.
Equation~(\ref{eq:generalbrems}) can be simplified to read
\begin{equation}
Q_{\nu\bar\nu}=
\frac{C_A^2 G_{\rm F}^2}{40\,\pi^4}\,n_B
\int_0^\infty d\omega\,\omega^6\,S(-\omega),
\end{equation}
where the Lorentzian structure function is
\begin{equation}
S(\omega)=\frac{2\Gamma}{\Gamma^2+\omega^2}\;
\frac{2}{1+e^{-\omega/T}}.
\end{equation}
Since $S(\omega)$ is multiplied with a high power of $\omega$
under the integral we may neglect $\Gamma^2$ in the denominator.
We then find
\begin{eqnarray}
Q_{\nu\bar\nu}&=&
\frac{9\,\zeta_5}{4\,\pi^4}\,C_A^2 G_{\rm F}^2 n_B\,T^5\,\Gamma
\nonumber\\
&=&2.37\times10^{35}~{\rm erg~cm^{-3}~s^{-1}}\nonumber\\
&\times&
\rho_{14}^2\,T_{10}^6\,\sqrt{\frac{3}{2+T_{10}}}\,,
\end{eqnarray}
where we have used Eq.~(\ref{eq:Gammanumerical}) for the spin
relaxation rate, and $T_{10}=T/10~{\rm MeV}$
and $\rho_{14}=\rho/10^{14}~{\rm g~cm^{-3}}$.

\subsection{Neutrino Absorption}

\begin{figure}
\columnwidth=6.5cm
\plotone{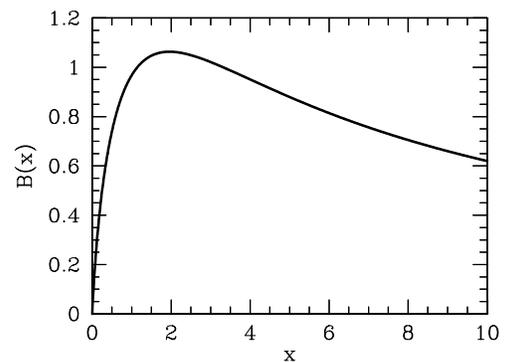}
\caption{\label{fig:Bcurve}
  Function $B(x)$ defined in Eq.~(\protect\ref{eq:Bdef}).}
\end{figure}

We next ask for the neutrino mean free path against the absorption
process $\nu\bar\nu NN\to NN$ (inverse bremsstrahlung). By means
of the structure function one easily finds for a neutrino
of energy $\epsilon$
\begin{eqnarray}
\lambda^{-1}_{\rm brems}&=&\frac{C_A^2 G_{\rm F}^2}{2}\, n_B\,
\frac{1}{2\epsilon}\nonumber\\
&\times&
\int\frac{d^3\bar{\bf k}}{2\bar\epsilon\,(2\pi)^3}\,
f(\bar\epsilon)\,24\,\epsilon\,\bar\epsilon\,
S(\epsilon{+}\bar\epsilon),
\nonumber\\
\end{eqnarray}
where over-barred quantities refer to the anti-neutrino which is
absorbed together with our test neutrino. Using Boltzmann statistics
and assuming thermal equilibrium, the distribution function is
$f(\bar\epsilon)=e^{-\bar\epsilon/T}$.  Ignoring $\Gamma^2$ downstairs
in the Lorentzian structure function, we find
\begin{eqnarray}\label{eq:bremsmfp}
\lambda^{-1}_{\rm brems}&\!=\!&
\frac{3C_A^2 G_{\rm F}^2\,n_B\,T^2\,\Gamma}{\pi^2}\,
\frac{2}{5\,\epsilon}\,B(\epsilon/T)\nonumber\\
&\!=\!&1.92~{\rm km}^{-1}\,\frac{\rho_{14}^2\,T_{10}^3}{\epsilon_{10}}\,
\sqrt{\frac{3}{2+T_{10}}}\;B(\epsilon/T).
\nonumber\\
\end{eqnarray}
The dimensionless function
\begin{equation}\label{eq:Bdef}
B(x)=\frac{5\,x}{2}\int_0^\infty d\bar x\,
\frac{{\bar x}^2}{(x+\bar x)^2}\;
\frac{2}{e^{-x}+e^{+\bar x}}
\end{equation}
is shown in Fig.~\ref{fig:Bcurve}. For neutrino energies of a few $T$,
corresponding to $x$-values of a few, this function is close to unity.
It varies slowly for most energies of interest. Therefore, for rough
estimates we may use that the bremsstrahlung absorption rate varies as
$\epsilon^{-1}$.

\eject

\subsection{Neutrino Emissivity}

We also need the ``source strength'' or emissivity of the medium
in neutrinos. In contrast with the energy-loss rate, we need
the production rate of neutrinos, not of pairs. Ignoring
Pauli-blocking (we always use Boltzmann statistics!), the
differential production rate can be found directly from an
expression like Eq.~(\ref{eq:generalbrems}) or else by 
detailed-balance arguments from the bremsstrahlung absorption
rate, leading to
\begin{equation}
\frac{d \dot n_\nu}{d\epsilon}=
\frac{\epsilon^2}{2\pi^2}\,e^{-\epsilon/T}\,
\lambda_{\rm brems}^{-1}(\epsilon).
\end{equation}
As the absorption rate scales roughly as $\epsilon^{-1}$, the
production rate varies approximately as $\epsilon\,e^{-\epsilon/T}$.


\section{Recoil and Inelastic Scattering Combined}

\label{sec:RecoilScatteringCombined}

The dynamical structure function for bremsstrahlung also accounts for
inelastic neutrino-nucleon scattering, $\nu+NN\to NN+\nu$, i.e.\
neutrinos can gain or lose energy in a collision with a nucleon due to
the simultaneous interaction of the nucleon with another nucleon. This
energy gain or loss is unrelated to nucleon recoil. One could ignore
nucleon recoils, and still take neutrino-nucleon energy transfers into
account by using the Lorentzian structure function
Eq.~(\ref{eq:lorentzian}) in the differential scattering cross section
Eq.~(\ref{eq:diffsigma}). However, while one can have recoil effects
without the bremsstrahlung-related inelastic energy transfer if the
medium is sufficiently dilute, one cannot have inelastic scattering
without recoil as both effects arise when the nucleon mass is not 
taken to be infinite.

In principle, there is one complete dynamical structure function which
includes all effects, but alas, it is not known. In principle, it
should be possible to calculate a dynamical structure function which
accounts for both bremsstrahlung and recoil effects. However, all
bremsstrahlung calculations were performed in the long-wavelength
limit where the momentum transfer of the neutrinos to the nucleon pair
was ignored. In this limit, recoil effects do not show up.

We presently derive a heuristic form for $S(\omega,k)$ as a
convolution of the expressions for recoil and bremsstrahlung.  The
most important property of the structure function is detailed
balancing Eq.~(\ref{eq:detailedbalance}). As this property would not
persist after convolving two different structure functions with each
other, we use the symmetric form $\bar S(\omega,k)$ as a starting
point.  The proper structure function is then trivially recovered by
Eq.~(\ref{eq:unsymmSfromS}).

The symmetric form of the recoil structure function is explicitly
\begin{eqnarray}
\bar S_{\rm recoil}(\omega,k)&=&
\sqrt{\frac{\pi}{\omega_k T}}\,\,e^{-\omega^2/4T\omega_k}
e^{-\omega_k/4T}\nonumber\\
&\times&\frac{e^{\omega/2T}+e^{-\omega/2T}}{2}.
\end{eqnarray}
In the non-relativistic limit of a very large nucleon mass $m$ we
have $\omega_k\ll T$ so that the energy transfers are typically
much smaller than $T$. Therefore, in this limit we may approximately
use $e^{-\omega_k/4T}=e^{\pm\omega/4T}=1$ so that the
structure function simplifies to
\begin{equation}\label{eq:recoil}
\bar S_{\rm recoil}(\omega,k)=\frac{1}{k}\,\sqrt{\frac{2\pi m}{T}}\,
\exp\left\{-\frac{m}{2T}\,\left(\frac{\omega}{k}\right)^2\right\}.
\end{equation}
In the limit of $m\to\infty$ the two forms are equivalent. The new
version gives an average energy transfer per collision which is 15\%
smaller than the original version for $T=10~\rm MeV$. The original
version was derived under the assumption of non-relativistic nucleons
even though for $T=10~\rm MeV$ a typical nucleon speed is about 0.17
of the speed of light. Therefore, at the level of approximation where
nucleons are treated as non-relativistic it is not even clear if the
simplified version or the original one are a better approximation to
the true state of affairs.  We have 
introduced the simplified symmetric
version primarily because it allows us to combine it easily with
inelastic neutrino-nucleon scattering.

Bremsstrahlung is characterized by the Loren\-tzian structure function
Eq.~(\ref{eq:lorentzian}), which is itself a heuristic result.  For
the present purpose it is more useful to replace it with the ansatz
\begin{equation}
\bar S_{\rm brems}(\omega)=\frac{2\Gamma}{\omega^2}
\left(1-e^{-(\pi/4)\,(\omega/\Gamma)^2}\right).
\end{equation}
Crucially, this function has the same behavior $2\Gamma/\omega^2$ for
$\omega\gg \Gamma$ and it obeys the same normalization.

After some tinkering one can derive an interpolation formula which
roughly represents the convolution of the two expressions,
\begin{eqnarray}\label{eq:stotal}
T \bar S_{\rm total}(\omega,k)
&=&\frac{2\sqrt\pi}{\kappa+\gamma e^{\gamma/\kappa}}\,
e^{-(x/\kappa)^2}\nonumber\\
&&\kern-7em{}+\frac{2\gamma}{x^2}\,
\left[1-\exp\left\{-\frac{\pi}{4}\,
\left(\frac{x}{\gamma+\kappa 
e^{-\gamma/\kappa}}\right)^2\right\}\right],\nonumber\\
\end{eqnarray}
which is properly normalized, even though it may not look that way.
The r.h.s.\ is expressed in terms of the dimensionless variables
\begin{eqnarray}
x&=&\frac{\omega}{T}\,,\nonumber\\
\kappa&=&\frac{k}{T}\,\sqrt{\frac{2T}{m}}\,,\nonumber\\
\gamma&=&\frac{\Gamma}{T}.
\end{eqnarray}
The interpolation formula takes on the relevant limiting cases for
$\gamma\to0$ (no bremsstrahlung) or $\kappa\to0$ (no recoil).

\begin{figure}[b]
\columnwidth=6.5cm
\plotone{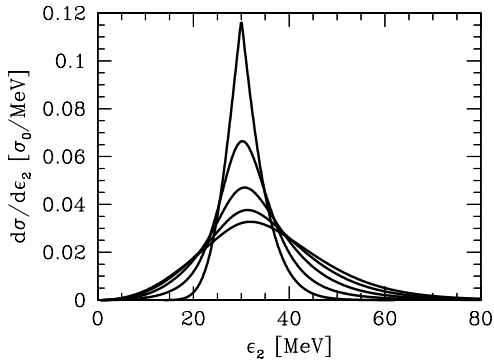}
\caption{\label{fig:stotal} Distribution of final-state energies
  $\epsilon_2$ of a neutrino with initial energy $\epsilon_1=30~{\rm
    MeV}$, scattering on non-degenerate nucleons in thermal equilibrium
  with $T=10~{\rm MeV}$. As a structure function, $S_{\rm total}$ of
  Eq.~(\protect\ref{eq:stotal}) was used, including both nucleon
  recoil and inelastic scattering. From top to bottom, the curves are
  for $\gamma=0$, 0.5, 1.0, 1.5 and~2.0.}
\end{figure}

The distribution $d\sigma/d\epsilon_2$ of final-state neutrino
energies is again obtained by integrating Eq.~(\ref{eq:diffsigma})
over $\cos\theta$ with Eq.~(\ref{eq:stotal}) for $S(\omega,k)$ and
observing that $\omega=\epsilon_1-\epsilon_2$ and
$k=(\epsilon_1^2+\epsilon_2^2-2\epsilon_1\epsilon_2\cos\theta)^{1/2}$.
In analogy to Fig.~\ref{fig:srecoil}, we show in Fig.~\ref{fig:stotal}
this distribution for $T=10~{\rm MeV}$ and an initial-state neutrino
energy $\epsilon_1=30~{\rm MeV}$ for several values of the assumed
spin relaxation rate $\gamma$. We note that $\gamma=1.0$ corresponds
roughly to a density of $10^{14}~\rm g~cm^{-3}$. Therefore, while the
bremsstrahlung-related broadening of the recoil structure function can
be quite significant, this is only the case for relatively large
densities.


\section{Numerical Code}

\label{sec:NumericalCode}

In order to study $\nu_\mu$ and $\nu_\tau$ transport we have written a
simple Monte Carlo code to solve the Boltzmann collision equation for
a plane-parallel geometry.  We prescribe a power-law density and
temperature profile, i.e.\ in contrast to full-scale SN codes there
are no ``radial'' zones. The medium consists of one species of
non-degenerate nucleons with an adjustable mass.  The only microscopic
processes are $\nu N$ scattering and $NN$ bremsstrahlung.  We treat
bremsstrahlung as described in Appendix~\ref{sec:Bremsstrahlung}. In
particular, to calculate the mfp against inverse bremsstrahlung
absorption $\nu\bar\nu NN\to NN$ we assume that the other neutrino is
in thermal equilibrium.  In order to include energy transfer in $\nu
N$ collisions we use the dynamical structure function described in
Appendix~\ref{sec:RecoilScatteringCombined} where the effect of
nucleon recoil and inelastic $\nu N$ scatterings are combined.  We can
switch recoil and inelastic scattering independently on or off.

To solve for the neutrino distribution function we follow the
trajectory of a single neutrino, allowing for random collisions or
absorptions. Whenever the neutrino is lost through the surface, into
the blackbody surface at the bottom, or when it has been absorbed by
inverse bremsstrahlung, a new neutrino is launched. Its starting point
can be somewhere in the medium according to the source strength of the
$NN$ bremsstrahlung process, or it can be launched at the bottom from
the blackbody surface.  If $NN$ bremsstrahlung is switched off, the
latter is the only option. When bremsstrahlung is on, we can switch
the blackbody source off, creating all neutrinos within the medium, or
we may replace neutrinos absorbed by the blackbody surface with
neutrinos emitted by it.  When $NN$ bremsstrahlung is on, the
neutrinos achieve full thermal equilibrium below the energy sphere;
the treatment of the lower boundary condition is then found to be
irrelevant as long as it is some distance from the energy sphere.

This approach makes it impractical to include neutrino phase-space
blocking effects as one would need the distribution function as input
information for the scattering rates. Since phase-space blocking
effects are not important for $\mu$- and $\tau$-neutrinos in the
atmosphere of a SN core, we have opted for ignoring Pauli blocking
entirely and to treat neutrinos with Maxwell-Boltzmann statistics. Put
another way, the equilibrium distribution found by our code is
$e^{-\epsilon/T}$, not a Fermi-Dirac distribution.

As the neutrinos propagate, we sample their energy and direction
whenever they cross any of a set of sampling radii that were chosen at
the beginning of the run.  This way we accumulate statistics for the
distribution function and several of its angular and energy moments at
the sampling radii.


\end{document}